\documentclass[acmsmall,nonacm]{acmart}
\setcopyright{none}
\setcitestyle{numbers,sort}

\AtBeginDocument{%
  }

\usepackage{amsmath,amssymb,amsfonts,amsthm,mathtools,bm}
\usepackage{booktabs}
\usepackage{multirow}
\usepackage{array}
\usepackage{graphicx}
\graphicspath{{./}}
\usepackage{subcaption}
\usepackage{algorithm}
\usepackage{algpseudocode}
\usepackage{enumitem}
\usepackage{tikz}
\usetikzlibrary{
  arrows.meta,
  backgrounds,
  calc,
  fit,
  positioning,
  decorations.pathreplacing,
}
\usepackage{quantikz}
\tikzset{open/.style={fill=white}}

\usepackage{xcolor}
\usepackage{float}

\ifPDFTeX\else
\renewcommand{\bm}[1]{\symbfit{#1}}
\fi

\hypersetup{
  colorlinks=true,
  linkcolor=blue!60!black,
  citecolor=blue!60!black,
  urlcolor=blue!60!black
}
\newcommand{\appWorkflowDescription}{Appendix~\ref{app:workflow_description}}
\newcommand{\appVQAFamily}{Appendix~\ref{sec:vqa_family}}
\newcommand{\appPropagator}{Appendix~\ref{sec:qsp_block_greens_appendix}}
\newcommand{\appOffline}{Appendix~\ref{sec:offline}}
\newcommand{\appResults}{Appendix~\ref{sec:results_discussion}}
\newcommand{\appResultsNum}{Appendix~\ref{sec:results_discussion}}
\newcommand{\appVQAParamNum}{Appendix~\ref{app:vqa_parameter_summaries}}
\newcommand{\appVQAPosteriorNum}{Appendix~\ref{app:vqa_posterior_engine_sensitivity}}
\newcommand{\appVQABenchmark}{Appendix~\ref{app:vqa_benchmark_supplement}}
\newcommand{\appResourceDetail}{Appendix~\ref{sec:resource_estimation}}

\newcommand{\C}{\mathbb{C}}
\newcommand{\ii}{\mathrm{i}}
\newcommand{\ee}{\mathrm{e}}

\newcommand{\Tr}{\mathrm{Tr}}

\newcommand{\norm}[1]{\left\lVert #1 \right\rVert}
\newcommand{\abs}[1]{\left| #1 \right|}
\providecommand{\ket}[1]{\left| #1 \right\rangle}
\providecommand{\bra}[1]{\left\langle #1 \right|}

\providecommand{\expect}[1]{\left\langle #1 \right\rangle}

\definecolor{cBlue}{HTML}{2B6CB0}
\definecolor{cOrange}{HTML}{E65100}
\definecolor{cPurple}{HTML}{6A1B9A}
\definecolor{cRed}{HTML}{C62828}
\definecolor{cGreen}{HTML}{2E7D32}
\definecolor{cTeal}{HTML}{00695C}
\definecolor{cBrown}{HTML}{4E342E}
\definecolor{cGray}{HTML}{616161}
\definecolor{cIndigo}{HTML}{283593}

\tikzset{
  module/.style={
    draw=#1, line width=2.1pt, rounded corners=6pt,
    fill=#1!6, inner sep=8pt
  },
  head/.style={
    font=\sffamily\bfseries\small, text=#1
  },
  step/.style={
    draw=#1!75, line width=0.6pt, rounded corners=3pt,
    fill=white, inner sep=4pt,
    minimum width=3.5cm, minimum height=0.75cm,
    align=center, font=\sffamily\scriptsize
  },
  stepwide/.style={
    draw=#1!75, line width=0.6pt, rounded corners=3pt,
    fill=white, inner sep=4pt,
    minimum width=4.2cm, minimum height=0.75cm,
    align=center, font=\sffamily\scriptsize
  },
  chip/.style={
    draw=cGray!80, line width=0.5pt, rounded corners=2pt,
    fill=cGray!8, inner sep=3pt, font=\sffamily\tiny
  },
  arr/.style={
    -{Stealth[length=5pt,width=4pt]},
    line width=0.9pt, color=cGray!80
  },
  arrstrong/.style={
    -{Stealth[length=6pt,width=5pt]},
    line width=1.2pt, color=#1
  },
  arrdash/.style={
    -{Stealth[length=5pt,width=4pt]},
    line width=0.8pt, dashed, color=#1
  },
  arrlbl/.style={
    font=\sffamily\tiny, fill=white, inner sep=1pt, text=cGray!90
  },
  phase/.style={
    font=\sffamily\bfseries\footnotesize, text=#1
  },
  note/.style={
    font=\sffamily\tiny, text=cGray!90, align=center
  },
  brace/.style={
    decorate, decoration={brace, amplitude=6pt, raise=2pt},
    line width=0.8pt, color=#1
  }
}
\title[]{Unified Uncertainty Quantification Framework Bridging Noisy Quantum Backends Across Variational Quantum Algorithms and Quantum Signal Processing}

\author{Priyabrata Senapati}
\authornote{Corresponding author.}
\affiliation{%
  \institution{Pacific Northwest National Laboratory}
  \city{Richland}
  \state{Washington}
  \country{USA}}
\additionalaffiliation{%
  \institution{Kent State University}
  \city{Kent}
  \state{Ohio}
  \country{USA}}
\email{priyabrata.senapati@pnnl.gov}
\email{psenapat@kent.edu}  

\author{Vibin Abraham}
\email{vibin.abraham@pnnl.gov}
\affiliation{%
  \institution{Pacific Northwest National Laboratory}
  \city{Richland}
  \state{Washington}
  \country{USA}}

\author{Qiang Guan}
\authornote{Corresponding author.}
\email{qguan@kent.edu}
\affiliation{%
  \institution{Kent State University}
  \department{Department of Computer Science}
  \city{Kent}
  \state{Ohio}
  \country{USA}}

\author{Bo Peng}
\authornote{Corresponding author.}
\email{peng398@pnnl.gov}
\affiliation{%
  \institution{Pacific Northwest National Laboratory}
  \city{Richland}
  \state{Washington}
  \country{USA}}

\begin{document}

\begin{abstract}

We present an uncertainty quantification (UQ) framework for application level benchmarking and characterization of noisy quantum backends. The framework compares two workload classes under one statistical pipeline: noisy intermediate scale quantum (NISQ) variational quantum algorithms (VQAs) and Quantum Singular Value Transformation (QSVT) based Green's function reconstruction. For the VQA branch, we evaluate ten benchmark families spanning chemistry, optimization, simulation, compiling, linear solving, partial differential equations, metrology, error correction, tomography, and channel fidelity estimation. For the QSVT branch, we reconstruct orbital resolved Green's functions and spectral peaks from a block encoded real time propagator. The workflow combines Bayesian optimization, posterior distribution refinement, sensitivity analysis, robust parameter density estimation, backend ranking, noise correlation, and resource estimation analysis. Instead of reporting only one best parameter vector, the framework identifies robust parameter regions, residual gaps to ideal behavior, backend specific failure modes, and calibration sensitive uncertainty. The result is a common benchmark for variational and non-variational workloads that measures how reliably each backend reaches useful task level behavior.

\end{abstract}



\maketitle

\section{Introduction}
\label{sec:intro_restructured}

Quantum hardware quality is often reported through hardware level diagnostics.
Common examples include gate fidelity, randomized benchmarking, quantum volume,
and quantum characterization, verification, and validation (QCVV)
protocols~\cite{cross2019qv,proctor2022mirror,blumekohout2020volumetric,hashim2025qcvv,blumekohout2025qcvv}.
These metrics characterize the device. They do not directly answer the
question a scientific user asks: for a concrete workload, can a backend produce
outputs whose uncertainty is small enough to support interpretation? Application
oriented benchmark suites make this point clearly
~\cite{lubinski2021application,tomesh2022supermarq}. A unified statistical
framework for both near term variational workloads and matrix function
workloads is still missing.

We address this gap with a backend aware UQ framework. The framework treats two workload classes as application level probes
of noisy hardware. The first class is VQAs,
which dominate near term quantum computing on NISQ devices~\cite{peruzzo2014vqe,farhi2014qaoa,cerezo2021vqa,mcclean2016theory}.
VQA performance depends strongly on the objective. A backend that supports one
ansatz or loss landscape well may fail on another. A single best objective value
therefore hides the structure needed for hardware comparison.

The second class is a non-variational algorithm type which is a matrix function execution through QSVT~\cite{gilyen2019qsvt,low2017hamiltonian}. We study this
branch through Green's function reconstruction
~\cite{endo2020greens,abraham2026core}. Unlike a variational ansatz, the QSVT
circuit has a fixed polynomial structure controlled by signal processing phase
angles. Its benchmark target is a physically meaningful spectral feature.
Reading QSVT alongside the VQA family tests whether the same backend can support
both variational optimization and matrix function execution.

The common thread is UQ
~\cite{sullivan2015uq,xu2022uqhardware,senapati2025uqvarqa, senapati2026uqvqe}. For both VQAs and
QSVT, our framework records each backend evaluation as a pair
$(\bm{\xi},y)$. Here $\bm{\xi}$ is the parameter vector of the quantum algorithms used in this work, and $y$ is the noisy
task level output of those quantum algorithms. The framework characterizes the full evaluation history,
not only the final point. A Gaussian process (GP) surrogate guides the search.
Posterior distribution of parameters refinement adds controlled stochastic exploration near acquisition
maxima. Global sensitivity analysis identifies the coordinates that control
backend success. Density estimation describes the robust parameter region. A
calibration aware benchmarking layer reports backend rankings together with
compiled resource cost. The resulting benchmark measures how often, how
quickly, and how stably a backend reaches useful task level behavior.

\subsection{Literature Review and Positioning}
\label{sec:positioning_restructured}
Our work connects four bodies of work: VQAs and uncertainty aware backend
benchmarking, QCVV and application oriented benchmark suites, statistical UQ
and Bayesian optimization, and quantum algorithms for Green's functions. We
review them in this order because the paper first defines the VQA branch, then
uses the same UQ machinery for the QSVT Green's function branch.

\textit{(i) VQAs and UQ for backend benchmarking.} VQAs now span chemistry,
combinatorial optimization, simulation, linear solving, compiling, metrology,
tomography, and error correction style
tasks~\cite{peruzzo2014vqe,farhi2014qaoa,cerezo2021vqa,khatri2019qaqc,yuan2019vqs,bravoprieto2023vqls,sato2021vqapde,meyer2021vqmet,xu2021vqec,xue2023vqptnu,cerezo2020vqfe,kandala2017hardware,mitarai2018qcl,cao2019quantum,tilly2022vqe}.
This literature motivates our workload set, but it also warns against treating
one variational task as a universal device proxy. A backend can behave well for
one ansatz or loss and fail on another. Trainability makes the problem sharper:
barren plateaus, cost concentration, noise induced flatness, and the formal
hardness of VQA training all reshape the optimization landscape seen by a noisy
backend~\cite{mcclean2016theory,cerezo2021costdep,bittel2021training,wang2021noise,holmes2022connecting,larocca2025barren}.
Recent studies have started to quantify this behavior through sensitivity
analysis, reproducibility audits, predictive modeling, visualization, and UQ of
VQA outputs on IBM style backends~\cite{xu2022uqhardware,senapati2023reproducibility,senapati2023visualization,senapati2024pqml,senapati2025uqvarqa,senapati2025visualization}.
Our VQA branch builds on that line of work by putting ten workload families
inside one shared backend aware score.

\textit{(ii) QCVV and application oriented benchmarking.} Randomized
benchmarking and cycle benchmarking established statistical tools for hardware
fidelity estimation~\cite{knill2008rb,magesan2011rb,erhard2019cycle}. Gate set
tomography adds a predictive characterization layer through self consistent
process maps~\cite{nielsen2021gst}. Quantum volume, mirror circuit benchmarks,
volumetric benchmarks, and large device scaling studies diagnose hardware and
compiler behavior~\cite{cross2019qv,proctor2022mirror,blumekohout2020volumetric,wright2019benchmarking,mooney2021qv}.
Application oriented benchmark suites and certification surveys then move the
comparison closer to workload performance~\cite{lubinski2021application,tomesh2022supermarq,eisert2020certification}.
Recent QCVV overviews organize this broader landscape across randomized,
tomographic, predictive, and system level protocols~\cite{hashim2025qcvv,blumekohout2025qcvv}.
These protocols answer essential device level questions. Our paper asks a more
specific application level question: how often a backend reaches a good region
for a scientific workload, how stable that region is, and which parameters and
noise features explain the result.

\textit{(iii) UQ, Bayesian optimization, and posterior inference.} The online
and offline parts of our workflow draw on standard UQ and statistical learning
tools~\cite{sullivan2015uq}. Gaussian process regression supplies the surrogate
model~\cite{rasmussen2006gp}; Bayesian optimization provides the acquisition
logic for expensive noisy evaluations~\cite{snoek2012practical}; trust region
ideas stabilize search in higher dimensional parameter spaces~\cite{eriksson2019turbo};
and variational inference provides a scalable posterior refinement route~\cite{blei2017variational}.
We use these methods as part of a backend characterization loop rather than as
standalone optimizer choices. The recorded evaluations feed sensitivity analysis,
robust region density estimation, backend ranking, and noise association, so
optimization becomes one stage of the benchmark rather than the final answer.

Quantum optimal control is a mature discipline, from broad surveys of open and
closed loop control~\cite{dong2010quantum} to analyses of why optimal controls
are often easy to locate on quantum control landscapes~\cite{chakrabarti2007quantum}.
Within it, Bayesian optimization has become a standard data efficient tool,
where a Gaussian process surrogate searches control parameters to reach a target
state under finite, noisy data. This Bayesian control line spans analog state
preparation in cold atom and optical lattice systems~\cite{mukherjee2020bayesian,xie2022bayesian},
optimal control under poor measurement statistics on NISQ hardware~\cite{sauvage2020optimal},
high dimensional multi fidelity control field design~\cite{lazin2023high}, and
robustness aware state preparation with local parameter sensitivity~\cite{blatz2024bayesian}.
We adopt a similar surrogate guided engine, but we add a posterior refinement
step and apply the loop for a different purpose. Rather than return a single
optimal control, the loop acts as a backend characterization instrument whose
evaluation history feeds global sensitivity, robust region, ranking, and resource
analyses across both variational \& non-variational workloads. We refine each
Bayesian optimization candidate with variational inference because it gives a
stable and scalable posterior approximation over the high performing parameters
and produces the robust region density summaries the benchmark uses, while
Metropolis-Hastings and Langevin variants serve as robustness checks. The control
studies above tune one protocol on a single simulated system and, where they
examine the parameters at all, use local sensitivity to prioritize calibration;
our loop instead compares the same hardware backends and runs global variance
based sensitivity analysis over the parameters, together with backend ranking and
routed resource cost. A trust region constraint in the spirit of TuRBO keeps this
guided search stable enough to act as a measurement instrument rather than a
single point optimizer, an extension the optimization only control studies above
do not employ.

\textit{(iv) Quantum algorithms for Green's functions.} Quantum Green's
function algorithms have developed through variational, real time propagation,
Lehmann representation, equation of motion, and block encoding routes. Early
near term proposals showed that retarded Green's functions can be estimated
through real time dynamics or excited state information~\cite{endo2020greens,kowalski2024manybody}.
Related work recovered quasiparticle spectra from directly measured Green's
functions~\cite{kosugi2020greens}, used Lanczos and cumulant strategies for
Green's functions and ground states~\cite{baker2021lanczos,greenediniz2024cumulant},
developed coupled cluster and hybrid quantum classical formulations~\cite{keen2022ccgf},
and used adaptive variational dynamics for real time many body Green's
functions~\cite{gomes2023greens}. Other work improved time domain propagation
for near term processors~\cite{libbi2022effective}, introduced local
variational quantum compilation for lattice Green's functions~\cite{kanasugi2023lvqc},
formulated equation of motion and imaginary time Green's function algorithms~\cite{rizzo2022eom,dhawan2024jctc},
and targeted spectroscopy applications such as X-ray absorption and electron
energy loss~\cite{fomichev2025xas,kunitsa2025eels}. A complementary study by
several of the present authors connects real time many body core spectra to a
fault tolerant quantum signal processing (QSP)/QSVT route for core hole Green's functions~\cite{abraham2026core}.
These studies establish Green's functions as meaningful quantum algorithm
outputs. We use the Green's function itself as the object of backend
benchmarking: the score depends on spectral peak recovery, phase sensitivity,
robust phase region geometry, and routed circuit cost.

Together, these strands define the position of this paper. VQA studies provide
a broad family of noisy variational workloads. QCVV and application benchmarks
provide device and workload diagnostics. UQ and Bayesian optimization provide
the statistical engine. Green's function algorithms provide a structured
non-variational spectral target. We combine them in one workflow so the VQA
branch and the QSVT branch become two instances of the same backend aware UQ
problem rather than two separate case studies.
Table~\ref{tab:lit_positioning_restructured} condenses this positioning by
listing, for each strand, what prior work establishes and what our
integrated benchmark adds on top of it.

\begin{table}[t]
\centering
\caption{Previous studies that motivate the proposed integrated benchmark.}
\label{tab:lit_positioning_restructured}
\begin{tabular}{>{\raggedright\arraybackslash}p{0.2\linewidth} >{\raggedright\arraybackslash}p{0.35\linewidth} >{\raggedright\arraybackslash}p{0.35\linewidth}}
\toprule
\textbf{Literature} & \textbf{Prior work shows} & \textbf{This work adds}\\
\midrule
VQAs and UQ~\cite{peruzzo2014vqe,farhi2014qaoa,cerezo2021vqa,kandala2017hardware,mitarai2018qcl,cao2019quantum,tilly2022vqe,mcclean2016theory,cerezo2021costdep,bittel2021training,wang2021noise,holmes2022connecting,larocca2025barren,xu2022uqhardware,senapati2023reproducibility,senapati2024pqml,senapati2025uqvarqa} & Diverse VQA families, trainability barriers, and recent UQ of VQA outputs on IBM style backends. & Ten VQA families inside one tuple framework with shared sensitivity, density, ranking, and resource layers across the same four backends.\\ && \\
QCVV and benchmarking~\cite{knill2008rb,magesan2011rb,erhard2019cycle,nielsen2021gst,cross2019qv,proctor2022mirror,blumekohout2020volumetric,wright2019benchmarking,mooney2021qv,lubinski2021application,tomesh2022supermarq,eisert2020certification,hashim2025qcvv,blumekohout2025qcvv} & Randomized benchmarking, cycle benchmarking, gate set tomography, volumetric and mirror circuit diagnostics, large device scaling studies, and application oriented benchmark suites. & Application level UQ with robust region geometry, workload selective rankings, sensitivity fingerprints, residual to ideal alignment, and routed resource cost.\\  && \\
UQ and optimization~\cite{sullivan2015uq,rasmussen2006gp,snoek2012practical,eriksson2019turbo,blei2017variational} & Surrogate modeling, Bayesian optimization, trust region search, and posterior inference for expensive noisy functions. & A closed backend characterization loop in which optimization feeds sensitivity analysis, density estimation, ranking, and noise association.\\ && \\
Quantum Green's function algorithms~\cite{endo2020greens,kosugi2020greens,baker2021lanczos,keen2022ccgf,gomes2023greens,libbi2022effective,rizzo2022eom,dhawan2024jctc,greenediniz2024cumulant,kanasugi2023lvqc,kowalski2024manybody,fomichev2025xas,kunitsa2025eels,abraham2026core} & Real time, Lanczos, equation of motion, imaginary time, local variational compilation, hybrid, adaptive variational, and QSP/QSVT routes for Green's functions and spectroscopy. & Green's function reconstruction as a UQ benchmark with spectral reliability, phase sensitivity, robust region density, and routed compilation cost.\\ && \\
Bayesian quantum control~\cite{mukherjee2020bayesian,xie2022bayesian,sauvage2020optimal,lazin2023high,blatz2024bayesian} & Gaussian process Bayesian optimization that finds control pulses or parameters to prepare target states under noisy, expensive data across analog and NISQ systems, including robustness aware protocols with local parameter sensitivity. & Similar surrogate engine used as a characterization instrument, with a trust region and posterior refinement with MCMC loop whose evaluation history drives global sensitivity, robust region density, ranking, and routed resource analysis and backend comparison across variational and non-variational workloads.\\
\bottomrule
\end{tabular}
\end{table}

\subsection{Contributions}

Our work makes four contributions. First, we cast variational and matrix
function quantum workloads into a single backend aware UQ workflow: every
backend evaluation is recorded as a $(\bm{\xi},y)$ tuple, and the offline
analysis pipeline (Bayesian optimization with posterior refinement,
sensitivity analysis, density estimation, backend ranking, and calibration
alignment) operates on the resulting histories without branch specific
modifications. Second, we instantiate the template on ten VQA families
(VQE, QAOA, VQC, VQTE, VQLS, VQAPDE, VQAMET, VQEC, VQPTNU, and VQCFE) and on
the QSVT construction of the Green's function of a hydrogen molecule with a 27 dimensional phase
vector, evaluated on the same four IBM quantum backends (Brisbane, Kawasaki, Kyoto, and Osaka). The QSVT path is mathematically grounded in block
encoding, QSVT phase synthesis, propagator degree segmentation, and an
explicit workflow consistency error decomposition. Third, we report
backend characterizations that go beyond single number rankings such as quantum volume (QV) or mirror quantum volume (MQV) within the QCVV framework. These characterizations include
backend rankings that vary across workloads, sensitivity fingerprints that identify dominant control coordinates per backend, density level
geometry of the robust parameter region, and a compilation cost layer that distinguishes workloads whose cost is dominated by one heavy circuit (VQE, QAOA) from workloads whose cost is dominated by many lighter characterization circuits (VQEC, VQCFE, VQPTNU). Fourth, by reading the VQAs and QSVT branches together we expose agreement and disagreement patterns that neither paradigm alone can produce. Here, agreement
signals backend behavior, while disagreement reveals workload selectivity and warns against using any single benchmark as a universal proxy.

\begin{figure}[t]
\centering
% \resizebox{\textwidth}{!}{\input{figures/uq_workflow_figure.tex}}
% \resizebox{\textwidth}{!}{\input{figures/uq_workflow_figure_compact.tex}}
\resizebox{\textwidth}{!}{\input{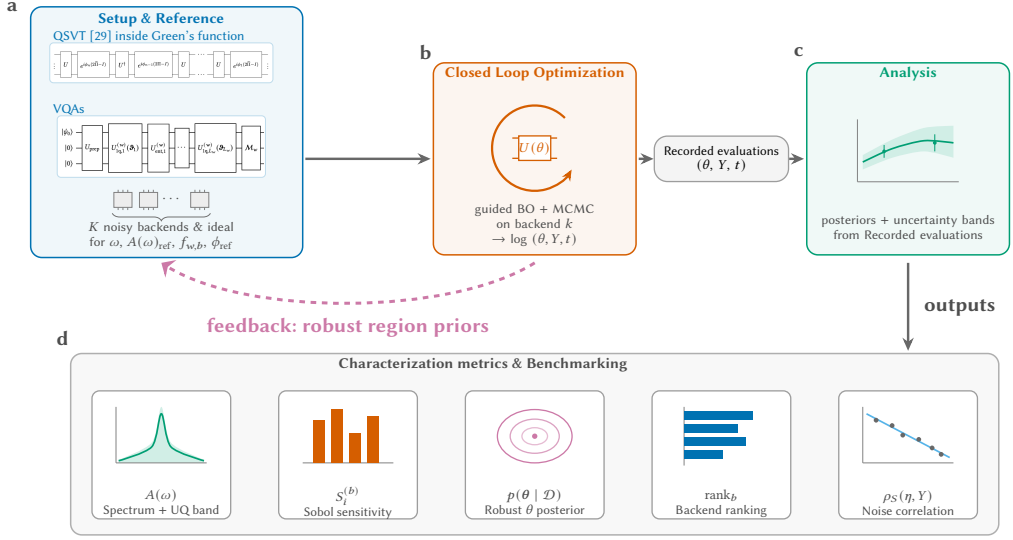}}
\caption{Shared UQ workflow for application level benchmarking across VQAs and QSVT Green's function construction. The pipeline starts from workload and backend setup for both VQAs and QSVT as shown in panel (a), performs backend aware closed loop optimization in (b), performs uncertainty analysis in (c), and produces spectra, sensitivity fingerprints, robust parameter regions, backend rankings, and noise performance correlations in (d). This common structure lets variational and non-variational workloads be compared under one UQ driven benchmarking framework. A phase by phase walkthrough of the diagram, including
the saved workflow outputs contract and the offline branches, appears in
\appWorkflowDescription{} together with the full workflow rendering of
Figure~\ref{fig:workflow_appendix}.}
\label{fig:workflow_main_restructured}
\end{figure}

Figure~\ref{fig:workflow_main_restructured} summarizes the shared workflow that
connects the VQA and Green's/QSVT branches to the same saved evaluation records
and offline UQ products.

\section{Unified Benchmarking Formulation}
\label{sec:formulation_restructured}
Both the VQA and the QSVT branch need a single statistical interface
so that backend comparisons are not confounded by branch specific notation. We
define that interface in three steps. We first state the backend conditioned
VQA objective, then the Green's function and its QSVT realization, and finally
the shared tuple model on which the UQ engine acts. Throughout the rest of the paper, every figure and every benchmark score is built from one of three primitives defined here: the backend conditioned VQA objective $f_{w,b}$, the Green's/QSVT spectral observable, and their output tuples $(\bm{\xi}, y)$.

\subsection{VQA branch}
For workload $w$ and backend $b$, the central variational quantity $f_{w,b}$ is the
objective induced on backend $b$,
\begin{equation}
  f_{w,b}(\bm{\vartheta})
  =
  \Tr\!\left[
    O_w\,\mathcal{E}_b\!\left(
      U_w(\bm{\vartheta})\rho_0 U_w(\bm{\vartheta})^\dagger
    \right)
  \right],
  \label{eq:vqa_main_restructured}
\end{equation}
where $\bm{\vartheta}\in\mathbb{R}^{p_w}$ is the trainable parameter vector of
workload $w$, $U_w(\bm{\vartheta})$ is the corresponding ansatz unitary,
$\rho_0$ is the prepared initial state, $\mathcal{E}_b$ is the quantum channel that backend $b$ imposes on the ideal
unitary (it absorbs transpilation, routing, sampling, decoherence, and
readout), and $O_w$ is the task observable (e.g., a molecular
Hamiltonian for VQE or a cost Hamiltonian for QAOA). Equation~\eqref{eq:vqa_main_restructured}
feeds the optimization loop in the tuple form
$y_n = -f_{w,b}(\bm{\vartheta}_n) + \varepsilon_n$, so that minimization of
the noisy task quantity becomes maximization of $y_n$ in the surrogate
defined in \S\ref{sec:formulation_restructured}.

We evaluate ten VQA families through this same template:
VQE, QAOA, VQC, VQTE, VQLS, VQAPDE, VQAMET, VQEC, VQPTNU, and VQCFE.
Table~\ref{tab:vqa_family_compact_restructured} lists what each family
optimizes; \appVQAFamily{} gives the per workload mathematics
and benchmark instances (Table~\ref{tab:vqa_instances_main} and the shared
circuit template of Figure~\ref{fig:vqa_template_main}), the compact circuit structures of each of the VQAs, and objective map
(Tables~\ref{tab:vqa_objective_map_appendix}
and~\ref{tab:vqa_compact_circuit_atlas}).

\begin{table}[t]
\centering
\caption{Ten VQA families covered by our benchmark. In this study, ``VQC''
refers to variational quantum compiling rather than to supervised
classification.}
\label{tab:vqa_family_compact_restructured}
\begin{tabular}{>{\raggedright\arraybackslash}p{0.14\linewidth} >{\raggedright\arraybackslash}p{0.78\linewidth}}
\toprule
\textbf{Workload} & \textbf{Main task family} \\
\midrule
VQE\cite{peruzzo2014vqe} & Ground state energy minimization in chemistry or spin models \\
QAOA\cite{farhi2014qaoa} & Combinatorial optimization via alternating cost and mixer layers \\
VQC\cite{khatri2019qaqc} & Variational quantum compiling / unitary synthesis \\
VQTE\cite{yuan2019vqs} & Variational time evolution and simulation \\
VQLS\cite{bravoprieto2023vqls} & Linear system solving \\
VQAPDE\cite{sato2021vqapde} & Partial differential equation solving through variational discretizations \\
VQAMET\cite{meyer2021vqmet} & Metrology and probe response optimization \\
VQEC\cite{xu2021vqec} & Variational code or encoder design for error correction \\
VQPTNU\cite{xue2023vqptnu} & Non-unitary process tomography \\
VQCFE\cite{cerezo2020vqfe} & Fidelity and channel characterization \\
\bottomrule
\end{tabular}
\end{table}

\subsection{QSVT/Green's Function branch}

Following the Green's function via quantum circuits route used in recent core
spectra studies~\cite{abraham2026core}, we estimate the `retarded' Green's function
$G^R_{jj}(\omega)$ by reconstructing its real time companion $G^{(h)}_{jj}(t)$
on a backend, damping it exponentially, and Fourier transforming to the
frequency axis. Let $H$ be the many body Hamiltonian and let $\ket{\Psi_0}$ denote the ground state with energy $E_0$. The retarded single particle Green's function is
\begin{equation}
  G^R_{ij}(t)
  =
  -\ii \,\Theta(t)\,
  \bra{\Psi_0}
  \left\{
    c_i(t), c_j^\dagger(0)
  \right\}
  \ket{\Psi_0},
  \label{eq:retarded_green}
\end{equation}
where $c_i$ and $c_j^\dagger$ are fermionic annihilation and creation operators and
\begin{equation}
  c_i(t) = \ee^{\ii H t} c_i \ee^{-\ii H t}.
\end{equation}
These definitions are standard in many body Green's function theory
\cite{mahan2000many,fetterwalecka}.
For occupied orbitals, the hole sector is especially relevant, and one may write the corresponding time domain contribution as
\begin{equation}
  G^{(h)}_{jj}(t)
  =
  -\ii\,
  \bra{\Psi_0}
  c_j^\dagger \ee^{-\ii (H-E_0 I)t} c_j
  \ket{\Psi_0},
  \qquad t \ge 0.
  \label{eq:hole_green}
\end{equation}

In the hole sector the time domain Green's function reads
\begin{equation}
  G_{jj}^{(h)}(t)
  =
  -\ii\,\ee^{\ii E_0 t}
  \bra{\Phi_j}\ee^{-\ii H t}\ket{\Phi_j},
  \qquad
  \ket{\Phi_j}=c_j\ket{\Psi_0},
  \qquad t\geq 0,
  \label{eq:green_main_restructured}
\end{equation}

where $H$ is the molecular Hamiltonian, $\ket{\Psi_0}$ and $E_0$ are its
ground state and the corresponding energy, $c_j$ removes an electron from orbital $j$,
and $\ket{\Phi_j}=c_j\ket{\Psi_0}$ is the resulting hole state. The phase
factor $\ee^{\ii E_0 t}$ projects out the ground state energy so that the
Fourier transform of $G^{(h)}_{jj}(t)$ peaks at the physical ionization
energies relative to $E_0$.

We realize the propagator $\ee^{-\ii H t}$ on a backend through a QSVT
pipeline~\cite{low2017hamiltonian,gilyen2019qsvt,low2019qubitization,martyn2021grand,dong2022qet}.
Let $U_H$ be a unitary block encoding of $H$ on $a$ ancilla qubits,
\begin{equation}
  (\bra{0^a}\otimes I)\,U_H\,(\ket{0^a}\otimes I)=\frac{H}{\lambda},
  \label{eq:block_main_restructured}
\end{equation}
with $I$ the identity on the system register and $\lambda\geq\norm{H}$ the
block encoding normalization that keeps $H/\lambda$ in the unit operator
norm ball. A polynomial transform of $H/\lambda$ then approximates the
propagator,
\begin{equation}
  \widetilde{E}_d(t)
  =
  P_d\!\left(\frac{H}{\lambda}\right)
  -\ii\, Q_d\!\left(\frac{H}{\lambda}\right)
  \approx
  \ee^{-\ii H t},
  \label{eq:qsvt_main_restructured}
\end{equation}
where $P_d$ and $Q_d$ are the QSP defined even and odd Chebyshev polynomials
of degree $d$, jointly parameterized by the phase vector
$\bm{\theta}\in\mathbb{R}^{d+1}$ used as the QSVT branch's control
variable. For propagation times $t$ at which a single QSP polynomial of
degree $d=\Theta(\lambda t + \log(1/\varepsilon))$ would be impractical, we
segment the total scaled time $\tau=\lambda t$ into
$r=\lceil\tau/\tau_{\max}\rceil$ equal pieces of length $\tau/r$, synthesize
the lower degree QSP polynomial $\widetilde{E}_d(\tau/r)$ once, and apply it
$r$ times in sequence; the degree segments tradeoff and the segment wise
error bound are shown in
\appPropagator. When the polynomial
$\widetilde{E}_d$ is composed with the block encoding $U_H$, the resulting
circuit is itself a linear combination of
unitaries~\cite{childs2012lcu}, and we synthesize and simulate it through
the PennyLane stack~\cite{bergholm2018pennylane}.

After reconstructing $G^{(h)}_{jj}(t)$ from Hadamard test style circuits over
a discrete time grid, we obtain the spectral function by exponential damping
and Fourier transform,
\begin{equation}
  A_j(\omega)
  =
  -\frac{1}{\pi}\,\Im \int_0^\infty
  \ee^{\,\ii \omega t-\eta t}\, G^{(h)}_{jj}(t)\,dt,
  \label{eq:spectral_main_restructured}
\end{equation}
with $\omega$ the spectral frequency and $\eta>0$ the broadening parameter that
controls the peak width. The present workflow uses discrete time samples $\{t_n\}_{n=0}^{N_t-1}$ and implements the Fourier transform numerically via FFT after exponential damping,
\begin{equation}
  \widetilde{G}_{jj}(\omega_m)
  \approx
  \sum_{n=0}^{N_t-1}
  G_{jj}(t_n)\,\ee^{-\eta |t_n|}\,\ee^{\ii \omega_m t_n}\,\Delta t.
  \label{eq:fft_discrete}
\end{equation}
The resulting peaks in $A_j(\omega_m)$ define the quantities that our framework benchmarks.
Concretely, the benchmark asks: on a given noisy backend, how reliably does the
QSVT circuit recover the peak at $\abs{\omega^\star} \approx 13.15\,\mathrm{eV}$
corresponding to the H$_2$ ionization energy?

From this spectrum we extract a peak tuple
\begin{equation}
  \bm{p}_{j,r}
  =
  \bigl(\omega_{j,r},A_{j,r}\bigr),
  \label{eq:peak_main_restructured}
\end{equation}
where $\omega_{j,r}$ is the frequency location and $A_{j,r}$ the height of
the $r$-th peak in orbital $j$. The backend's task is to recover the peak
tuples induced by Eqs.~\eqref{eq:green_main_restructured}-\eqref{eq:peak_main_restructured}
under its own noise channel, so we measure backend quality on this branch
against a physically interpretable spectrum rather than against a
circuit level fidelity proxy.

For the hydrogen molecule H$_2$ case study run on the same four IBM fake backends, we
instantiate this pipeline as follows. The system register holds four qubits
encoding the four orbital Hamiltonian. The block encoding $U_H$
of \eqref{eq:block_main_restructured} acts on the system register together
with three QSVT ancillas (a selector qubit, a real or imaginary projector
qubit, and the block encoding ancilla). The Hadamard test wrapper of
\eqref{eq:green_main_restructured} adds one control qubit and one
annihilation ancilla for $c_j$, giving a total circuit width of nine
qubits. The time grid samples $26$ equally spaced points up to $t=12.5$
atomic units, the broadening uses $\eta=0.15$, and the QSVT polynomial
approximation targets $\varepsilon\approx 10^{-4}$ at each segment. We
report the numerical resource cost of the compiled nine qubit circuit on
each backend in \S\ref{sec:results_restructured}.4, and the asymptotic
cost statement together with a projection to LiH at ten system qubits
appears in \S\ref{sec:greens_resource_extrapolation}. Appendix~\ref{sec:qsp_block_greens_appendix} collects the circuit level
estimator and block encoding details, with the estimator circuits drawn in
Figure~\ref{fig:alphaj_overview}.

\subsection{Shared UQ engine}
The VQA branch of \S\ref{sec:formulation_restructured}.1 and the QSVT
branch of \S\ref{sec:formulation_restructured}.2 meet at the level of the
objective history. Both produce, at each evaluation $n$, a noisy task
outcome that we model in the Bayesian optimization
form~\cite{snoek2012practical,shahriari2016bo,balandat2020botorch}
\begin{equation}
  y_n = g_b(\bm{\xi}_n) + \varepsilon_n,
  \qquad
  \varepsilon_n\sim\mathcal{N}(0,\sigma_n^2),
  \label{eq:generic_tuple_restructured}
\end{equation}
where $\bm{\xi}_n$ is the control vector at step $n$ (i.e., parameter
$\bm{\vartheta}$ for the VQA branch and the phase vector $\bm{\theta}$
for the QSVT branch), and $g_b$ is the scalar objective induced on
backend $b$ by Eq.~\eqref{eq:vqa_main_restructured} or the guided loss
defined below. Conceptually, $g_b$ already absorbs the backend's noise
channel $\mathcal{E}_b$ of Eq.~\eqref{eq:vqa_main_restructured} into its
definition,
$g_b(\bm{\xi})=\mathbb{E}\!\left[\text{noisy task outcome at }\bm{\xi}\text{ on backend }b\right]$,
so different backends define different $g_b$. The remaining stochasticity,
captured by $\varepsilon_n$, is the per evaluation fluctuation of the
empirical estimate around that backend induced expectation.

The true backend noise channel is generally non-Gaussian: shot statistics,
decoherence, readout error, and coherent biases all contribute, and we make
no distributional claim about that physical channel. Each recorded outcome
$y_n$ is instead an empirical estimate of $g_b(\bm{\xi}_n)$ over
$N_{\mathrm{shots}}$ measurements. By the central limit theorem, the
fluctuation of that empirical estimate around $g_b(\bm{\xi}_n)$ is
approximately Gaussian with a heteroscedastic variance $\sigma_n^2$ that
the optimizer estimates from the per evaluation shot statistics.
Equation~\eqref{eq:generic_tuple_restructured} records this effective
shot averaged Gaussian model that the GP surrogate uses to fit and update
its posterior in Eqs.~\eqref{eq:gp_main_restructured}-\eqref{eq:gp_post_restructured};
it is a regression side approximation, not a claim about the hardware
noise distribution itself.

For the VQA branch, the generic template specializes to workload
conditioned tuples
\begin{equation}
  y_n^{(w)}= J_{w,b}(\bm{\vartheta}_n)+\varepsilon_n,
  \label{eq:tuple_main_restructured}
\end{equation}
where $J_{w,b}$ is the workload's reported task quantity built from
$f_{w,b}$ (energy, residual, fidelity, etc.\ depending on workload $w$).  Let
\begin{equation}
  \bm{p}_{j,r}^{\star}
  =
  \left(
    \omega_{j,r}^{\star},
    A_{j,r}^{\star}
  \right)
\end{equation}
For QSVT, the backend objective $g_b$ instead specializes to the
guided peak loss
\begin{equation}
  \begin{aligned}
    f_b(\bm{\theta})
    &=
     -\sum_{(j,r)\in\mathcal{T}} w_{j,r}\, \ell\!\left(
      \widehat{\bm{p}}_{j,r}(\bm{\theta};b),
      \bm{p}_{j,r}
    \right),
  \end{aligned}
  \label{eq:guided_main_restructured}
\end{equation}
where $\mathcal{T}$ is the target peak set, $w_{j,r}\geq 0$ are per peak
weights, $\bm{p}_{j,r}$ is the ideal peak tuple from
Eq.~\eqref{eq:peak_main_restructured}, $\widehat{\bm{p}}_{j,r}(\bm{\theta};b)$
is the corresponding peak reconstructed under phases $\bm{\theta}$ on backend
$b$, and $\ell(\cdot,\cdot)$ is a position and height aware peak distance.

We model these tuples with a backend specific Gaussian process surrogate~\cite{rasmussen2006gp}
\begin{equation}
  g_b \sim \mathcal{GP}(m_b,k_b),
  \label{eq:gp_main_restructured}
\end{equation}
with prior mean function $m_b$ and covariance kernel $k_b$. After $n$
evaluations $\{(\bm{\xi}_i,y_i)\}_{i=1}^n$, we update the posterior mean by
the standard GP regression formula~\cite{rasmussen2006gp}
\begin{equation}
  \mu_n(\bm{\xi})
  =
  m_b(\bm{\xi}) +
  \bm{k}_n(\bm{\xi})^\top
  \bigl(K_n+\sigma_n^2 I\bigr)^{-1}
  \bigl(\bm{y}_n-\bm{m}_n\bigr),
  \label{eq:gp_post_restructured}
\end{equation}
where $K_n\in\mathbb{R}^{n\times n}$ is the kernel matrix on observed inputs
$\{\bm{\xi}_i\}_{i=1}^n$, $\bm{k}_n(\bm{\xi})\in\mathbb{R}^{n}$ is the cross
covariance vector from $\bm{\xi}$ to those inputs, $\bm{m}_n$ stacks the
prior means at observed inputs, $\bm{y}_n$ stacks the observed responses,
and $\sigma_n^2 I$ models the heteroscedastic observation noise. We apply parameter acquisition (for VQAs) or phase vector acquisition
(for QSVT), posterior refinement through variational
inference~\cite{blei2017variational} or stochastic gradient
Langevin dynamics~\cite{welling2011langevin}, and evaluation logging
in the same way on both branches. We implement
Eqs.~\eqref{eq:generic_tuple_restructured}-\eqref{eq:gp_post_restructured}
on top of the BoTorch framework~\cite{balandat2020botorch}, which supplies
the acquisition functions, the GP surrogate, and the posterior refinement
modes (variational inference, Metropolis-Hastings, and Langevin) used
throughout our evaluation. This common engine is the statistical reason our
framework can compare a variational task family and a QSVT spectral
reconstruction without pretending their underlying physics is the same.

Operationally, this same engine converts noisy backend evaluations into
parameter updates: it learns which variational parameters or perturbed phase
vectors best preserve the ideal VQA target or QSVT spectrum on a
given backend. The resulting tuples drive a single UQ stack: surrogate
guided search, posterior refinement, timestamped records, global sensitivity
analysis, density estimation for robust regions, backend ranking, and
optional alignment with calibration metadata.

Taken together, Eqs.~\eqref{eq:generic_tuple_restructured}-\eqref{eq:gp_post_restructured}
give the VQA family and the QSVT branch a common tuple interface,
and every downstream analysis in this paper (sensitivity, density, ranking,
resource) acts on that interface rather than on branch specific
reimplementations. Algorithm~\ref{alg:unified_uq_workflow} in
Appendix~\ref{app:workflow_description} states the full online and offline loop
built on this interface.

\subsection{Benchmark and characterization metrics}
\label{sec:metrics_restructured}
Before turning to the experimental design, we collect the metrics that
accompany every result in \S\ref{sec:results_restructured}. The two
branches use complementary metric families because their observables
differ in structure: the VQA branch produces a continuous task quantity
per evaluation, while the QSVT branch produces a binary good or
bad classification per evaluation. Both families ask the same question,
namely how often does the optimizer reach a good region and how quickly,
under the natural reduction for each observable type. We define the VQA
quality and ranking metrics first, then the QSVT reliability
metrics, then the shared phase sensitivity and robust region metrics,
and finally the routed resource metrics. Table~\ref{tab:metric_summary_restructured}
collects symbols and pointers as a quick lookup; the paragraphs below
carry the full interpretation.

\paragraph{VQA quality and ranking.}
We rescale each VQA workload's recorded task quantity into
$\mathrm{quality\_norm}\in[0,1]$ through a per workload min-max mapping,
with $1$ marking the best result and $0$ the worst across the four
backends. From this we derive the within workload ordinal
$\mathrm{rank\_in\_task}\in\{1,2,3,4\}$ as the position of each backend
after sorting by descending $\mathrm{quality\_norm}$ on that workload,
with $1$ denoting the best backend. The aggregate views collapse the ten
workloads per backend into three scalars: Mean Quality averages
$\mathrm{quality\_norm}$ across workloads (higher is stronger), Mean Rank
averages $\mathrm{rank\_in\_task}$ across workloads (lower is more
consistent), and Mean Eval95 averages the number of evaluations needed
to reach $95\%$ of the per workload best value (lower is more
efficient). A backend can lead Mean Quality without leading Mean Rank if
it dominates a few workloads while underperforming on others, so we
report both aggregates side by side. The $95\%$ threshold in Mean Eval95
marks the practical event of reaching the good regime while staying
robust to single evaluation noise around the per workload best.

\paragraph{QSVT reliability.}
We classify each QSVT evaluation as a \emph{good tuple} when the
guided peak loss in Eq.~\eqref{eq:guided_main_restructured} reaches its
lower target and as a \emph{bad tuple} otherwise. From this binary
classification we derive two reliability summaries per backend. Hit Rate
is the fraction of the $100$ evaluations classified as good tuples on a
backend, so a higher hit rate means the optimizer revisits the good
spectral regime more often. Time-to-Good is the evaluation index at
which the first good tuple is recorded on a backend, so a lower value
means the optimizer reaches the good regime sooner; we use Time-to-Good
as the secondary discriminator whenever two backends share the same Hit
Rate.

\paragraph{Shared phase sensitivity and robust region geometry.}
For phase sensitivity we report the Morris elementary effect mean
$\mu^*_i$~\cite{morris1991ee,saltelli2010sobol} per parameter coordinate
$i$, renormalized to sum to one across coordinates so that the top
contributors read as a sensitivity mass distribution. A high $\mu^*_i$
identifies a coordinate that moves the objective strongly, so the
optimizer has to control it carefully; a low $\mu^*_i$ marks a
coordinate that we can fix without changing backend behavior. For robust
region geometry we estimate the density of good tuples in the parameter
space~\cite{silverman1986kde} and define the robust region as the
$\alpha=0.9$ level set of that density. This convention matches the
highest density region used in posterior summarization and yields robust
regions that contain about $90\%$ of the density mass; a backend with a
narrow robust region (small level set volume) is selective in phase
space, while a backend with a broad robust region (large level set
volume) is forgiving to phase perturbations.

\paragraph{Routed resource cost.}
For resource cost we report three quantities that we compute after transpiling every
circuit to the shared four backends at Qiskit
\texttt{optimization\_level=3}. Mean compiled depth averages the per
circuit transpiled depth across the work and tracks how heavy one
circuit becomes once it is routed. Total two qubit gate count per
objective evaluation sums two qubit gates across every circuit produced
by one objective evaluation, which credits workloads that expand a
single evaluation into a family of measurement circuits rather than into
one deep circuit. The overhead factor is the ratio of the routed metric
to the ideal abstract metric for each backend and each polynomial
degree; an overhead factor of $5\times$ means routing inflated the
metric by a factor of five relative to the ideal compiled circuit.

\begin{table}[t]
\centering
\caption{Quick lookup for the metrics defined in
\S\ref{sec:metrics_restructured}. Branch column: V = VQA, G = QSVT,
R = resource. Morris and the density estimate are the main text
representatives; \appOffline{} reports additional sensitivity indices
($S_i$, $S_T$, FAST, Delta, SHAP) and additional density families
(Gaussian mixture, copula, and normalizing flow, fit with Pyro) for completeness.}
\label{tab:metric_summary_restructured}
\small
\setlength{\tabcolsep}{3pt}
\begin{tabular}{>{\raggedright\arraybackslash}p{0.21\linewidth} c >{\raggedright\arraybackslash}p{0.36\linewidth} >{\raggedright\arraybackslash}p{0.28\linewidth}}
\toprule
\textbf{Metric} & \textbf{Branch} & \textbf{Symbol / shorthand} & \textbf{Defined / used in}\\
\midrule
$\mathrm{quality\_norm}$ & V & $(x-x_{\min})/(x_{\max}-x_{\min})$ per workload & \S\ref{sec:metrics_restructured}, \S\ref{sec:results_restructured}.1\\
$\mathrm{rank\_in\_task}$ & V & $\mathrm{argsort}(\downarrow \mathrm{quality\_norm})$; $1=$ best & \S\ref{sec:metrics_restructured}, \S\ref{sec:results_restructured}.1, \S\ref{sec:results_restructured}.5\\
Mean Quality & V & $\overline{\mathrm{quality\_norm}}_{\,w}$; higher is stronger & \S\ref{sec:metrics_restructured}, \S\ref{sec:results_restructured}.5\\
Mean Rank & V & $\overline{\mathrm{rank\_in\_task}}_{\,w}$; lower is more consistent & \S\ref{sec:metrics_restructured}, \S\ref{sec:results_restructured}.1, \S\ref{sec:results_restructured}.5\\
Mean Eval95 & V & $\overline{\min\{n:y_n\geq 0.95\,y^\star\}}_{\,w}$; lower is faster & \S\ref{sec:metrics_restructured}, \S\ref{sec:results_restructured}.5\\
Hit Rate & G & $\#\{\text{good tuples}\}/100$; higher is more reliable & \S\ref{sec:metrics_restructured}, \S\ref{sec:results_restructured}.2, \S\ref{sec:results_restructured}.5\\
Time-to-Good & G & $\min\{n: y_n\text{ is a good tuple}\}$; lower is faster & \S\ref{sec:metrics_restructured}, \S\ref{sec:results_restructured}.2, \S\ref{sec:results_restructured}.5\\
Morris $\mu^*_i$ & V, G & Mean of $\abs{\Delta_i f}$ per coordinate, sum to one normalized~\cite{morris1991ee,saltelli2010sobol} & \S\ref{sec:metrics_restructured}, \S\ref{sec:results_restructured}.3\\
Density / robust region & V, G & $\alpha=0.9$ level set of good tuple density~\cite{silverman1986kde} & \S\ref{sec:metrics_restructured}, \S\ref{sec:results_restructured}.3\\
Mean compiled depth & R & $\overline{\text{depth}}$ over evaluations at \texttt{opt\_level=3} & \S\ref{sec:metrics_restructured}, \S\ref{sec:results_restructured}.4\\
Total 2q gates / eval & R & $\sum_c \#\mathrm{CX}_c$ over all circuits in one evaluation & \S\ref{sec:metrics_restructured}, \S\ref{sec:results_restructured}.4\\
Overhead factor & R & routed metric / ideal metric, per backend and per degree & \S\ref{sec:metrics_restructured}, \S\ref{sec:results_restructured}.4\\
\bottomrule
\end{tabular}
\end{table}

\section{Experimental Design}
\label{sec:design_restructured}

Three design choices control how our results should be read. First, we share
a single set of four backends across both branches so that
cross workload comparisons are not confounded by backend identity. Second,
we score every backend through the full recorded history of its evaluations
rather than through a single best point, so that the benchmark measures how
often good regions are reached and how stable they are. Third, we treat
transpilation, routing, and noise models as part of the workload rather than
as preconditions outside it, so that compiled cost and application level
reliability live on the same score.

\subsection{VQA setup}
We evaluate the ten VQA families of
Table~\ref{tab:vqa_family_compact_restructured} on the same four IBM fake
backends: Brisbane, Kawasaki, Kyoto, and Osaka. Each
workload runs the Bayesian optimization plus posterior refinement engine defined in
\S\ref{sec:formulation_restructured}.3. We aggregate the four posterior
refinement settings (\texttt{none} (without any Bayesian inference parameter/phase refinement), Langevin, Metropolis-Hastings (MH), and variational inference (VI)) for the cross workload
summaries whenever the saved run records preserve that distinction, and we
fix VI as the baseline for the parameter density summaries so that the
density panels remain directly comparable across workloads. \appVQAFamily{} reports the configured benchmark instances.

\subsection{QSVT setup}
The QSVT case study uses the H$_2$ Hamiltonian and a $27$ dimensional
QSVT phase vector. We run a guided Bayesian optimization plus VI loop on the same four backends for $100$ evaluations per backend. All transpilation, routing, basis
gate decomposition, and Aer based noisy execution use the Qiskit
stack~\cite{javadi2024qiskit} with transpilation optimization level of 3, applied to
both branches. Because every Bayesian optimization evaluation only updates the QSP phase vector
$\bm{\theta}$, we transpile the parametrized circuit \emph{once per backend}
and bind new phases at runtime; since the four IBM fake backends used here
are static calibration snapshots (\S\ref{sec:limitations_restructured}), the
layout, routing, and basis decomposition chosen at transpile time stay noise
faithful across all $100$ evaluations on a given backend, so the cached
transpilation is an amortization optimization rather than an approximation.
\appResourceDetail{} documents the cache invariants and the parameter
binding scope. The resource study reports compiled
depth, qubit width, and entangling gate burden after routing each circuit
onto the coupling map of the corresponding backend, so that the routed cost
view in \S\ref{sec:results_restructured} is built from the same backends as
the spectral reliability view.

We treat the H$_2$ molecule as a tractable testbed for the unified UQ
pipeline rather than as a production level chemistry calculation. The four
orbital encoding keeps the QSVT phase vector at $27$ entries
and the routed circuits inside what the four shared backends can run within
a reasonable wall clock for $100$ evaluations on each backend. Future work
will apply the same statistical workflow to larger Hamiltonians, starting
from LiH at ten system qubits and then moving to larger small molecule
instances. Section~\ref{sec:greens_resource_extrapolation} states the
asymptotic cost of the circuit and projects the LiH numbers from the H$_2$
molecule so that the gap between the present case study and those targets is
quantitative.

\begin{table}[t]
\centering
\caption{Unified experimental and benchmarking matrix. The shared tuple
interface $(\bm{\xi},y)$ and the downstream UQ stages are common across both
branches; task semantics differ by workload family.}
\label{tab:integrated_matrix_main_restructured}
\scriptsize
\setlength{\tabcolsep}{2pt}
\begin{tabular}{>{\raggedright\arraybackslash}p{0.13\linewidth} >{\raggedright\arraybackslash}p{0.16\linewidth} >{\raggedright\arraybackslash}p{0.2\linewidth} >{\raggedright\arraybackslash}p{0.2\linewidth} >{\raggedright\arraybackslash}p{0.22\linewidth}}
\toprule
\textbf{Branch} & \textbf{Parameter object} & \textbf{Benchmark target} & \textbf{Backend and UQ setup} & \textbf{Outputs} \\
\midrule
VQA family & Workload ansatz or channel parameters across ten families & Energy, cost, residual, fidelity, tomography, or sensing objective on backend $b$ & Four backends; VI baseline Bayesian optimization with aggregated posterior refinement settings & Quality landscape across workloads, backend ranking, sensitivity and density summaries, routed resource views \\ &&&&\\
QSVT & $27$ dimensional QSVT phase vector & Guided spectral peak loss and Green's function reconstruction & Same four backends; $100$ evaluations per backend with guided Bayesian optimization and VI & Backend spectra, hit rate ranking, sensitivity profiles, compiled resource trends \\
\bottomrule
\end{tabular}
\end{table}

Table~\ref{tab:integrated_matrix_main_restructured} condenses this
experimental design into one matrix: both branches share the tuple
interface, the four backends, and the downstream UQ stages, while the
parameter objects and benchmark targets stay workload specific. The shared
engine makes our cross paradigm comparison
interpretable as a single coherent benchmark rather than as two unrelated
case studies. We do not observe noisy degradation passively: the Bayesian optimization plus
posterior refinement loop uses backend induced noisy evaluations to learn
perturbed control parameters and then treats the remaining gap to the ideal
reference as a backend characterization signal. Algorithm~\ref{alg:unified_uq_workflow}
in Appendix~\ref{app:workflow_description} gives the merged online and offline UQ workflow that generates
backend local tuple histories, robust parameter regions, rankings, and noise
alignment summaries. Its statistical components follow GP regression,
Bayesian optimization, trust region search, and posterior refinement through
VI or Langevin dynamics~\cite{rasmussen2006gp,snoek2012practical,shahriari2016bo,balandat2020botorch,eriksson2019turbo,blei2017variational,welling2011langevin}.

\section{Results and Discussion}
\label{sec:results_restructured}
We present results in five steps. We start with the cross workload VQA
benchmark, then move to the QSVT Green's function spectral benchmark, then read the
robustness and sensitivity structure across both branches, then the
compiled resource view, and finally a cross paradigm interpretation that
ties the two scores together.

\subsection{Cross workload VQA benchmark}
Backend quality is strongly workload dependent across the VQA family. Among our four quantum backends, VQAMET is uniformly easy across all four backends,
whereas VQCFE is highly selective and reaches fully normalized quality only on
Kyoto in the aggregated summary. Between those extremes sit chemistry,
optimization, simulation, PDE, linear solving, and tomography tasks, each with
its own backend preference.

\begin{figure}[t]
\centering
\begin{subfigure}[t]{0.49\textwidth}
\centering
\includegraphics[width=\linewidth]{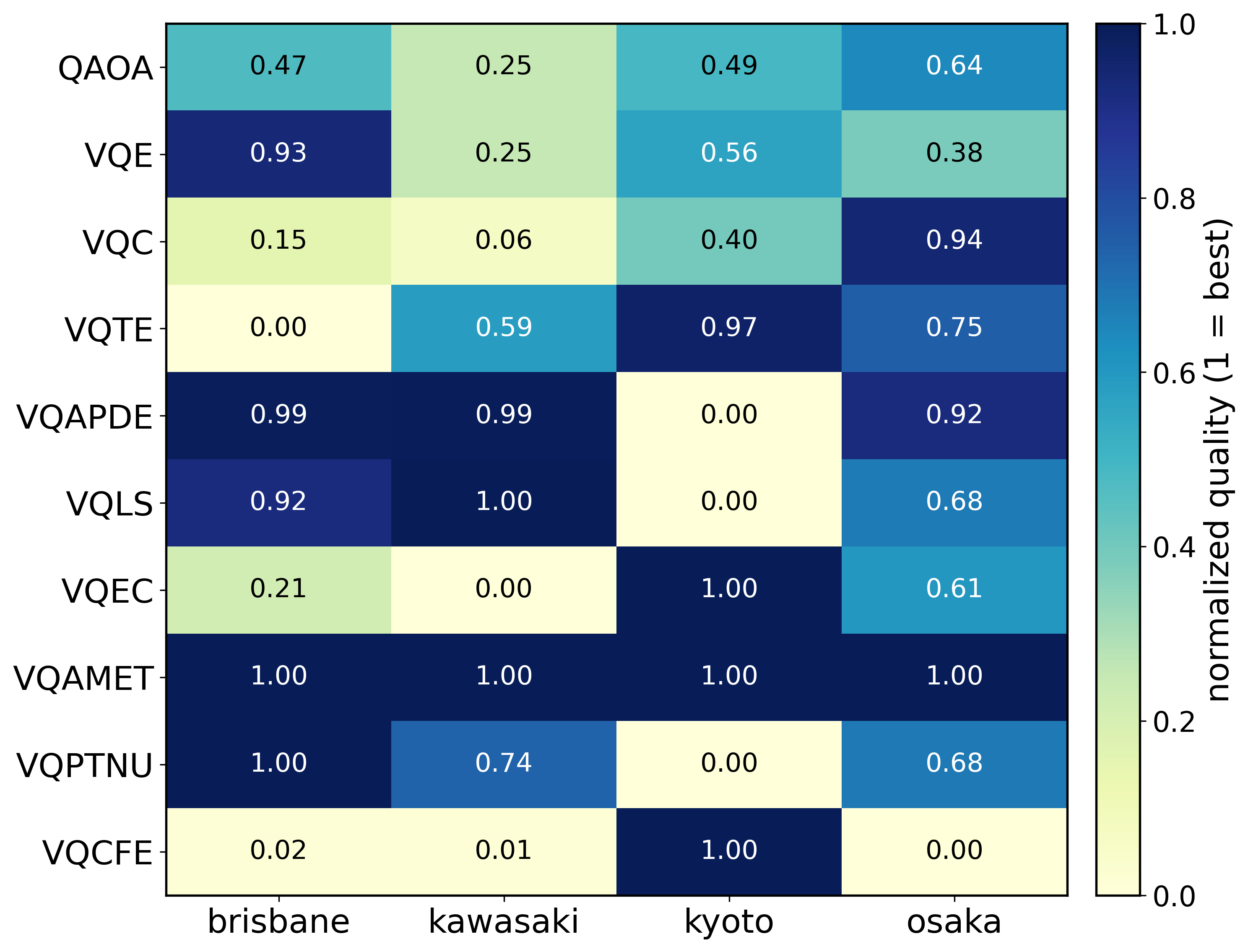}
\caption{Cross workload quality landscape.}
\end{subfigure}
\hfill
\begin{subfigure}[t]{0.50\textwidth}
\centering
\includegraphics[width=\linewidth]{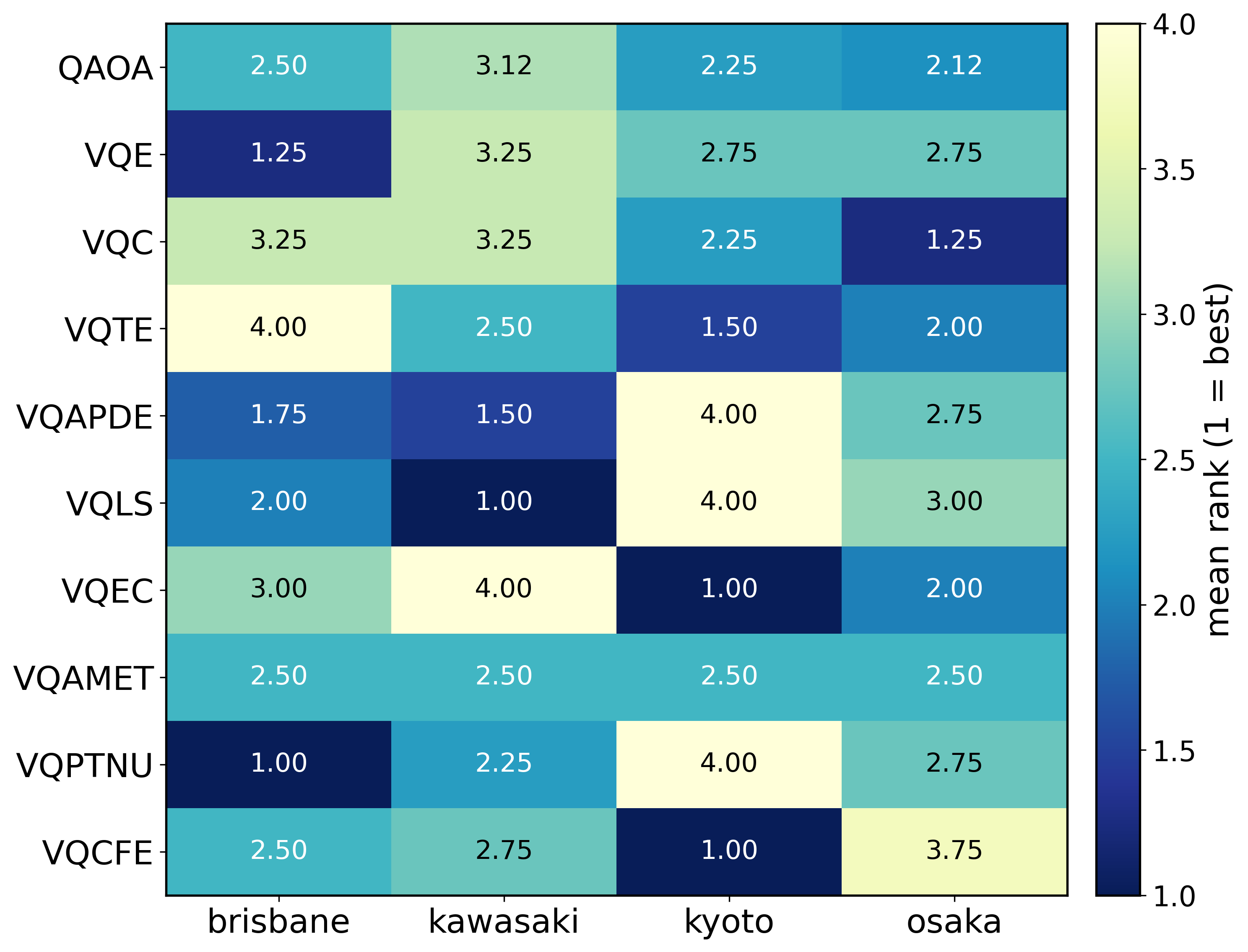}
\caption{Workload wise and overall backend ranks.}
\end{subfigure}
\caption{Four backends VQA benchmark views built from \S\ref{sec:design_restructured}.1: each of the ten VQA workloads is run on
Brisbane, Kawasaki, Kyoto, and Osaka through the Bayesian optimization plus posterior
refinement engine of \S\ref{sec:formulation_restructured}.3, with four
refinement settings (none, Langevin, MH, VI) per (workload, backend) cell.
The left panel plots $\mathrm{quality\_norm}\in[0,1]$, defined per workload
as a min-max rescaling of the achieved task quantity so that $1$ marks the
best result and $0$ the worst across the four backends, then averaged across
the four refinement settings. The right panel plots the within workload rank of each backend, with $1$ denoting the best backend for that workload. The average ranks across the ten workloads are close: Brisbane $2.38$, Osaka $2.49$, Kyoto $2.52$, and Kawasaki $2.61$. Read jointly, the two panels
show no single backend dominates the family, which is why backend quality on the VQA side
has to be reported at the application level rather than through a single scalar.}
\label{fig:vqa_main_restructured}
\end{figure}

The four workload means make the difficulty separation explicit:
VQAMET has mean quality $1.000$, VQAPDE reaches $0.726$, VQLS reaches
$0.649$, VQPTNU reaches $0.605$, VQTE reaches $0.576$, VQE reaches $0.533$,
QAOA and VQEC sit near $0.46$, VQC drops to $0.389$, and VQCFE is the hardest
workload at $0.257$. The workload winners are also selective: Osaka leads QAOA
and VQC, Brisbane leads VQE and VQPTNU, Kyoto leads VQTE, VQEC, and VQCFE,
Kawasaki leads VQAPDE and VQLS, and VQAMET is effectively tied across the four
backends. That pattern is the main VQA benchmarking result: a backend that is
good for one task family is not automatically good for the others.

The overall average rank view is more conservative. Under that aggregate
criterion, Brisbane is the most consistent backend in the four backends,
followed closely by Osaka, Kyoto, and Kawasaki; by mean normalized quality,
Osaka is the strongest aggregate backend. The detailed VQA results show
that VQAMET reaches its good regime fastest, whereas VQE, QAOA, and VQC carry
the longest optimization tails; that QAOA and VQEC concentrate parameter
sensitivity into a few coordinates more strongly than VQAPDE or VQLS; and that
VQPTNU and VQEC display the strongest parameter to output coupling in the
VI based density analysis. For the supporting per workload views, Figure~\ref{fig:app_vqa_parameter_summary}
in \appVQAParamNum{} reports the sensitivity and density summaries,
Figure~\ref{fig:vqa_objective_distributions} and
Table~\ref{tab:vqa_posterior_engine_summary} in \appVQAPosteriorNum{} report
the posterior engine views together with the standalone ranking and Pareto
panels of Figure~\ref{fig:vqa_results_benchmark}, and
Figure~\ref{fig:app_vqa_benchmark_noise} in \appVQABenchmark{} reports the
benchmark trajectory and noise views.

\subsection{QSVT application level benchmark}
Backend quality on the QSVT branch is set by how often the UQ loop
revisits a good spectral reconstruction regime, not by whether one lucky
phase vector once hit the target. Across the $100$ evaluations,
the hit rates are $51\%$ for Brisbane, $48\%$ for Osaka, and $44\%$ for both
Kawasaki and Kyoto. We break the tie between Kawasaki and Kyoto by time to
target, because Kyoto first reaches the good regime only at evaluation $15$.
This gives the application level ordering Brisbane $>$ Osaka $>$ Kawasaki
$>$ Kyoto. Table~\ref{tab:greens_backend_rank} in \appResults{} states this
ranking with the reliability numbers that produce it, and
Figure~\ref{fig:results_benchmark} collects the Pareto front and pairwise
win probability views that support the ordering and the tie break between
Kawasaki and Kyoto.

\begin{figure}[t]
\centering
\begin{subfigure}[t]{0.48\textwidth}
\centering
\includegraphics[width=\linewidth]{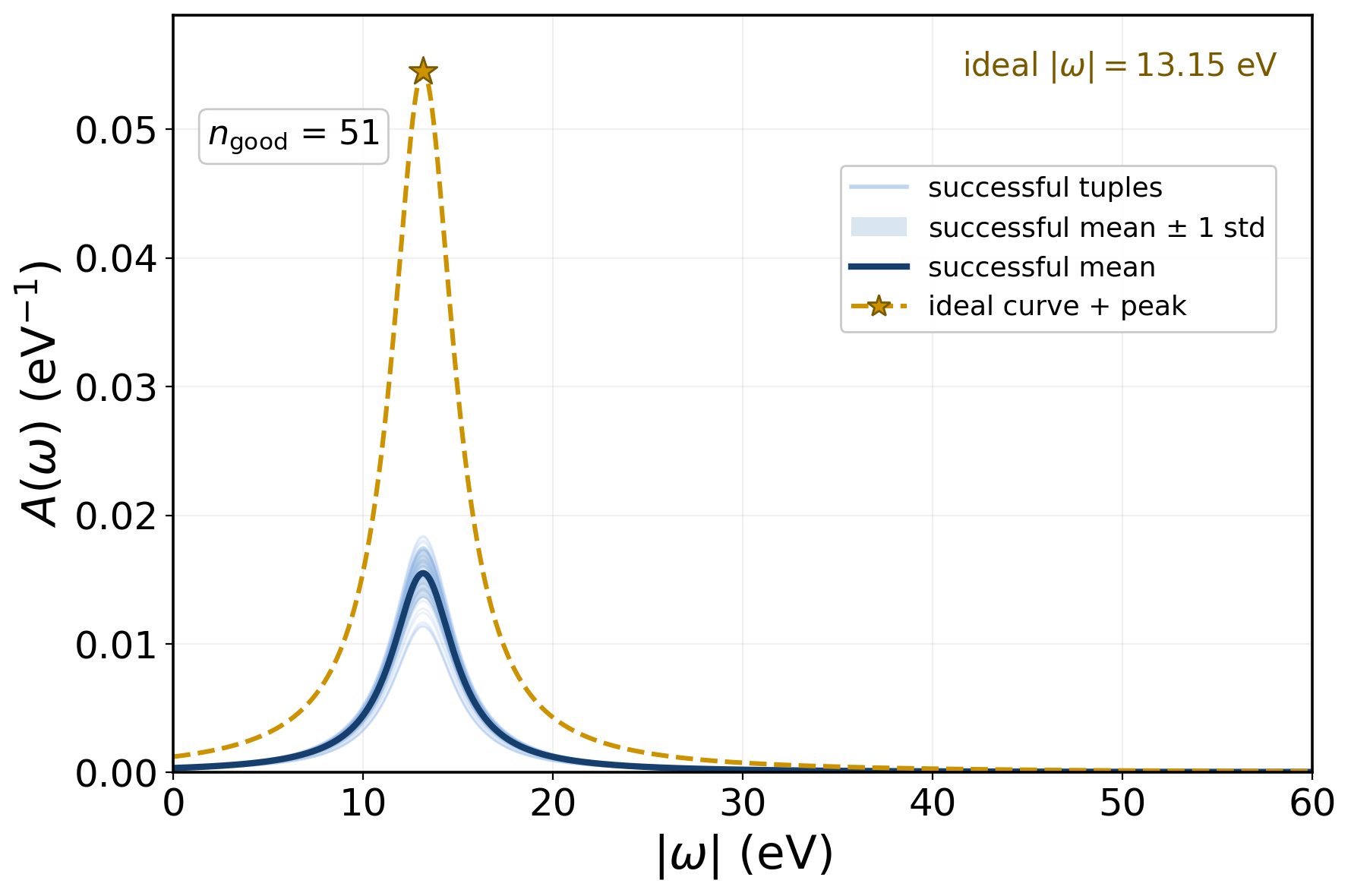}
\caption{\texttt{Brisbane}}
\end{subfigure}
\hfill
\begin{subfigure}[t]{0.48\textwidth}
\centering
\includegraphics[width=\linewidth]{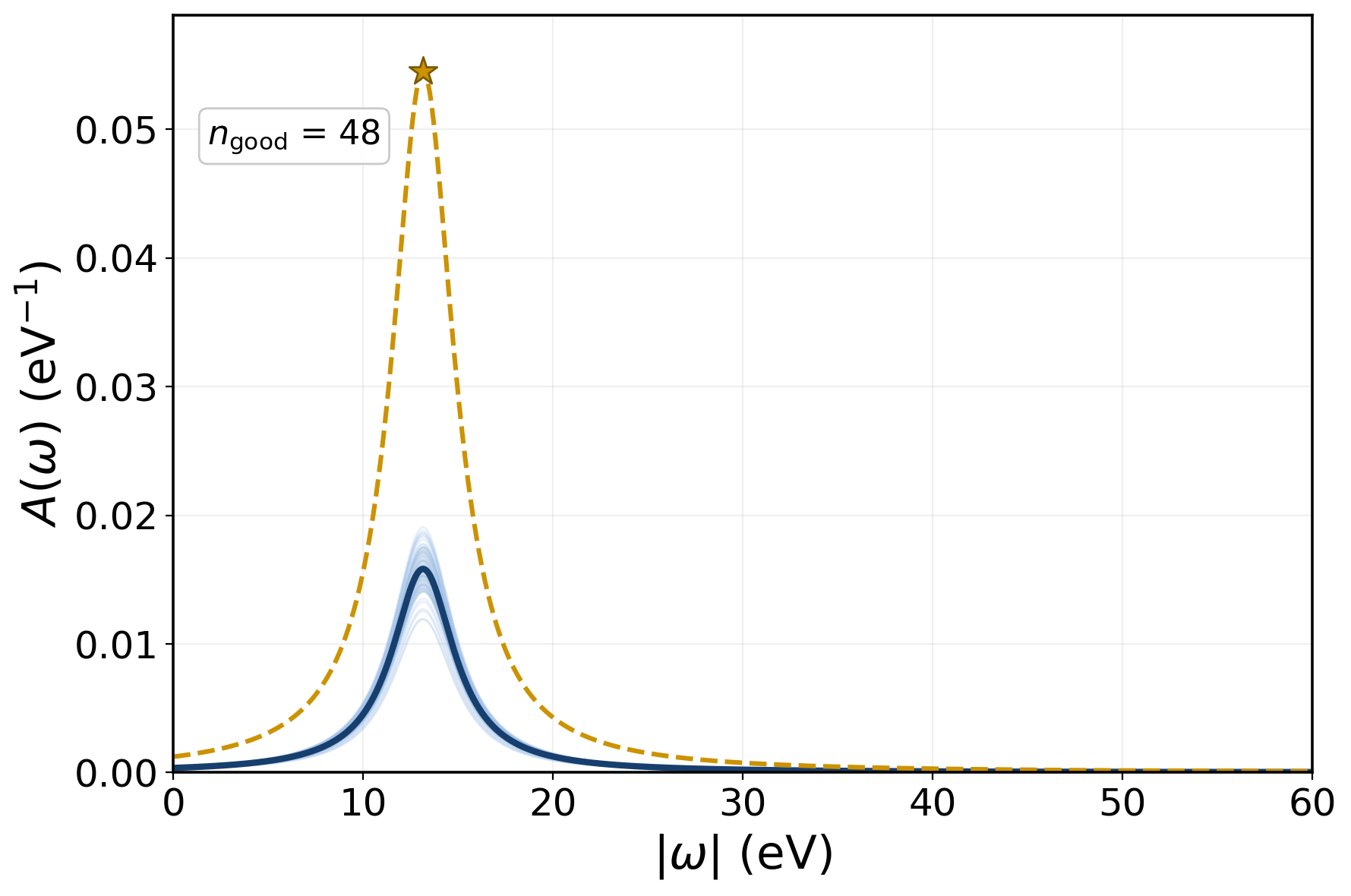}
\caption{\texttt{Osaka}}
\end{subfigure}
\\[0.7em]
\begin{subfigure}[t]{0.48\textwidth}
\centering
\includegraphics[width=\linewidth]{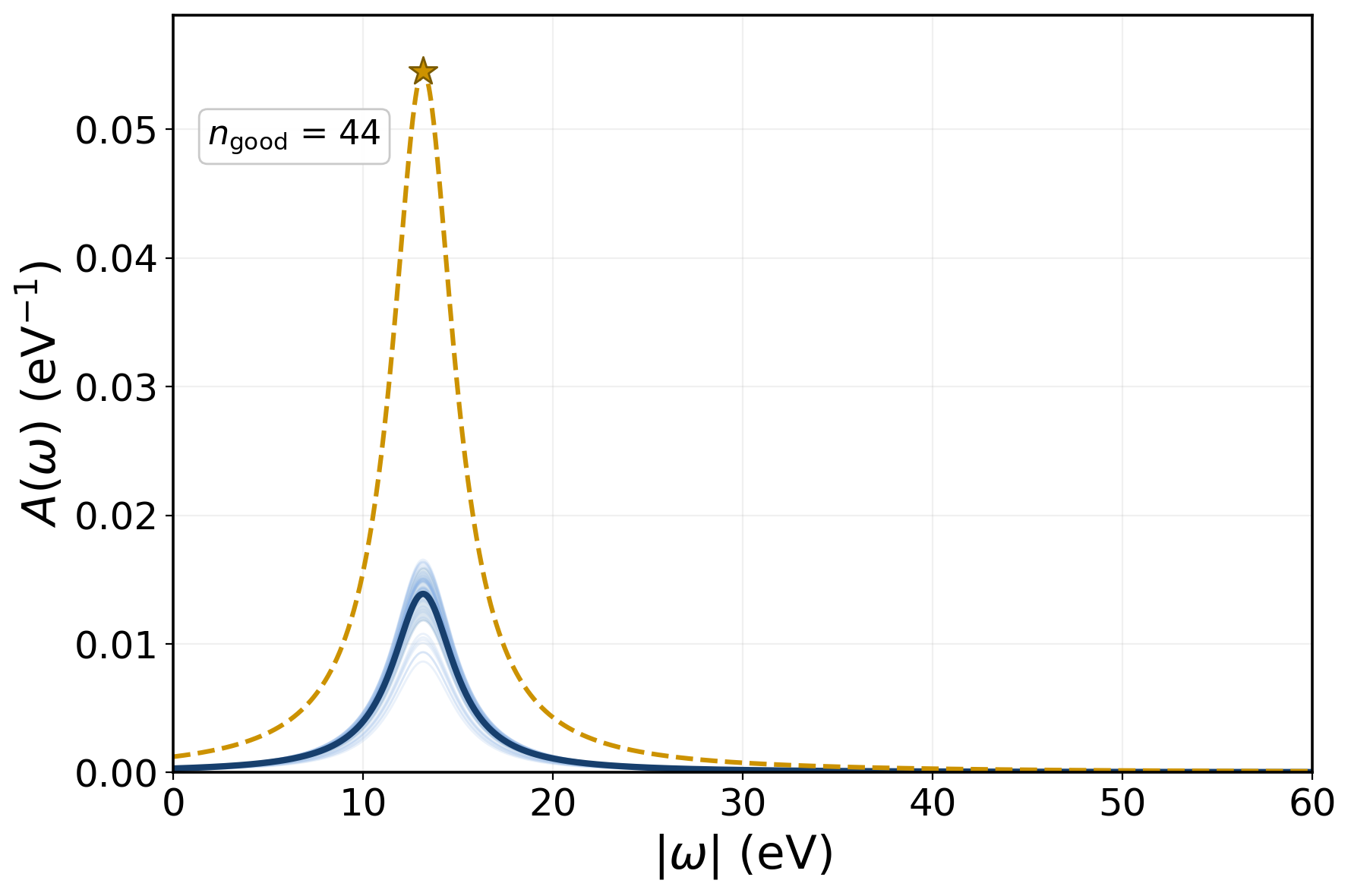}
\caption{\texttt{Kawasaki}}
\end{subfigure}
\hfill
\begin{subfigure}[t]{0.48\textwidth}
\centering
\includegraphics[width=\linewidth]{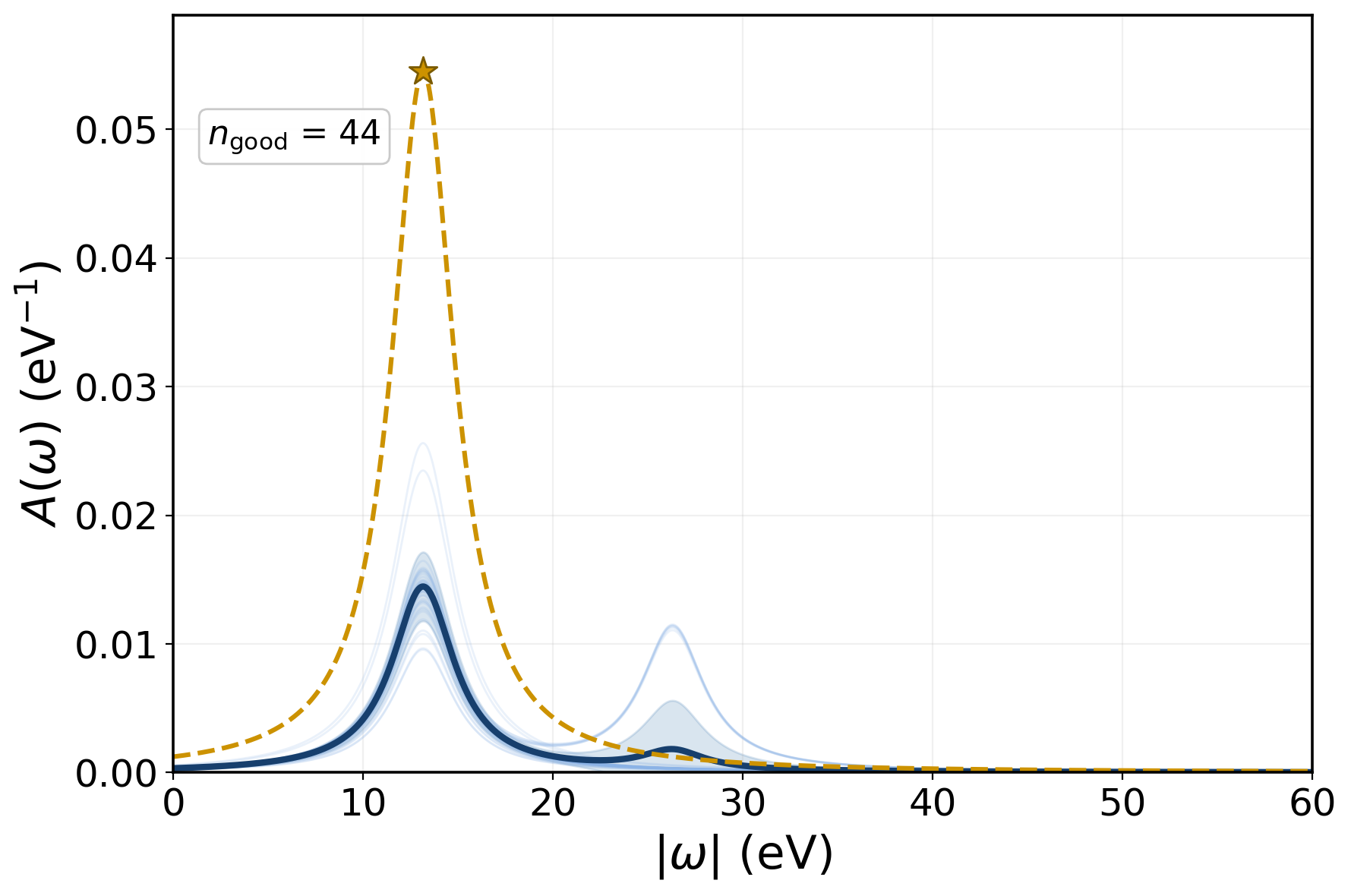}
\caption{\texttt{Kyoto}}
\end{subfigure}
\caption{QSVT spectral function reconstructions across the four backends,
built from the $100$ evaluations of Bayesian optimization+VI of
\S\ref{sec:design_restructured}.2 on the H$_2$ $27$ dimensional QSVT phase
vector. Each successful phase vector $\bm{\theta}$ produces a backend
reconstruction of the time domain Green's function
$G_{jj}^{(h)}(t;b,\bm{\theta})$ through the QSVT pipeline of
Eqs.~\eqref{eq:green_main_restructured}-\eqref{eq:qsvt_main_restructured};
we apply exponential damping at $\eta=0.15$~a.u.\ and Fourier transform to
the frequency axis through Eq.~\eqref{eq:spectral_main_restructured} to
obtain the spectral function $A_j(\omega)$ drawn in each panel. Light curves
are individual good tuple reconstructions, the dark curve and the band show
the mean and the one standard deviation envelope after sign aligning each
successful spectrum to the target peak magnitude, and the gold reference
shows the ideal spectral function together with the ideal peak marker at
$\abs{\omega}=13.15\,\mathrm{eV}$. The ideal Green's function peak is
symmetric, so we score a phase vector as a good tuple when the reconstructed
peak lands on either $+\omega_{\mathrm{ideal}}$ or $-\omega_{\mathrm{ideal}}$,
that is, we use
$\min(\abs{\omega-\omega_{\mathrm{ideal}}},\abs{\omega+\omega_{\mathrm{ideal}}})$
in Eq.~\eqref{eq:guided_main_restructured}; this is why the panels report
$\abs{\omega}$ rather than $\omega$ directly. The inset $n_{\mathrm{good}}$
records how many good tuples contribute to each backend panel. The
corresponding bad tuple spectra appear in
Figure~\ref{fig:results_spectral_bad_uq} of \appResults.
The benchmarking signal is not that all four backends produce at least one
good spectrum (they do) but that they revisit the good spectral regime with
materially different reliability.}
\label{fig:greens_spectral_main_restructured}
\end{figure}

We classify each recorded evaluation as a \emph{good tuple} when the guided peak
loss of Eq.~\eqref{eq:guided_main_restructured} reaches its lower target on
the peak set, that is, when the optimizer hits all guided spectral
peak targets, and as a \emph{bad tuple} otherwise. Each panel in
Figure~\ref{fig:greens_spectral_main_restructured} shows only the good tuples
for that backend, and $n_{\mathrm{good}}$ reports how many they are
(Brisbane $51$, Osaka $48$, Kawasaki $44$, Kyoto $44$). \appResults{} collects the corresponding bad tuple spectra in
Figure~\ref{fig:results_spectral_bad_uq}. We split good and bad spectra across two
figures because plotting them together would compress the bad region scatter
so heavily that the structure of the successful reconstructions becomes
invisible.

In each panel of Figure~\ref{fig:greens_spectral_main_restructured}, the pale blue traces
are the individual successful spectra found by the UQ loop. The dark blue
curve is the mean of those successful spectra, and the blue band is the
corresponding one standard deviation envelope. The gold dashed curve and gold
star are the ideal H$_2$ core orbital reference spectrum and its ideal peak
height. The inset $n_{\mathrm{good}}$ reports how many successful tuples were
available for that backend. If the blue mean is aligned with the gold peak in
$|\omega|$ but remains well below the gold curve in height, then the backend is
recovering the correct ionization magnitude but not the full ideal spectral
weight. If the blue traces are tightly clustered, the successful region is
stable; if they spread widely, the backend reaches the good region less
uniformly.

Figure~\ref{fig:greens_spectral_main_restructured} shows why the benchmark is
application level rather than hardware level. Kyoto is the most instructive
contrast case: when Kyoto succeeds, it can produce a sharp spectrum near the
correct ionization feature, but it reaches that regime later and less
reliably than Brisbane or Osaka. The method therefore distinguishes
``conditionally sharp'' from ``reliably accessible,'' which is exactly the kind
of distinction that generic device scores are not designed to capture. The
gold ideal curve in each panel makes a second point visible at a glance:
backend conditioned phase perturbations can restore the correct H$_2$
ionization magnitude, but they do not fully restore the ideal spectral
weight. The remaining amplitude deficit is part of the benchmark signal rather
than a cosmetic plotting detail.

\appOffline{} and \appResults{} give the longer
sensitivity and density analysis, which shows that the robust QSVT region is backend specific rather
than universal: the dominant phase coordinates differ across backends, and the
geometry of the good region is tighter for some backends than for others. Those
results are the mechanism that turns the QSVT branch from a circuit
study into a backend characterization study.

\subsection{Robustness, sensitivity, and region geometry}
Our benchmarking claim does not rest only on the best achieved objective
values. It also rests on what the UQ machinery reveals about the shape of
the good region. The VQA branch shows structural selectivity at the
workload level. QAOA and VQEC concentrate sensitivity into a few coordinates,
which is consistent with sharper, more selective good regions. VQAPDE and
VQLS distribute their sensitivity more broadly, which is consistent with
broader robust regions. The VI based density summaries show that VQPTNU and
VQEC have the strongest parameter to output coupling, while VQAMET stays
simple and nearly backend invariant in our work.
Figure~\ref{fig:app_vqa_parameter_summary} in \appVQAParamNum{} reports the
density diagnostics behind this coupling claim. This distinction
matters because hard and expensive are not the same thing: some workloads
are hard because their good region is narrow, others are expensive because
each evaluation expands into many circuits.

On the QSVT branch the same selectivity appears at the phase index
level rather than at the workload level. The sensitivity analysis shows that
the robust phase region is backend specific. Coordinates $X_1$, $X_5$,
$X_{23}$, and $X_{24}$ recur across multiple backends, and so do $X_{26}$
and $X_{27}$, although their ordering and relative importance change. The
top five parameter block carries roughly two thirds of the Morris
sensitivity mass ($\mu^*_i$ as defined in
\S\ref{sec:metrics_restructured} and summarized in
Table~\ref{tab:metric_summary_restructured}, normalized to sum to one), so
the optimization is effectively low dimensional in a backend specific way
rather than in a universal one. Figure~\ref{fig:results_sensitivity} in
\appResults{} compares the Morris, SHAP, and Sobol views behind this claim,
and Figure~\ref{fig:results_density} shows the parameter means and PCA
embedding of the robust phase regions.

These robustness views are what make our benchmark a genuine characterization
study. We do not judge a backend only by its final score: we also judge it by
whether it exposes a broad or a narrow good region, whether that region is
stable across repeated evaluations, and whether the dominant control
directions are scientifically interpretable. \appResultsNum{}, \appVQAParamNum{}, and~\appVQAPosteriorNum{} provide the detailed plots
behind these claims.

\subsection{Resource perspective across branches}
Execution cost changes the meaning of backend quality, so we read the
benchmark together with the resource study. Figures~\ref{fig:resource_degree_summary}, \ref{fig:resource_time_summary},
and \ref{fig:resource_appendix_vqa_extra} in \appResourceDetail{} report the
detailed degree and time sweeps and the supplemental VQA resource panels, and states the cache policy.

\begin{figure}[t]
\centering
\begin{subfigure}[t]{0.48\textwidth}
\centering
\includegraphics[width=\linewidth]{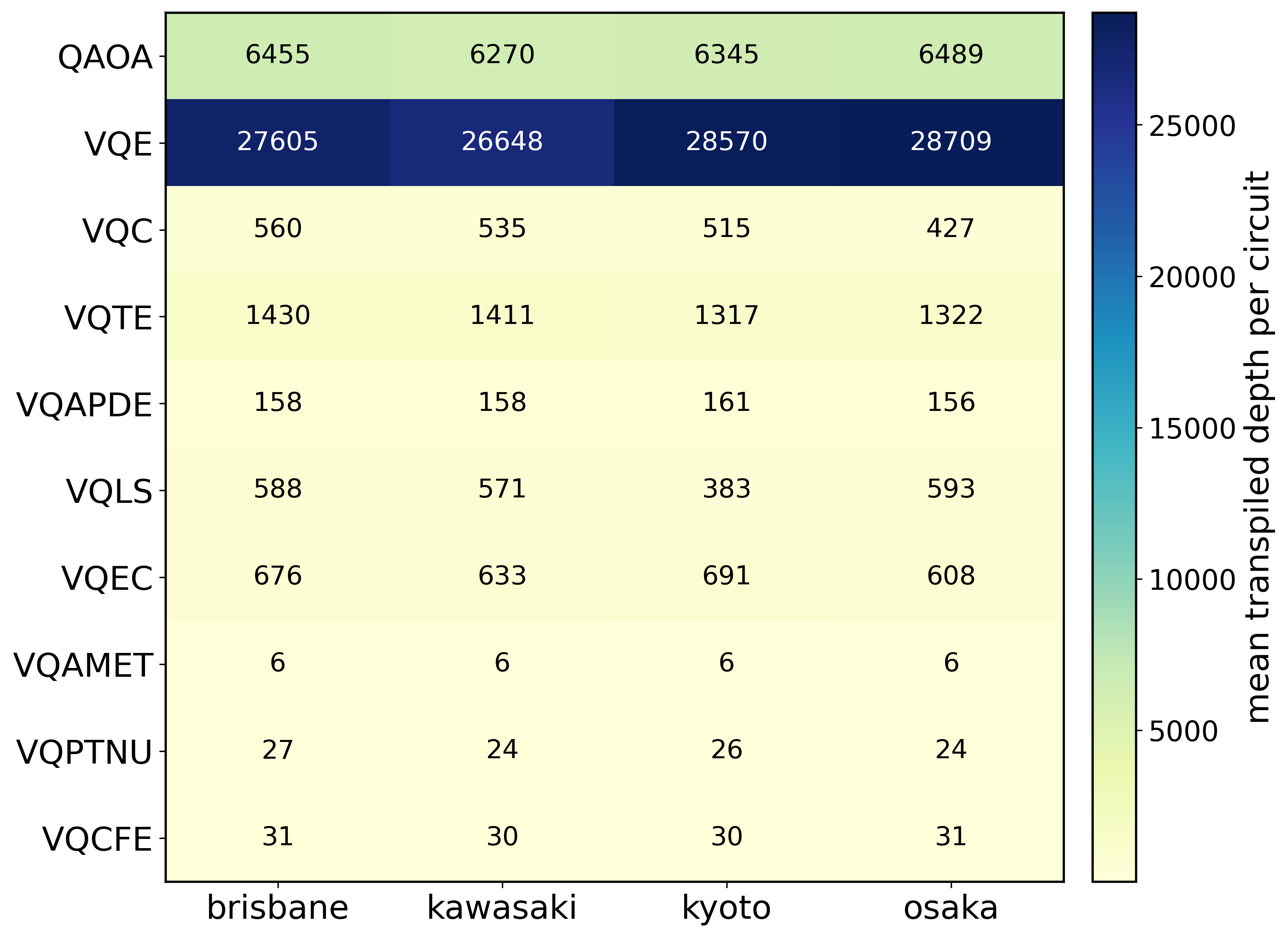}
\caption{Mean transpiled depth per circuit.}
\end{subfigure}
\hfill
\begin{subfigure}[t]{0.48\textwidth}
\centering
\includegraphics[width=\linewidth]{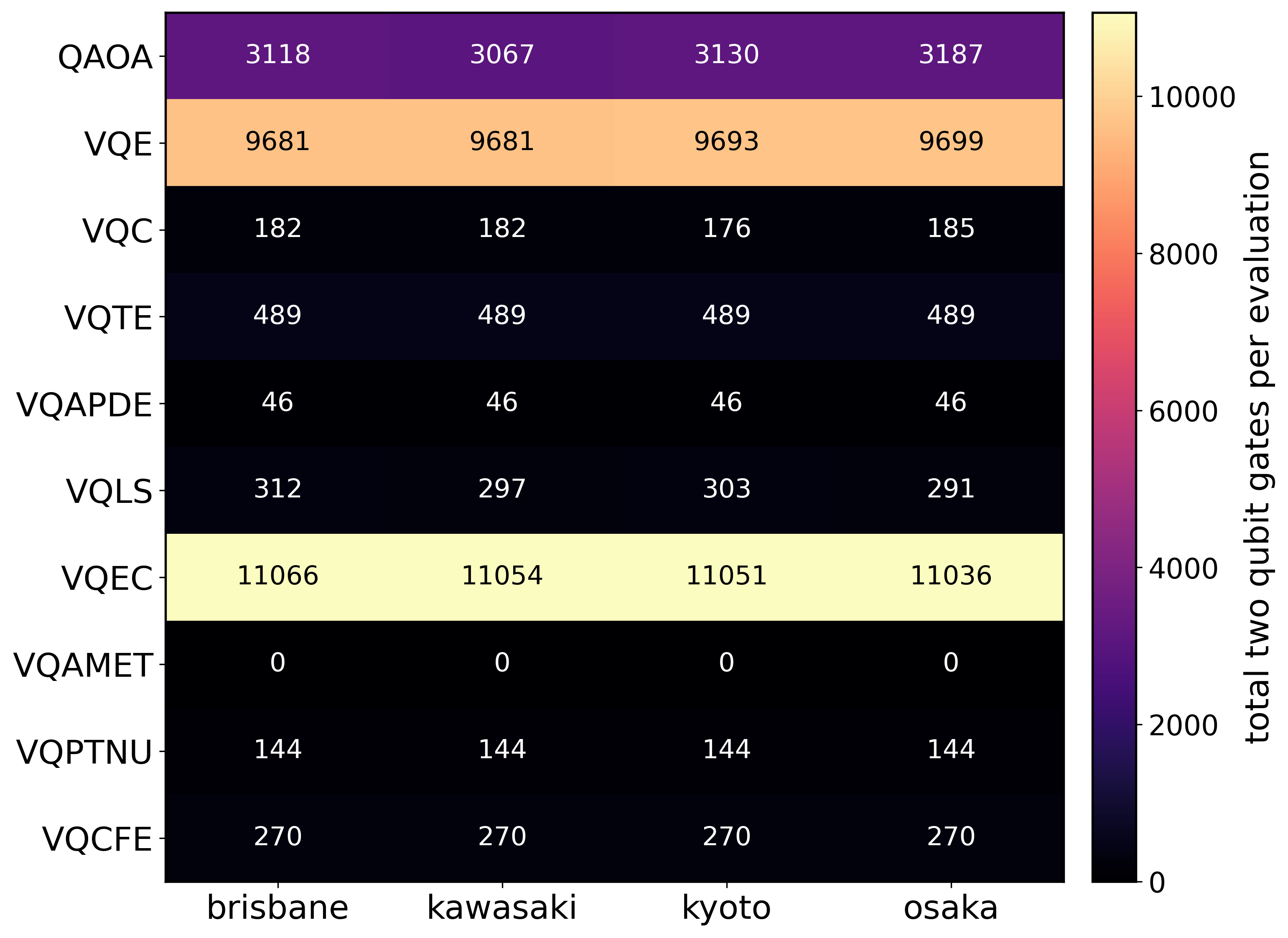}
\caption{Total two qubit burden per evaluation.}
\end{subfigure}
\caption{VQA resource summary built from the same runs as
Figure~\ref{fig:vqa_main_restructured}. Every circuit produced during the
$100$ evaluation Bayesian optimization+VI loop is transpiled onto the coupling map and basis
gate set of each backend at Qiskit's transpiler \texttt{optimization\_level=3}, and we
then aggregate the two heatmap quantities across that history. The left
panel reports the mean compiled depth per circuit, that is, the mean of
the per circuit transpiled depths across the backends, and therefore tracks
how heavy one circuit becomes once it is routed. The right panel reports
the total two qubit gate count per objective evaluation, summing over every
circuit a single evaluation produces, which credits workloads that expand
one evaluation into a family of measurement circuits (tomography style or
randomized characterization style) rather than just into one deep circuit.
The two views must be read together: a workload that is light per circuit
but expands into many circuits can still be expensive overall, while a
workload that is deep per circuit but uses one circuit per evaluation is
expensive in a different way. This separation is what makes the VQA
resource benchmark application level rather than circuit level.}
\label{fig:vqa_resource_main_restructured}
\end{figure}

Figure~\ref{fig:vqa_resource_main_restructured} reveals three regimes on the
VQA side. VQE is the heaviest workload at the single circuit level, with
mean compiled depths between $26648$ and $28709$ and total two qubit counts
between $9681$ and $9699$ per evaluation on the four backends. QAOA
follows, with depths in the range $6270$ to $6489$. VQTE, VQC, and VQLS sit
in the middle regime. VQAMET and VQAPDE are light, and VQAMET is especially
notable because it remains a depth $6$ circuit with zero two qubit gates
across all four routed backends. VQEC is a special case: it is not expensive
because each of its circuits is deep, but because one objective evaluation
expands into a large sampled error family that costs roughly $1.1\times
10^4$ two qubit gates per evaluation even though every individual circuit
has depth below $700$. The resource benchmark therefore distinguishes
workloads whose cost is dominated by one heavy circuit from workloads whose
cost is dominated by many lighter circuits.

\begin{figure}[t]
\centering
\includegraphics[width=1.0\textwidth]{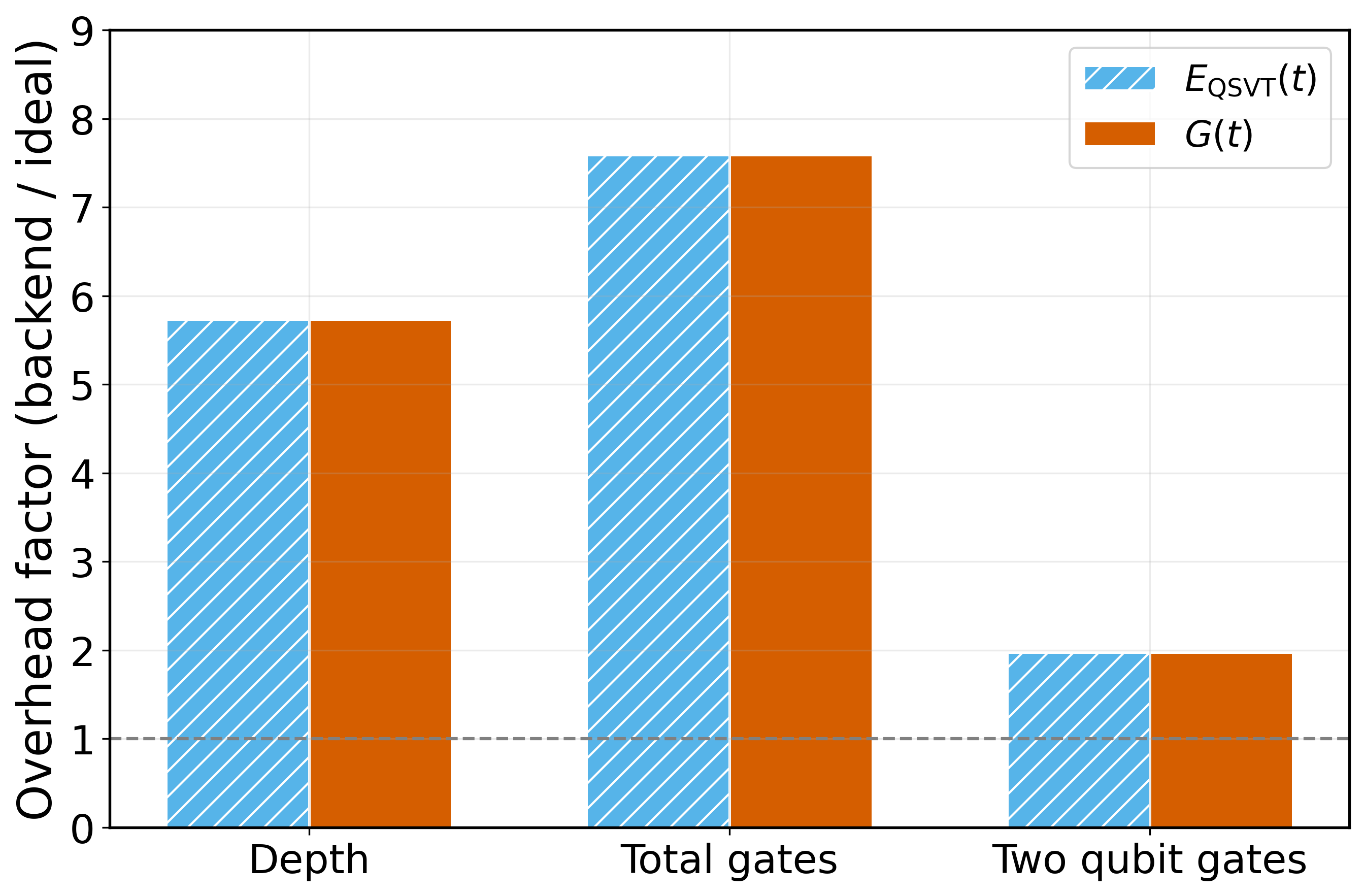}
\caption{QSVT resource summary built from the QSVT block encoding
circuit of \S\ref{sec:formulation_restructured}.2 evaluated across a sweep
of polynomial degrees $d$. For every degree, we first construct the
abstract QSVT circuit using the ideal basis (no routing constraint), then
transpile it onto each of the four backends' coupling maps and basis gate
sets at Qiskit's transpiler \texttt{optimization\_level=3} and record routed
depth and gate counts. Each bar reports the overhead factor
$\mathrm{backend}/\mathrm{ideal}$ averaged over the four matched backends.
We collapse the
backends into one bar set because their overheads agree closely: routed
depth inflates by $5.6\times$ to $5.8\times$, total gate count by
$7.4\times$ to $7.7\times$, and two qubit count by about $2\times$ on every
backend and for both circuit families. The fold change between depth and
total gates forces a joint reading of spectral reliability and compiled
execution cost: a backend that is slightly cheaper in routed depth is not
automatically cheaper in routed total gates, so reliability cannot be
interpreted in isolation from the compilation layer.}
\label{fig:greens_resource_main_restructured}
\end{figure}

Figure~\ref{fig:greens_resource_main_restructured} gives the matching
Green's/QSVT overhead view by comparing routed circuits against the ideal
reference circuit over the degree sweep.

\begin{figure}[t]
\centering
\begin{subfigure}[t]{0.325\textwidth}
\centering
\includegraphics[width=\linewidth]{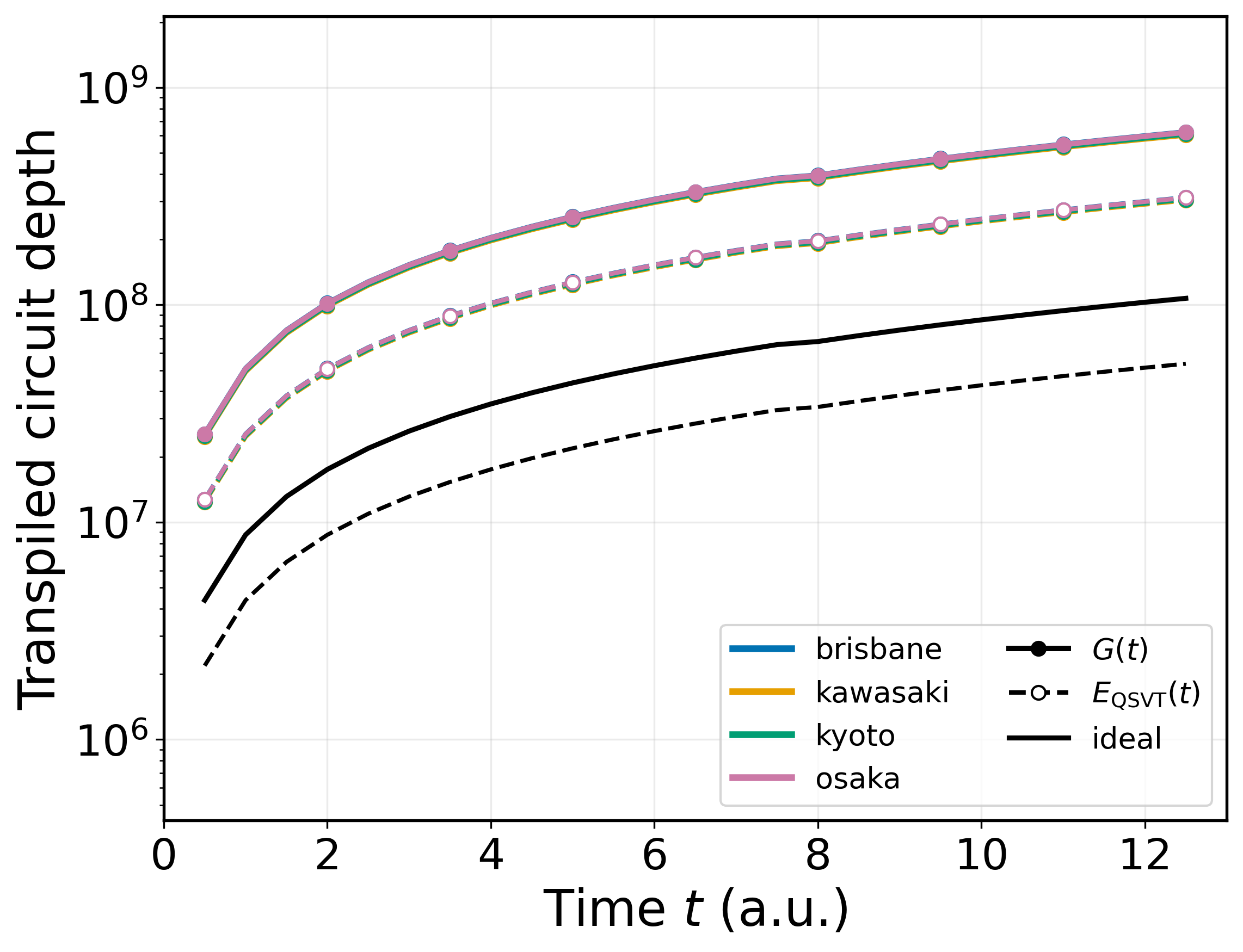}
\caption{Transpiled circuit depth.}
\end{subfigure}
\hfill
\begin{subfigure}[t]{0.325\textwidth}
\centering
\includegraphics[width=\linewidth]{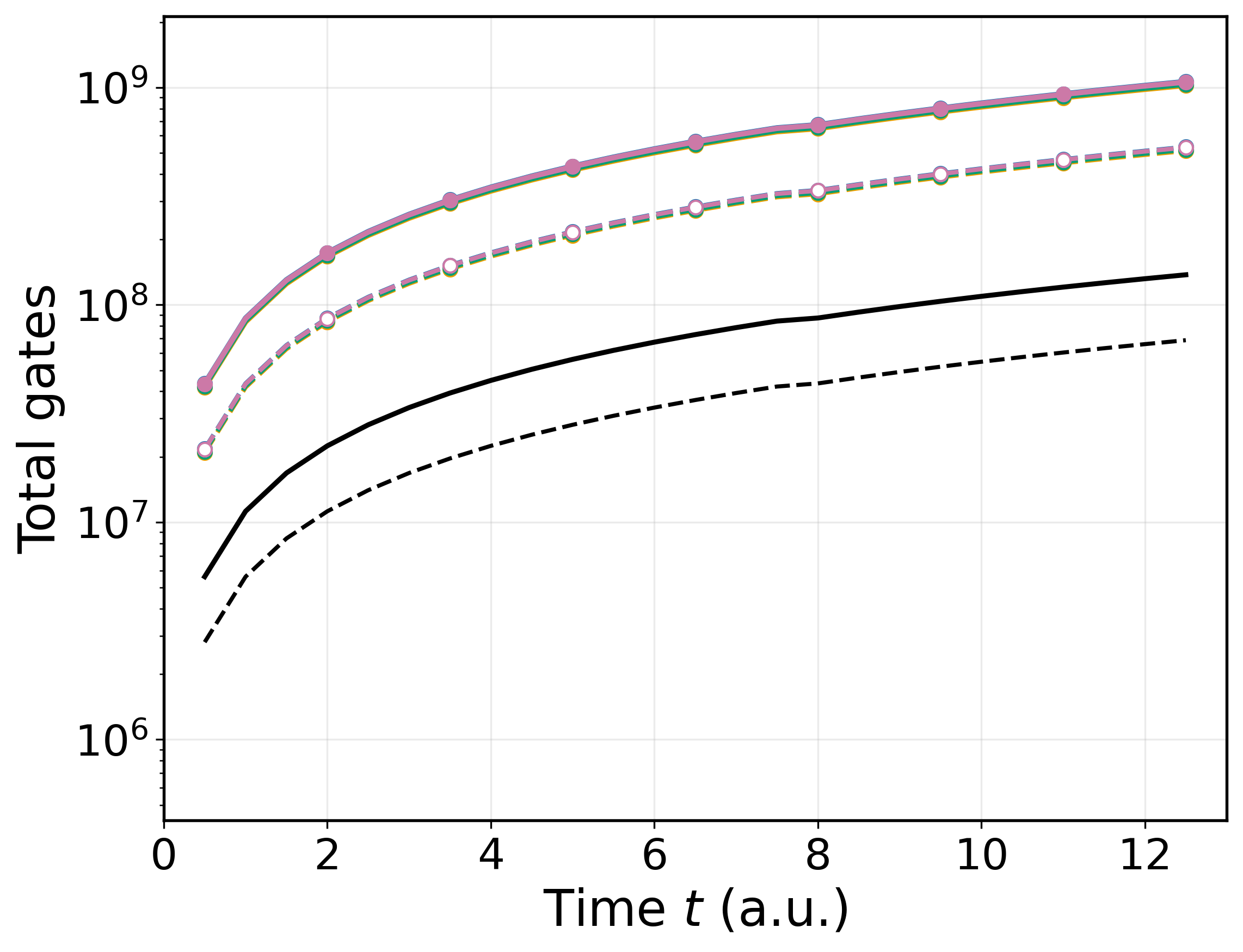}
\caption{Total gate count.}
\end{subfigure}
\hfill
\begin{subfigure}[t]{0.325\textwidth}
\centering
\includegraphics[width=\linewidth]{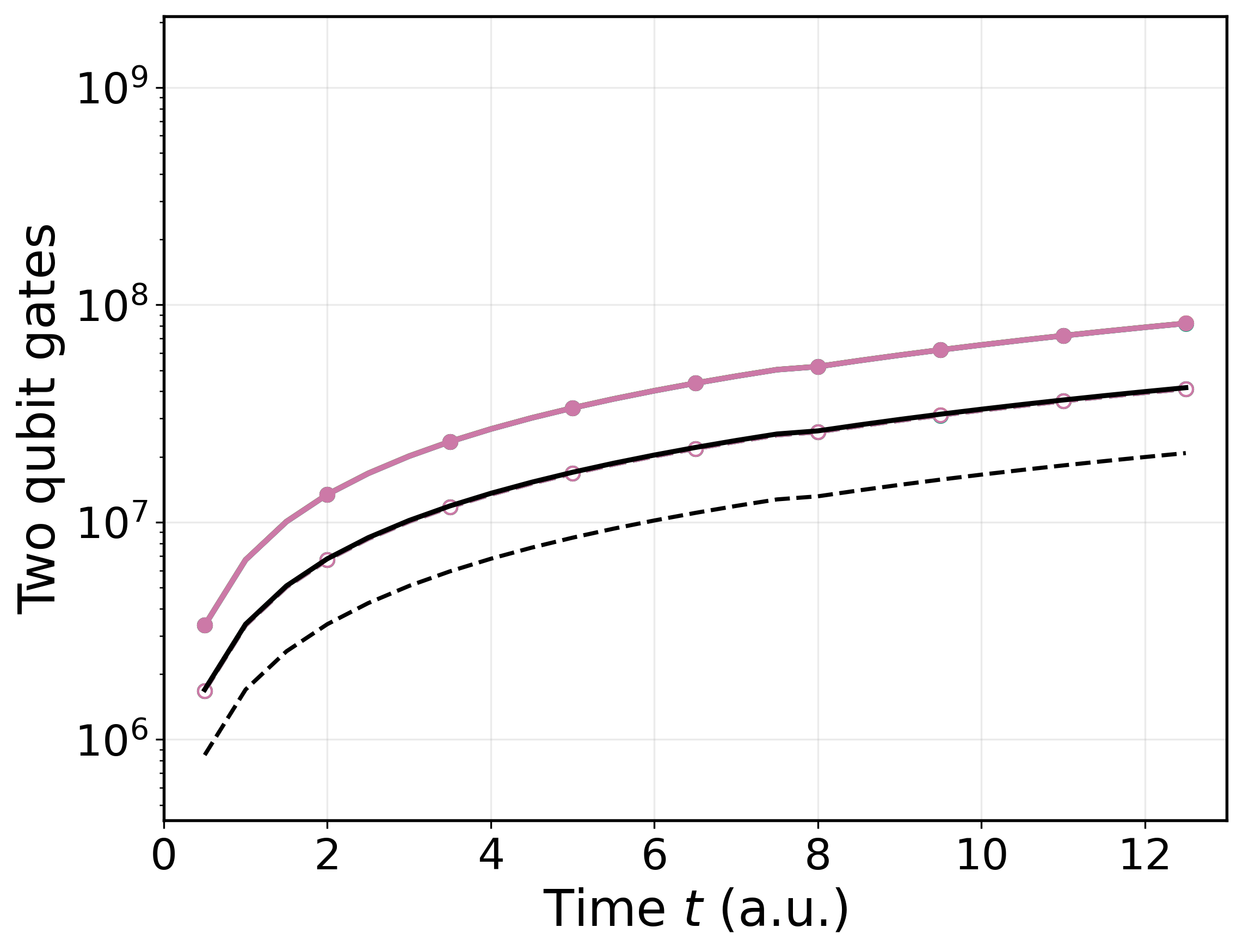}
\caption{Two qubit gate count.}
\end{subfigure}
\caption{Compact QSVT multimetric resource summary for the four routed
backends and the ideal reference circuits at degree pair $(12,13)$. Panels
(a)-(c) show transpiled circuit depth, total gate count, and two qubit
gate count against the propagation time on identical axis ranges, with
solid filled lines for the Green's function circuit $G(t)$ and dashed open
lines for the propagator block $E_{\mathrm{QSVT}}(t)$; black lines mark the
ideal references, and the legend in panel (a) applies to all three panels.
The four backend curves nearly coincide on this scale, which matches the
collapsed overhead bars of
Figure~\ref{fig:greens_resource_main_restructured}, and the curves remain
broadly parallel as time increases, so the routed cost penalty behaves like
a persistent multiplier rather than an isolated outlier at one time point.
QSVT reliability must therefore be read together with depth, total gate
volume, and entangling gate burden because each metric captures a different
part of the execution cost.}
\label{fig:greens_resource_multimetric_main}
\end{figure}

On the QSVT side, transpilation onto the routed backend dominates
the abstract circuit cost: routed depth inflates by a factor of about
$5.6\times$ to $5.8\times$ and routed total gates inflate by about
$7.4\times$ to $7.7\times$ over the four backends. The bars in
Figure~\ref{fig:greens_resource_main_restructured} sit inside this compact
band, with Brisbane and Osaka at the upper edge, Kyoto slightly lower, and
Kawasaki the lightest of the four. Figure~\ref{fig:greens_resource_multimetric_main}
then shows that this near coincidence persists across depth, total gates,
and two qubit gates over the time sweep. Even within this controlled subset,
compiled cost differs materially enough that backend quality cannot be
interpreted through spectral reliability alone, so we report application
level reliability and compiled cost jointly rather than separately.

\subsection{Asymptotic cost of the QSVT circuit and a projection to LiH}
\label{sec:greens_resource_extrapolation}
We close the resource view with a compact scaling statement. The H$_2$
QSVT study in this paper is a controlled methodological testbed, not a
production level chemistry calculation. It lets us measure the full routed
cost of the implemented workflow and then state what changes when the same
construction is projected to a larger molecule such as LiH. The equations below give the polynomial degree argument, the segmentation
count, and a worked example at scaled time $\lambda t=5$.

For the implemented Green's function circuit, the total width is
\begin{equation}
  n_C = n_{\mathrm{sys}} + 5.
  \label{eq:width_main_restructured}
\end{equation}
Here $n_{\mathrm{sys}}$ is the number of system qubits in the Hamiltonian
register. The five additional qubits are the Hadamard test control, the
annihilation ancilla used for $c_j$, and the three QSVT registers used by the
present block encoding and real or imaginary projection circuit. This count is
therefore specific to the circuit template used in this paper. For H$_2$, the
Hamiltonian acts on a $16$ dimensional space, so $n_{\mathrm{sys}}=4$ and
$n_C=9$. For a LiH instance represented on ten system qubits, the same template
would use
\begin{equation}
  n_C^{\mathrm{LiH}} = 10 + 5 = 15
  \label{eq:lih_width_main_restructured}
\end{equation}
qubits before backend routing. The qubit width remains modest; the binding
resource is the routed gate volume.

The QSVT degree and segment count are controlled by the scaled time
$\tau=\lambda t$ and the target approximation error $\varepsilon$, not directly
by the molecule name. With block encoding normalization
$\lambda\geq\lVert H\rVert$, the standard real time QSVT scaling gives
\begin{equation}
  d = \mathcal{O}\!\left(\lambda t + \log(1/\varepsilon)\right),
  \label{eq:qsvt_degree_main_restructured}
\end{equation}
and the segmented implementation uses
\begin{equation}
  r = \left\lceil \frac{\lambda t}{\tau_{\max}} \right\rceil
  \label{eq:segments_main_restructured}
\end{equation}
segments when each segment is capped at scaled duration $\tau_{\max}$. Each segment uses $\tau_{\max}=1$. For the cosine and sine QSVT branches, the block
encoding query count in this circuit template is
\begin{equation}
  N_{\mathrm{oracle}}
  = 2\,(d_{\cos}+d_{\sin})\,r
  = \mathcal{O}\!\left(\lambda t\,\log(\lambda t/\varepsilon)\right).
  \label{eq:oracle_main_restructured}
\end{equation}
The factor of two counts the alternating $U_H$ and $U_H^\dagger$ calls in the
implemented QSVT round. At fixed $\lambda t$ and fixed $\varepsilon$, Eqs.
\eqref{eq:qsvt_degree_main_restructured}-\eqref{eq:oracle_main_restructured}
are unchanged when we replace H$_2$ by LiH. What changes is the cost of
compiling each block encoding query.

We summarize the routed total gate count as
\begin{equation}
  G_b(t)
  = \mathcal{O}\!\Bigl(
      \mu_b\,(d_{\cos}+d_{\sin})\,r\,
      G_{\mathrm{block}}(n_{\mathrm{sys}})
    \Bigr),
  \qquad
  G_{\mathrm{block}}(n_{\mathrm{sys}})
  = \mathcal{O}\!\left(4^{n_{\mathrm{sys}}}\right).
  \label{eq:greens_gate_main_restructured}
\end{equation}
The factor $\mu_b$ represents backend routing and basis translation overhead
for backend $b$. The term $G_{\mathrm{block}}$ bounds the cost of synthesizing
the block encoding query for the Hamiltonian register in the generic dense
unitary model. Structured and sparse block encodings can lower this constant,
but the exponential dependence marks the conservative scaling envelope for the
present direct embedding approach.

Under that envelope, moving from H$_2$ with $n_{\mathrm{sys}}=4$ to LiH with
$n_{\mathrm{sys}}=10$ multiplies the per query block cost by
\begin{equation}
  \frac{G_{\mathrm{block}}(10)}{G_{\mathrm{block}}(4)}
  = 4^{10-4}
  = 4^{6}
  \approx 4.1\times 10^{3}.
  \label{eq:lih_inflation_main_restructured}
\end{equation}
Using the empirical H$_2$ time sweep as a linear proxy gives the projected LiH
routed gate count
\begin{equation}
  G_b^{\mathrm{LiH}}(\tau)
  \approx 4.1\times 10^{3}\,G_b^{\mathrm{H}_2}(\tau),
  \qquad \tau=\lambda t.
  \label{eq:lih_projection_main_restructured}
\end{equation}
For example, the H$_2$ fit gives roughly $2.5\times10^{8}$ routed gates at
$\lambda t=5$ on the four backends, so the corresponding LiH
projection is about $1.0\times10^{12}$ routed gates per Green's function
evaluation. At $\lambda t=12.5$, the same linear model gives roughly
$2.6\times10^{12}$ routed gates. These LiH numbers are projections, not
measured routed LiH circuits. They show why the H$_2$ results should be read as
an application level benchmark and a resource calibrated starting point: the
fifteen qubit LiH width is feasible in register size, but the routed gate
volume is far beyond the present noisy circuit regime without better block
encodings, lower routing overhead, or error corrected execution.

\subsection{Cross paradigm interpretation}
Read together, the two branches support one coherent claim. The VQA branch
shows that our UQ language scales to a broad family of variational tasks and
exposes workload dependent backend selectivity. The QSVT branch
shows that the same UQ language benchmarks a non-variational matrix
function workload by tracking spectral reliability, phase sensitivity,
robust region geometry, and compiled cost. Our framework therefore performs
robust UQ across backends and across workloads rather than isolated per
algorithm tuning. The same noise aware loop uses backend responses to learn
perturbed parameters and quantify how close each backend can be driven
toward its ideal variational or spectral output.

Comparison across branches is scientifically useful in both directions.
Agreement signals backend behavior. Disagreement is equally useful,
because it exposes selectivity across workloads and warns against using any
single benchmark as a universal proxy for noisy quantum systems. Our
framework is constructive in this sense rather than only diagnostic: it
uses backend specific noisy evaluations to learn perturbed control
parameters that compensate for the noise channel, and treats the remaining
gap to the ideal reference as the backend characterization signal. This
distinguishes our score from generic QCVV style device level metrics,
which serve as the predictive layer for hardware and compiler stacks
~\cite{hashim2025qcvv,blumekohout2025qcvv}, while our score answers the
narrower application facing question of which backend most reliably supports
the VQA and QSVT workloads studied here under uncertainty, compiled
cost, and robust region geometry.

\begin{figure}[t]
\centering
\includegraphics[width=1.0\textwidth]{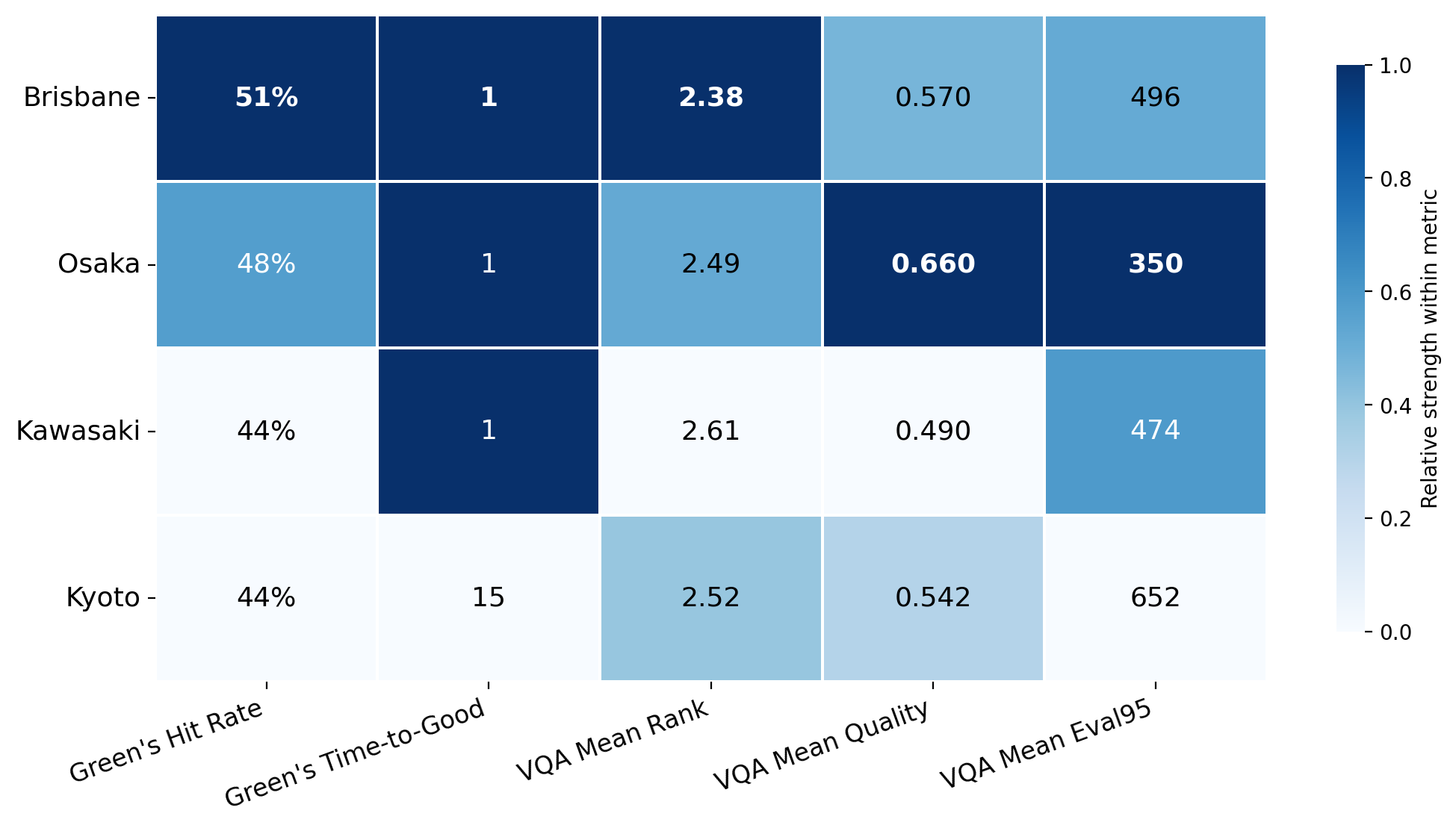}
\caption{Cross paradigm backend summary panel for the four backends of \S\ref{sec:design_restructured}. The five columns are computed
as follows. (i) Green's Hit Rate is the fraction of the $100$ QSVT
evaluations on each backend that landed in the good tuple regime defined in
Eq.~\eqref{eq:guided_main_restructured}, that is, the fraction of phase
vectors that met every guided spectral peak target. (ii) Green's
Time-to-Good is the evaluation index at which the first good tuple was
recorded on each backend, so lower is faster. (iii) VQA Mean Rank averages
the within workload rank of each backend across the ten VQA workloads, so
lower is more consistent (the ranks themselves come from the same
\textsf{rank\_in\_task} field used in Figure~\ref{fig:vqa_main_restructured}).
(iv) VQA Mean Quality averages the per workload $\mathrm{quality\_norm}$
across the ten VQA workloads, so higher is stronger. (v) VQA Mean Eval95 is
the mean number of evaluations needed to reach $95\%$ of the per workload
best value, averaged across the ten VQAs, so lower is more efficient.
Within each column we apply an independent min-max rescaling so darker
cells indicate stronger relative performance, with column specific
orientation (high or low better); cell annotations preserve the raw values.
This summary panel condenses our integrated benchmark at a glance: Brisbane is
strongest on QSVT reliability, Osaka is strongest on aggregate VQA
quality and efficiency, Kyoto is selective rather than uniformly strong,
and Kawasaki is competitive on specific VQA workloads without leading the four backends aggregate views. Read as a characterization signal, the
panel shows that backend ranking is metric and workload dependent
rather than universal.}
\label{fig:cross_paradigm_score}
\end{figure}

Figure~\ref{fig:cross_paradigm_score} provides the compact backend to
backend comparison panel typically expected of a benchmark study, while
keeping the comparison in the application level language used throughout
this paper rather than reverting to generic circuit level diagnostics. The
underlying claim still rests on spectral reliability, the VQA quality
landscape, convergence, sensitivity, and routed cost rather than on abstract
width or depth alone.

\section{Scope and Limitations}
\label{sec:limitations_restructured}
Two boundaries scope our claims. First, all reported results use IBM
fake backend noise models and Aer execution rather than live QPUs. The
advantage is that we can produce every evaluation without queue delays or
calibration drift, and the trade off is that day to day live hardware
variation does not enter our score. Second, the VQA side noise
alignment layer uses static properties extracted from the four IBM
fake backends rather than from live calibration snapshots. The resulting
noise to performance correlation analysis is therefore a qualitative
monotonic association, not a live QPU calibration claim.

The QSVT side of our work uses an H$_2$ Hamiltonian and therefore
serves as a tractable testbed for the unified UQ pipeline rather than as a
production level chemistry calculation. The statistical workflow can be reused
for LiH at ten system qubits and for larger molecules, but the circuit resource
model changes sharply: the projection in
\S\ref{sec:greens_resource_extrapolation} shows that LiH remains modest in
qubit width while its routed gate volume rises by orders of magnitude under the
present direct embedding approach. Larger QSVT chemistry therefore
requires improved block encodings, lower routing overhead, or error corrected
execution.

The five benchmarking metrics introduced in
\S\ref{sec:metrics_restructured} are designed to guide future, more capable
quantum simulations. They compare implementations through task accuracy,
uncertainty, robust parameter geometry, routed resource cost, and noise
association against the same ideal reference, so future QSVT and VQA
work can use the same score as hardware and algorithms improve.
Appendix~\ref{app:extended_scope} records additional threats to validity
that scope these claims.

\section{Conclusion}
\label{sec:conclusion_restructured}

Our uncertainty quantification framework provides a common statistical language for benchmarking and characterizing noisy quantum systems across both a broad VQA family and QSVT workloads. On the VQA branch, our framework measures whether a backend can reliably support distinct variational objective geometries across chemistry, optimization, simulation, compiling, metrology, tomography, and error correction style tasks. On the QSVT branch, our framework measures whether a backend can reliably recover a physically meaningful spectral observable. Resource estimation strengthens both readings by showing that reliability and compiled cost must be interpreted together.

Operationally, by evaluating both algorithm families under backend specific noise models, our loop learns perturbed control parameters that move VQA objectives and QSVT spectra as close as possible to their ideal references. The benchmark therefore acts simultaneously as a characterization tool and as a practical recipe for recovering useful task level behavior from noisy backends whenever a robust region exists. Our integrated framework is about robust UQ across backends and across workloads, not about two unrelated case studies.

{\sloppy
Our integrated benchmark also yields concrete backend conclusions. On the four backends VQA branch, the main result is not a single universal winner but a pattern of strengths that vary across workloads: Osaka leads QAOA/\allowbreak VQC, Brisbane leads VQE/\allowbreak VQPTNU, Kyoto leads VQTE/\allowbreak VQEC/\allowbreak VQCFE, and Kawasaki leads VQAPDE/\allowbreak VQLS, while Brisbane is the strongest aggregate backend by average rank and Osaka is strongest by mean normalized quality. On the QSVT side of our work, Brisbane is the strongest backend under the application level criterion used here, followed by Osaka, Kawasaki, and Kyoto. These outcomes demonstrate that backend quality is application dependent, and uncertainty quantification makes that dependence measurable through reliability, convergence, robust region geometry, and
compiled cost.\par}

\section{ACKNOWLEDGMENTS}

This work is supported by the NSF grants 2230111, 2238734, 2311950, and 2513432. B.P. acknowledges the support from the Early Career Research Program by the U.S. Department of Energy, Office of Science, under Grant No. FWP 83466.

\bibliographystyle{ACM-Reference-Format}
\bibliography{main_bibliography}

\appendix
\section*{Appendices}

\section{Detailed Variational Quantum Workload Family and Backend Aware Objective Maps}
\label{sec:vqa_family}
The integrated benchmark has two branches: a broad family of variational
quantum algorithms and the Green's/QSVT workflow. We discuss the VQA
branch first because it defines the backend conditioned evaluation language
used throughout the paper. This section records the configured instances,
the shared variational template, and a compact objective map for the ten VQA
families.

Table~\ref{tab:vqa_instances_main} records the concrete configured instance
behind each workload. It pins the physical or mathematical problem (molecule,
graph, target unitary, Hamiltonian, linear system, PDE grid, sensing phase,
code family, process model, or channel stack) together with what the optimizer
drives toward, so every quality number in \S\ref{sec:results_restructured}.1
traces back to a fixed, reproducible task rather than to a workload label. The
table also explains the difficulty and cost ordering observed in the main
text: the heavy chemistry and combinatorial instances (VQE, QAOA) carry deep
circuits, while the sensing and PDE instances (VQAMET, VQAPDE) stay shallow,
which lines up with the resource regimes of
Figure~\ref{fig:vqa_resource_main_restructured}.

\begin{table}[tbp]
\centering
\scriptsize
\setlength{\tabcolsep}{2pt}
\caption{Concrete VQA benchmark instances used in our work; the guided loop configuration then adds the
cross backend BOpt/UQ controls on top of the same per workload task definitions.}
\label{tab:vqa_instances_main}
\begin{tabular}{>{\raggedright\arraybackslash}p{0.11\linewidth} >{\raggedright\arraybackslash}p{0.46\linewidth} >{\raggedright\arraybackslash}p{0.27\linewidth}}
\toprule
Workload & Active configured instance in our work & What the optimizer is trying to achieve \\
\midrule
VQE & LiH at fixed geometry (Li-H distance $1.6\,\text{\AA}$), STO-3G basis, frozen core, parity mapping, Hartree-Fock + UCCSD ansatz & Minimize the measured molecular ground state energy toward the classical reference value \\
QAOA & MaxCut on the default complete graph on eight nodes, with depth $p=20$ & Tune the alternating angles so the output distribution concentrates on high quality cuts \\
VQC & Variational compiling with \texttt{unitary\_type = E}, which in the implementation corresponds to a QFT target, using one compilation round & Maximize overlap with the target unitary so the learned circuit reproduces the desired transformation \\
VQTE & Real time evolution under a fixed eight qubit nine term Pauli Hamiltonian, with total time $1.0$ and three time slices & Match the exact reference evolution as measured through the backend energy of the evolved state \\
VQLS & Seeded 13 variable random invertible linear system, decomposed into Hermitian and anti Hermitian quadratic costs & Minimize a residual inspired linear system cost toward the classical optimum \\
VQAPDE & Four dimensional Poisson type problem on a $2\times2\times2\times2$ grid with Dirichlet boundary conditions and $\alpha=-1$ & Minimize the discretized PDE energy toward the exact small instance ground energy \\
VQAMET & Ten qubit Ramsey style sensing benchmark with encoded phase $\phi=\pi/4$ & Reproduce the target phase response through averaged $Z$ measurements across the sensing register \\
VQEC & Code family \texttt{((5,2,3))[5]} with $L_2$ loss and a $25\%$ sampled error set & Learn an encoder that keeps logical information distinguishable after the allowed error family \\
VQPTNU & Twelve system qubit Stinespring process model with one ancilla, germs built from $G_x/G_y$, and Z/X/Y measurements & Match target process statistics across the tomography experiment set with minimal MSE mismatch \\
VQCFE & Three qubit GHZ benchmark with channel stack [amplitude damping, toy depolarizing, amplitude damping] and strengths [0.2, 0.1, 0.3] & Maximize the reconstructed fidelity with the ideal GHZ state after noisy evolution \\
\bottomrule
\end{tabular}
\end{table}

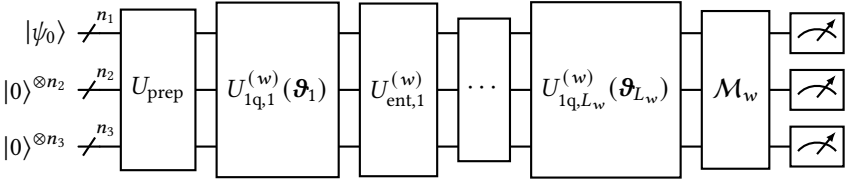
\begin{figure}[h]
\centering
\begin{quantikz}[row sep={0.75cm,between origins}, column sep=0.28cm]
\lstick{$\ket{\psi_0}$} & \qwbundle{n_1} & \gate[3]{U_{\mathrm{prep}}} & \gate[3]{U^{(w)}_{\mathrm{1q},1}(\bm{\vartheta}_1)} & \gate[3]{U^{(w)}_{\mathrm{ent},1}} & \gate[3]{\cdots} & \gate[3]{U^{(w)}_{\mathrm{1q},L_w}(\bm{\vartheta}_{L_w})} & \gate[3]{\mathcal{M}_w} & \meter{} \\
\lstick{$\ket{0}^{\otimes n_2}$} & \qwbundle{n_2} & \ghost{U_{\mathrm{prep}}} & \ghost{U^{(w)}_{\mathrm{1q},1}(\bm{\vartheta}_1)} & \ghost{U^{(w)}_{\mathrm{ent},1}} & \ghost{\cdots} & \ghost{U^{(w)}_{\mathrm{1q},L_w}(\bm{\vartheta}_{L_w})} & \ghost{\mathcal{M}_w} & \meter{} \\
\lstick{$\ket{0}^{\otimes n_3}$} & \qwbundle{n_3} & \ghost{U_{\mathrm{prep}}} & \ghost{U^{(w)}_{\mathrm{1q},1}(\bm{\vartheta}_1)} & \ghost{U^{(w)}_{\mathrm{ent},1}} & \ghost{\cdots} & \ghost{U^{(w)}_{\mathrm{1q},L_w}(\bm{\vartheta}_{L_w})} & \ghost{\mathcal{M}_w} & \meter{}
\end{quantikz}
\caption{Shared variational workload template. Each VQA begins from a task
specific preparation or encoding block $U_{\mathrm{prep}}$, applies trainable
one body and entangling layers, and ends with a workload specific measurement
or tomography map $\mathcal{M}_w$ that produces the scalar objective used by
the UQ analysis. Slashed wires denote multi qubit registers rather than
single qubits; the register sizes $n_1$, $n_2$, and $n_3$ are workload
specific.}
\label{fig:vqa_template_main}
\end{figure}

% \subsection{Shared VQA Formulation}
% \label{sec:vqa_formulation}
Each variational workload chooses a trainable circuit family whose reachable
states or channels form a restricted manifold
\begin{equation}
  \mathcal{M}_w
  =
  \left\{
    U_w(\bm{\vartheta}) \rho_0 U_w(\bm{\vartheta})^\dagger
    \;:\;
    \bm{\vartheta}\in\Theta_w
  \right\},
  \label{eq:vqa_manifold_main}
\end{equation}
and then optimizes over $\bm{\vartheta}$ so that the prepared state, unitary,
or channel best fits the scientific task. Across the implementation, the
trainable family almost always has a layered form
\begin{equation}
  U_w(\bm{\vartheta})
  =
  \prod_{\ell=1}^{L_w}
  U^{(w)}_{\mathrm{ent},\ell}\,
  U^{(w)}_{\mathrm{1q},\ell}\!\left(\bm{\vartheta}_\ell\right),
  \label{eq:vqa_layered_ansatz_main}
\end{equation}
where $U^{(w)}_{\mathrm{1q},\ell}$ collects one qubit or one body trainable
blocks and $U^{(w)}_{\mathrm{ent},\ell}$ encodes the workload specific
entangling structure. Figure~\ref{fig:vqa_template_main} gives the common
circuit template behind the ten configured workloads.

Operationally, each VQA instance generates evaluations of the form
$(\bm{\vartheta},y)$ by executing a parameterized circuit on a chosen backend,
measuring an expectation value, fidelity, tomography mismatch, or sensing cost,
and passing those records into the same BOpt+MCMC/VI engine used elsewhere in
the paper. The sign convention is workload dependent: $J$ denotes a minimized
loss or cost, whereas $S$ and $F$ denote maximized overlap or fidelity scores.
This convention affects only how the optimizer scores the records.

% \subsection{Compact VQA Circuit and Objective Map}
% \label{app:vqa_circuit_map}
Table~\ref{tab:vqa_objective_map_appendix} summarizes the ten VQA circuits as
compact TeX circuit flows. Each row records the input state or probe, the main
trainable block, the measurement stage, the scalar backend score, the
optimization direction, and the canonical citation for the workload.

\begin{table}[tbp]
\centering
\scriptsize
\setlength{\tabcolsep}{1.6pt}
\renewcommand{\arraystretch}{1.15}
\caption{Compact circuit and objective map for the ten VQA workloads. The
circuit flow column gives the workload specific circuit structure in symbolic
TeX form, while the score column gives the backend level objective used by the
optimizer.}
\label{tab:vqa_objective_map_appendix}
\begin{tabular}{>{\raggedright\arraybackslash}p{0.10\linewidth} >{\raggedright\arraybackslash}p{0.34\linewidth} >{\raggedright\arraybackslash}p{0.42\linewidth} >{\raggedright\arraybackslash}p{0.08\linewidth}}
\toprule
Workload & Compact TeX circuit flow & Backend score or objective equation & Direction \\
\midrule
VQE~\cite{peruzzo2014vqe} & $\ket{\psi_{\mathrm{HF}}}\to U_{\mathrm{UCCSD}}(\bm{\vartheta})\to$ Pauli term measurements & $J_{\mathrm{VQE},b}(\bm{\vartheta})=\langle H_{\mathrm{mol}}\rangle_b=\sum_\ell c_\ell\langle P_\ell\rangle_b$ & Min. \\
QAOA~\cite{farhi2014qaoa} & $\ket{+}^{\otimes n}\to\prod_{\ell=1}^{p}U_M(\beta_\ell)U_C(\gamma_\ell)\to$ computational basis samples & $J_{\mathrm{QAOA},b}(\bm{\gamma},\bm{\beta})=\langle H_C\rangle_{\bm{\gamma},\bm{\beta},b}$ for the encoded MaxCut cost Hamiltonian & Min. \\
VQC~\cite{khatri2019qaqc} & $U_{\mathrm{in}}\to V_1(\bm{\vartheta}_1)\cdots V_R(\bm{\vartheta}_R)\to$ local or Hilbert-Schmidt overlap test & $S_{\mathrm{VQC},b}=|\mathrm{Tr}(U_{\mathrm{tar}}^\dagger V(\bm{\vartheta}))|/2^n$ or local overlap proxy & Max. \\
VQTE~\cite{yuan2019vqs} & $\ket{\psi(0)}\to\prod_m\exp(-i\vartheta_m P_m)\to H$ measurement over time slices & McLachlan update $A(\bm{\vartheta})\dot{\bm{\vartheta}}=\bm{C}(\bm{\vartheta})$; backend score compares evolved energy with the exact trajectory & Min. \\
VQLS~\cite{bravoprieto2023vqls} & $\ket{0}\to U_{\mathrm{ans}}(\bm{\vartheta})\to$ Hermitian/anti Hermitian residual cost blocks & $J_{\mathrm{VQLS},b}(\bm{\vartheta})\approx\|A\ket{x(\bm{\vartheta})}-\ket{b}\|^2$ & Min. \\
VQAPDE~\cite{sato2021vqapde} & $\ket{0}\to$ HEA layers $(R_y,\mathrm{ring},R_z)\to H_{\mathrm{PDE}}$ and source measurements & $J_{\mathrm{VQAPDE},b}=\frac12\langle H_{\mathrm{PDE}}\rangle_b-\Re\langle g|\psi(\bm{\vartheta})\rangle_b$ & Min. \\
VQAMET~\cite{meyer2021vqmet} & $\ket{+}^{\otimes n}\to R_z(\varphi)\to$ trainable sensing rotations $\to Z$ measurements & $J_{\mathrm{VQAMET},b}(\bm{\vartheta})=(\overline{\langle Z\rangle}_{\bm{\vartheta},b}-\cos\varphi)^2$ & Min. \\
VQEC~\cite{xu2021vqec} & $\ket{k_L}\to U_{\mathrm{enc}}(\bm{\vartheta})\to E\in\mathcal{E}_{d-1}\to$ logical readout & $J_{\mathrm{VQEC},b}=\frac{1}{K|\mathcal{E}_{d-1}|}\sum_{k,E}\Delta_F(p_{k,E,b})$ & Min. \\
VQPTNU~\cite{xue2023vqptnu} & $\rho_f\to$ fiducial preparation $\to g^e\to\Lambda_{\bm{\vartheta}}\to M^{(\mu)}$ & $J_{\mathrm{VQPTNU},b}=\sum_m\|\widehat{\bm{p}}_m(\bm{\vartheta};b)-\bm{p}^{\star}_m\|_2^2$ & Min. \\
VQCFE~\cite{cerezo2020vqfe} & $\ket{0}^{\otimes n}\to U_{\mathrm{GHZ}}\to\mathcal{N}_{\chi}\to U_{\mathrm{corr}}(\bm{\vartheta})\to$ tomography & $F_{\mathrm{GHZ},b}(\bm{\vartheta})=F(\rho_{\mathrm{GHZ}}^{\mathrm{ideal}},\rho_{\bm{\vartheta},b}^{\mathrm{recon}})$, equivalently $J=1-F$ & Max. \\
\bottomrule
\end{tabular}
\end{table}
\renewcommand{\arraystretch}{1.0}

Table~\ref{tab:vqa_objective_map_appendix} makes the shared tuple interface of
Eq.~\eqref{eq:generic_tuple_restructured} concrete: every row ends in one
scalar backend score, and the direction column records whether the optimizer
maximizes or minimizes that scalar. Reading down the score column also shows
why the min-max normalization of \S\ref{sec:metrics_restructured} is
necessary, because the raw objectives live on incompatible scales (energies,
residuals, fidelities, and mismatch sums). The row structure identifies which
claims each workload can support: expectation and residual driven rows probe
objective estimation quality, while the tomography and fidelity rows probe
measurement channel behavior.

Table~\ref{tab:vqa_compact_circuit_atlas} gives a visual block-level companion
to Table~\ref{tab:vqa_objective_map_appendix}. The schematics are intentionally
coarse: they show preparation, trainable evolution, noise/channel or error
insertion where relevant, and the final measurement map without expanding each
ansatz into elementary gates.

\begin{table}[tbp]
\centering
\scriptsize
\setlength{\tabcolsep}{2pt}
\renewcommand{\arraystretch}{1.15}
\caption{Compact circuit level representation for the ten VQA workloads. Each of them is a small circuit showing the same workload structure summarized
by the circuit flow column of Table~\ref{tab:vqa_objective_map_appendix}.
Slashed wires denote multi qubit registers rather than single qubits: each
schematic circuit shows the workload's qubits into two registers whose sizes $n$
and $m$ are workload specific.}
\label{tab:vqa_compact_circuit_atlas}
\begin{tabular}{>{\raggedright\arraybackslash}p{0.10\linewidth} >{\centering\arraybackslash}p{0.84\linewidth}}
\toprule
VQAs & Compact circuit representation\\
\midrule
VQE~\cite{peruzzo2014vqe} &
\resizebox{0.65\linewidth}{!}{%
\begin{quantikz}[row sep={0.68cm,between origins}, column sep=0.24cm]
\lstick{$\ket{\psi_{\mathrm{HF}}}$} & \qwbundle{n} & \gate[2]{U_{\mathrm{UCCSD}}(\bm{\vartheta})} & \gate[2]{\text{measure }P_\ell} & \meter{} \\
\lstick{$\ket{0}^{\otimes m}$} & \qwbundle{m} & \ghost{U_{\mathrm{UCCSD}}(\bm{\vartheta})} & \ghost{\text{measure }P_\ell} & \meter{}
\end{quantikz}} \\
QAOA~\cite{farhi2014qaoa} &
\resizebox{0.65\linewidth}{!}{%
\begin{quantikz}[row sep={0.68cm,between origins}, column sep=0.24cm]
\lstick{$\ket{+}^{\otimes n}$} & \qwbundle{n} & \gate[2]{U_C(\gamma_1)} & \gate[2]{U_M(\beta_1)} & \gate[2]{\cdots} & \gate[2]{U_M(\beta_p)} & \meter{} \\
\lstick{$\ket{+}^{\otimes m}$} & \qwbundle{m} & \ghost{U_C(\gamma_1)} & \ghost{U_M(\beta_1)} & \ghost{\cdots} & \ghost{U_M(\beta_p)} & \meter{}
\end{quantikz}} \\
VQC~\cite{khatri2019qaqc} &
\resizebox{0.65\linewidth}{!}{%
\begin{quantikz}[row sep={0.68cm,between origins}, column sep=0.24cm]
\lstick{$\ket{0}^{\otimes n}$} & \qwbundle{n} & \gate[2]{U_{\mathrm{in}}} & \gate[2]{V(\bm{\vartheta})} & \gate[2]{\text{local / HS test}} & \meter{} \\
\lstick{$\ket{0}^{\otimes m}$} & \qwbundle{m} & \ghost{U_{\mathrm{in}}} & \ghost{V(\bm{\vartheta})} & \ghost{\text{local / HS test}} & \meter{}
\end{quantikz}} \\
VQTE~\cite{yuan2019vqs} &
\resizebox{0.65\linewidth}{!}{%
\begin{quantikz}[row sep={0.68cm,between origins}, column sep=0.24cm]
\lstick{$\ket{\psi(0)}$} & \qwbundle{n} & \gate[2]{\ee^{-\ii\vartheta_1P_1}} & \gate[2]{\cdots} & \gate[2]{\ee^{-\ii\vartheta_mP_m}} & \gate[2]{\text{measure }H} & \meter{} \\
\lstick{$\ket{0}^{\otimes m}$} & \qwbundle{m} & \ghost{\ee^{-\ii\vartheta_1P_1}} & \ghost{\cdots} & \ghost{\ee^{-\ii\vartheta_mP_m}} & \ghost{\text{measure }H} & \meter{}
\end{quantikz}} \\
VQLS~\cite{bravoprieto2023vqls} &
\resizebox{0.65\linewidth}{!}{%
\begin{quantikz}[row sep={0.68cm,between origins}, column sep=0.24cm]
\lstick{$\ket{0}^{\otimes n}$} & \qwbundle{n} & \gate[2]{U_{\mathrm{ans}}(\bm{\vartheta})} & \gate[2]{\text{cost}(H,S)} & \gate[2]{\text{residual proxy}} & \meter{} \\
\lstick{$\ket{0}^{\otimes m}$} & \qwbundle{m} & \ghost{U_{\mathrm{ans}}(\bm{\vartheta})} & \ghost{\text{cost}(H,S)} & \ghost{\text{residual proxy}} & \meter{}
\end{quantikz}} \\
VQAPDE~\cite{sato2021vqapde} &
\resizebox{0.65\linewidth}{!}{%
\begin{quantikz}[row sep={0.68cm,between origins}, column sep=0.24cm]
\lstick{$\ket{0}^{\otimes n}$} & \qwbundle{n} & \gate[2]{\text{HEA}(R_y,\mathrm{ring},R_z)} & \gate[2]{\text{measure }H_{\mathrm{PDE}},g} & \meter{} \\
\lstick{$\ket{0}^{\otimes m}$} & \qwbundle{m} & \ghost{\text{HEA}(R_y,\mathrm{ring},R_z)} & \ghost{\text{measure }H_{\mathrm{PDE}},g} & \meter{}
\end{quantikz}} \\
VQAMET~\cite{meyer2021vqmet} &
\resizebox{0.65\linewidth}{!}{%
\begin{quantikz}[row sep={0.68cm,between origins}, column sep=0.24cm]
\lstick{$\ket{+}^{\otimes n}$} & \qwbundle{n} & \gate[2]{R_z(\varphi)} & \gate[2]{U_{\mathrm{sens}}(\bm{\vartheta})} & \gate[2]{Z\text{ readout}} & \meter{} \\
\lstick{$\ket{+}^{\otimes m}$} & \qwbundle{m} & \ghost{R_z(\varphi)} & \ghost{U_{\mathrm{sens}}(\bm{\vartheta})} & \ghost{Z\text{ readout}} & \meter{}
\end{quantikz}} \\
VQEC~\cite{xu2021vqec} &
\resizebox{0.65\linewidth}{!}{%
\begin{quantikz}[row sep={0.68cm,between origins}, column sep=0.24cm]
\lstick{$\ket{k_L}$} & \qwbundle{n} & \gate[2]{U_{\mathrm{enc}}(\bm{\vartheta})} & \gate[2]{E\in\mathcal{E}_{d-1}} & \gate[2]{\text{logical readout}} & \meter{} \\
\lstick{$\ket{0}^{\otimes m}$} & \qwbundle{m} & \ghost{U_{\mathrm{enc}}(\bm{\vartheta})} & \ghost{E\in\mathcal{E}_{d-1}} & \ghost{\text{logical readout}} & \meter{}
\end{quantikz}} \\
VQPTNU~\cite{xue2023vqptnu} &
\resizebox{0.65\linewidth}{!}{%
\begin{quantikz}[row sep={0.68cm,between origins}, column sep=0.24cm]
\lstick{$\rho_f$} & \qwbundle{n} & \gate[2]{\text{fiducial}} & \gate[2]{g^e} & \gate[2]{\Lambda_{\bm{\vartheta}}} & \gate[2]{M^{(\mu)}} & \meter{} \\
\lstick{$\ket{0}^{\otimes m}$} & \qwbundle{m} & \ghost{\text{fiducial}} & \ghost{g^e} & \ghost{\Lambda_{\bm{\vartheta}}} & \ghost{M^{(\mu)}} & \meter{}
\end{quantikz}} \\
VQCFE~\cite{cerezo2020vqfe} &
\resizebox{0.65\linewidth}{!}{%
\begin{quantikz}[row sep={0.68cm,between origins}, column sep=0.24cm]
\lstick{$\ket{0}^{\otimes n}$} & \qwbundle{n} & \gate[2]{U_{\mathrm{GHZ}}} & \gate[2]{\mathcal{N}_{\chi}} & \gate[2]{U_{\mathrm{corr}}(\bm{\vartheta})} & \gate[2]{\text{tomography}} & \meter{} \\
\lstick{$\ket{0}^{\otimes m}$} & \qwbundle{m} & \ghost{U_{\mathrm{GHZ}}} & \ghost{\mathcal{N}_{\chi}} & \ghost{U_{\mathrm{corr}}(\bm{\vartheta})} & \ghost{\text{tomography}} & \meter{}
\end{quantikz}} \\
\bottomrule
\end{tabular}
\end{table}
\renewcommand{\arraystretch}{1.0}

This compact map preserves the central distinction among the workloads. VQE,
QAOA, VQTE, VQLS, and VQAPDE are expectation or residual driven; VQAMET is a
sensing response benchmark; VQEC asks whether logical information remains
separable after an error family; VQPTNU is a process distribution matching
problem; and VQCFE is a channel fidelity benchmark. The appendix therefore
retains both the circuit level interpretation and the application level meaning
of each objective without requiring separate circuit figures.

\section{QSP/QSVT, Block Encoding, Green's Function Circuit Level Estimator, and Real Time Segmentation of QSVT Propagator \texorpdfstring{$\ee^{-\ii H t}$}{e^{-iHt}}}
\label{sec:qsp_block_greens_appendix}
This section collects the Green's function target, block encoding model,
QSP/QSVT construction, phase synthesis details, real time propagator
approximation, and reconstruction estimator used by the benchmarking workflow.
The Green's function and its spectral peaks are the
\emph{observables} that define what a ``successful'' backend run means.
Block encoding and QSVT are the \emph{algorithmic primitives} that convert
a Hamiltonian into a quantum circuit whose phase angles become the
27 dimensional parameter space the optimizer and sensitivity analysis explore.
The order is deliberate: the appendix first defines the observable and signal
block, then gives the QSP/QSVT phase synthesis machinery, then specializes the
polynomial transform to real time propagation, and finally connects the
propagator to the post-selected Green's function estimator and its error chain.

Let $A \in \C^{N\times N}$ and let $\alpha > 0$. A unitary $U_A$ acting on an ancilla register of size $a$ and a system register of size $\log_2 N$ is an $(\alpha,a)$ block encoding of $A$ if
\begin{equation}
  \left( \bra{0^a}\otimes I \right)
  U_A
  \left( \ket{0^a}\otimes I \right)
  =
  \frac{A}{\alpha}.
  \label{eq:block_encoding}
\end{equation}

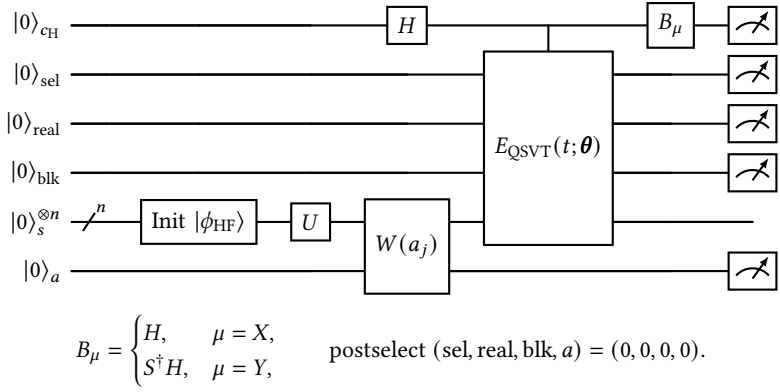
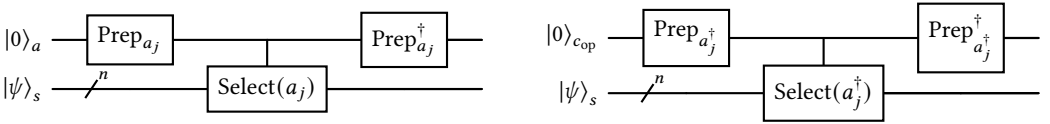
\begin{figure}[tbp]
\centering
\small

\begin{subfigure}{\textwidth}
\centering
\textbf{(a) Noisy Hadamard estimator path (for actual quantum computers)}
\[
\begin{quantikz}[row sep={0.65cm,between origins}, column sep=0.46cm]
\lstick{$\ket{0}_{c_{\mathrm H}}$}
  & \qw & \qw & \qw & \gate{H} & \ctrl{1} & \gate{B_\mu} & \meter{} \\
\lstick{$\ket{0}_{\mathrm{sel}}$}
  & \qw & \qw & \qw & \qw & \gate[wires=4]{E_{\mathrm{QSVT}}(t;\pmb{\theta})} & \qw & \meter{} \\
\lstick{$\ket{0}_{\mathrm{real}}$}
  & \qw & \qw & \qw & \qw & \ghost{E_{\mathrm{QSVT}}(t;\pmb{\theta})} & \qw & \meter{} \\
\lstick{$\ket{0}_{\mathrm{blk}}$}
  & \qw & \qw & \qw & \qw & \ghost{E_{\mathrm{QSVT}}(t;\pmb{\theta})} & \qw & \meter{} \\
\lstick{$\ket{0}^{\otimes n}_{s}$}
  & \qwbundle{n} & \gate{\mathrm{Init}\,\ket{\phi_{\mathrm{HF}}}} & \gate{U} & \gate[wires=2]{W(a_j)} & \ghost{E_{\mathrm{QSVT}}(t;\pmb{\theta})} & \qw & \qw \\
\lstick{$\ket{0}_{a}$}
  & \qw & \qw & \qw & \ghost{W(a_j)} & \qw & \qw & \meter{}
\end{quantikz}
\]
\[
B_\mu=
\begin{cases}
H, & \mu=X,\\
S^\dagger H, & \mu=Y,
\end{cases}
\qquad
\text{postselect }(\mathrm{sel},\mathrm{real},\mathrm{blk},a)=(0,0,0,0).
\]

\vspace{0.5em}
\textbf{(b) Ideal simulator path}
\[
\begin{quantikz}[row sep={0.65cm,between origins}, column sep=0.46cm]
\lstick{$\ket{0}_{a}$}
  & \qw & \qw & \qw & \gate[wires=2]{W(a_j)} & \qw & \qw & \qw & \qw \\
\lstick{$\ket{0}^{\otimes n}_{s}$}
  & \qwbundle{n} & \gate{\mathrm{HFprep}} & \gate{U} & \ghost{W(a_j)} & \gate{E_{\mathrm{QSVT}}(t;\pmb{\theta})} & \gate[wires=2]{W(a_j^\dagger)} & \gate{U^\dagger} & \gate{\mathrm{HFprep}^\dagger} \\
\lstick{$\ket{0}_{c_{\mathrm{op}}}$}
  & \qw & \qw & \qw & \qw & \qw & \ghost{W(a_j^\dagger)} & \qw & \qw
\end{quantikz}
\]
\caption{Compact comparison of the two implementations targeting $\alpha_j(t)$.
(a) Noisy Hadamard estimator: one explicit operator block $W(a_j)$, then X/Y
interference readout on control.
(b) Ideal explicit chain: direct composition
$U\!\to\!W(a_j)\!\to\!E_{\mathrm{QSVT}}\!\to\!W(a_j^\dagger)\!\to\!U^\dagger$,
with separate ancillas to avoid block mixing contamination.}
\label{fig:alphaj_two_paths}
\end{subfigure}

\vspace{1em}

\begin{subfigure}{\textwidth}
\centering
\textbf{(c) Operator block encodings (LCU form)}
\[
\begin{quantikz}[row sep={0.65cm,between origins}, column sep=0.46cm]
\lstick{$\ket{0}_{a}$}
  & \gate{\mathrm{Prep}_{a_j}}
  & \ctrl{1}
  & \gate{\mathrm{Prep}_{a_j}^{\dagger}}
  & \qw \\
\lstick{$\ket{\psi}_{s}$}
  & \qwbundle{n}
  & \gate{\mathrm{Select}(a_j)}
  & \qw
  & \qw
\end{quantikz}
\qquad
\begin{quantikz}[row sep={0.73cm,between origins}, column sep=0.46cm]
\lstick{$\ket{0}_{c_{\mathrm{op}}}$}
  & \gate{\mathrm{Prep}_{a_j^\dagger}}
  & \ctrl{1}
  & \gate{\mathrm{Prep}_{a_j^\dagger}^{\dagger}}
  & \qw \\
\lstick{$\ket{\psi}_{s}$}
  & \qwbundle{n}
  & \gate{\mathrm{Select}(a_j^\dagger)}
  & \qw
  & \qw
\end{quantikz}
\]

\vspace{0.35em}
\textbf{(d) One segment inside $\pmb{E_{\mathrm{QSVT}}}$}
\[
\begin{quantikz}[row sep={0.65cm,between origins}, column sep=0.52cm]
\lstick{$\ket{0}_{\mathrm{sel}}$}
  & \gate{H}
  & \octrl{1}
  & \ctrl{1}
  & \gate{S}
  & \gate{H}
  & \qw \\
\lstick{$\ket{0}_{\mathrm{real}}$}
  & \qw
  & \gate[wires=3]{G_c(\pmb{\theta}_{\cos})}
  & \gate[wires=3]{G_s(\pmb{\theta}_{\sin})}
  & \qw
  & \qw
  & \qw \\
\lstick{$\ket{0}_{\mathrm{blk}}$}
  & \qw
  & \ghost{G_c(\pmb{\theta}_{\cos})}
  & \ghost{G_s(\pmb{\theta}_{\sin})}
  & \qw
  & \qw
  & \qw \\
\lstick{$\ket{0}^{\otimes n}_{s}$}
  & \qwbundle{n}
  & \ghost{G_c(\pmb{\theta}_{\cos})}
  & \ghost{G_s(\pmb{\theta}_{\sin})}
  & \qw
  & \qw
  & \qw
\end{quantikz}
\]
Apply this segment $R=\texttt{segments}$ times.

\caption{Component level breakdown used by the Green's function circuits:
LCU block encodings for $W(a_j)$, $W(a_j^\dagger)$, and one QSVT segment
inside $E_{\mathrm{QSVT}}$.}
\label{fig:alphaj_components}
\end{subfigure}

\caption{Green's function circuit overview and component breakdown. Panels
(a) and (b) compare the active noisy Hadamard estimator path with the ideal
explicit full chain construction for $\alpha_j(t)$. Panels (c) and (d) show the
operator block encodings and a single QSVT segment used inside
$E_{\mathrm{QSVT}}$. In every panel the slashed wire marks the $n$ qubit
system register $s$; all other wires are single ancilla qubits, which
matches the width count of Eq.~\eqref{eq:width_main_restructured}.}
\label{fig:alphaj_overview}
\end{figure}

Thus projecting the ancilla onto $\ket{0^a}$ exposes $A/\alpha$ as the
signal block. The normalization $\alpha$ keeps the signal spectrum inside the
range required by QSP/QSVT polynomial transformations.
This definition does not depend on the circuit construction that realizes the
signal block. The QSP/QSVT literature offers several such constructions,
including direct embedding, LCU based PREP-SELECT-PREP, qubitization style walk
encodings, and FABLE style factorized embeddings
\cite{gilyen2019qsvt,childs2012lcu,low2019qubitization,camps2022fable}. Our
numerical pipeline uses only the direct embedding construction of
Eq.~\eqref{eq:block_encoding}; we treat the other three as background and as
future extension paths.

% \subsection{Green's Function Circuit Level Estimator and Real Time Segmentation of QSVT Propagator \texorpdfstring{$\ee^{-\ii H t}$}{e^{-iHt}}}
% \label{app:real_time_segmentation_notes}

The Green's function estimator used in the noisy path can be decomposed into
four logical blocks. First, state preparation produces the Hartree-Fock
reference and the occupied orbital removal state. Second, a controlled QSVT
segment implements the real time propagator approximation
\(E_{\mathrm{QSVT}}(t;\bm{\theta})\). Third, Hadamard test readout in the
\(X\) and \(Y\) bases estimates the real and imaginary parts of
\(\alpha_j(t)\). Fourth, success branch postselection on the block encoding and
operator ancillas yields the conditional estimator in
Eq.~\eqref{eq:conditional_xy}. The ideal path is the corresponding explicit
operator chain
\begin{equation}
  U_{\mathrm{prep}}
  \;W(a_j)\;
  E_{\mathrm{QSVT}}(t;\bm{\theta})\;
  W(a_j^\dagger)\;
  U_{\mathrm{prep}}^\dagger,
  \label{eq:alphaj_operator_chain}
\end{equation}
whereas the noisy path measures the same target overlap through controlled
interference and success filtering. 

Segmentation keeps the per segment degree manageable by trading the degree of
each QSVT polynomial segment against the number of segments $r$. Let
\begin{equation}
  B := \frac{H}{\lambda},
  \qquad
  \sigma(B)\subseteq[-1,1].
  \label{eq:normalized_B}
\end{equation}
Let
$E=\ee^{-\ii(\tau/r)B}$ be the exact short time propagator and let
$\widetilde{E}=E+\Delta$ be its QSVT approximation with
$\lVert\Delta\rVert\le\varepsilon_{\mathrm{seg}}$. The segmented approximation
is $\widetilde{E}^{r}$, while the exact propagator is $E^r=\ee^{-\ii\tau B}$.
The telescoping identity
\begin{equation}
  \widetilde{E}^{r}-E^{r}
  =
  \sum_{m=0}^{r-1}
  \widetilde{E}^{r-1-m}(\widetilde{E}-E)E^m
  \label{eq:appendix_segmentation_telescoping}
\end{equation}
gives
\begin{equation}
  \left\|\widetilde{E}^{r}-E^{r}\right\|
  \le
  r\,\varepsilon_{\mathrm{seg}}
  (1+\varepsilon_{\mathrm{seg}})^{r-1}.
  \label{eq:appendix_segmentation_error_bound}
\end{equation}
Thus the total error scales as $\mathcal{O}(r\varepsilon_{\mathrm{seg}})$ when
$r\varepsilon_{\mathrm{seg}}$ is small. To reach total error $\varepsilon$, we
choose $\varepsilon_{\mathrm{seg}}\approx\varepsilon/r$. Each segment has scaled
time $\tau/r$, so we have
\begin{equation}
  d_{\mathrm{seg}}
  =
  \mathcal{O}\!\left(\frac{\tau}{r}+\log(r/\varepsilon)\right).
  \label{eq:appendix_segment_degree}
\end{equation}
With our choice $\tau_{\max}=1$, we take
$r=\lceil\tau/\tau_{\max}\rceil$. At the illustrative scaled time
$\lambda t=5$, this gives $r=5$ segments. For $\varepsilon=10^{-4}$, the
per segment degree scale is
$\mathcal{O}(1+\log(5\times10^{4}))$, which is consistent with the
choice $(d_{\cos},d_{\sin})=(12,13)$. The corresponding circuit template uses
\begin{equation}
  N_{\mathrm{oracle}}
  = 2(d_{\cos}+d_{\sin})r
  = 2(12+13)5
  = 250
  \label{eq:appendix_lambda_t_five_oracles}
\end{equation}
block encoding calls for one Green's function evaluation at that scaled time.

The QSVT propagator splits into even and odd polynomial parities, and the implementation
represents these two parity families by distinct cosine and sine phase vectors,
since even polynomials encode the cosine branch and odd polynomials the sine
branch.

To estimate the complex quantity $\alpha_j(t)$ as shown in Figure \ref{fig:alphaj_overview}(a), the implementation uses a
controlled version of the QSVT time evolution block inside an ancilla assisted
Hadamard test. Let $U_t$ denote the controlled circuit that realizes the
relevant matrix element in Eq.~\eqref{eq:alphaj_operator_chain}. Then ancilla measurements
in the $X$ and $Y$ bases give
\begin{align}
  \Re\,\alpha_j(t)
  &= \expect{X_{\mathrm{anc}}}_{U_t}, \\
  \Im\,\alpha_j(t)
  &= \expect{Y_{\mathrm{anc}}}_{U_t},
\end{align}
up to the circuit dependent sign convention. Equivalently,
\begin{equation}
  \alpha_j(t)
  =
  \expect{X_{\mathrm{anc}}}_{U_t}
  +
  \ii\,\expect{Y_{\mathrm{anc}}}_{U_t}.
  \label{eq:hadamard_alpha}
\end{equation}
Equivalently, up to normalization the controlled circuit prepares
\begin{equation}
  \ket{\Psi_{\mathrm{HT}}}
  =
  \frac{1}{\sqrt{2}}
  \left(
    \ket{0}\ket{\Phi_j}
    +
    \ket{1}E(t)\ket{\Phi_j}
  \right),
  \label{eq:ht_state}
\end{equation}
so that with
\begin{equation}
  \rho_{\mathrm{HT}}
  =
  \ket{\Psi_{\mathrm{HT}}}\!\bra{\Psi_{\mathrm{HT}}},
\end{equation}
one has
\begin{align}
  \Tr[(X\otimes I)\rho_{\mathrm{HT}}]
  &= \Re\!\left(\bra{\Phi_j}E(t)\ket{\Phi_j}\right),\\
  \Tr[(Y\otimes I)\rho_{\mathrm{HT}}]
  &= \Im\!\left(\bra{\Phi_j}E(t)\ket{\Phi_j}\right),
\end{align}
up to the sign convention induced by the $Y$ basis readout circuit. This
shows explicitly why the control qubit interference pattern encodes the real
and imaginary parts of the target overlap.
This reduces the Green's function problem to repeated estimation of controlled QSVT expectation values over a discrete time grid.

In the noisy path, the estimator is postselected. Let
\begin{equation}
  \Pi_{\mathrm{succ}}
  =
  \ket{0}\!\bra{0}_{a}
  \otimes
  \ket{000}\!\bra{000}_{q}
  \otimes I_s
  \label{eq:success_projector}
\end{equation}
denote the success projector on the annihilation ancilla $a$, the internal
QSVT ancillas $q$, and the system register $s$. Then the experimentally
relevant quantities are conditional expectations
\begin{align}
  \langle X\rangle_{\mathrm{succ}}
  &=
  \frac{\Tr[(X_c\otimes \Pi_{\mathrm{succ}})\rho]}
       {\Tr[(I_c\otimes \Pi_{\mathrm{succ}})\rho]},
  \\
  \langle Y\rangle_{\mathrm{succ}}
  &=
  \frac{\Tr[(Y_c\otimes \Pi_{\mathrm{succ}})\rho]}
       {\Tr[(I_c\otimes \Pi_{\mathrm{succ}})\rho]},
  \label{eq:conditional_xy}
\end{align}
which we estimate from success filtered shot counts. In other words, the
noisy Green's function pipeline is not just a bare Hadamard test; it is a
postselected Hadamard estimator for the success branch of the block encoded
QSVT circuit. This distinction matters when interpreting backend dependent
reliability and sampling variance, because the measured overlap is explicitly a
conditional estimator rather than a raw amplitude readout.

Figure~\ref{fig:alphaj_overview} collects the circuit views used in this
estimator. Figure~\ref{fig:alphaj_two_paths} compares the noisy Hadamard path
with the ideal explicit chain, and Figure~\ref{fig:alphaj_components} shows the
operator block encodings and one QSVT segment. Panels (a) and (b) matter for
the reliability reading of \S\ref{sec:results_restructured}.2 because they
show that the noisy and ideal paths target the same overlap $\alpha_j(t)$
through different circuit mechanics: the noisy path pays for controlled
interference and postselection, so its per evaluation variance depends on the
backend's success branch statistics, while the ideal path composes the
operator chain of Eq.~\eqref{eq:alphaj_operator_chain} directly. Panel (c)
shows the two LCU blocks whose ancillas enter the success projector of
Eq.~\eqref{eq:success_projector}, and panel (d) shows the repeated QSVT
segment whose cosine and sine phase vectors form the 27 dimensional space
that the optimizer, the sensitivity analysis of
Figure~\ref{fig:results_sensitivity}, and the density analysis of
Figure~\ref{fig:results_density} all act on. The figure therefore ties every
phase level claim in the paper to a concrete circuit location.

\section{Workflow and Online UQ Engine}
\label{app:workflow_description}
\begin{figure}[tbp]
\centering
\resizebox{\textwidth}{!}{\input{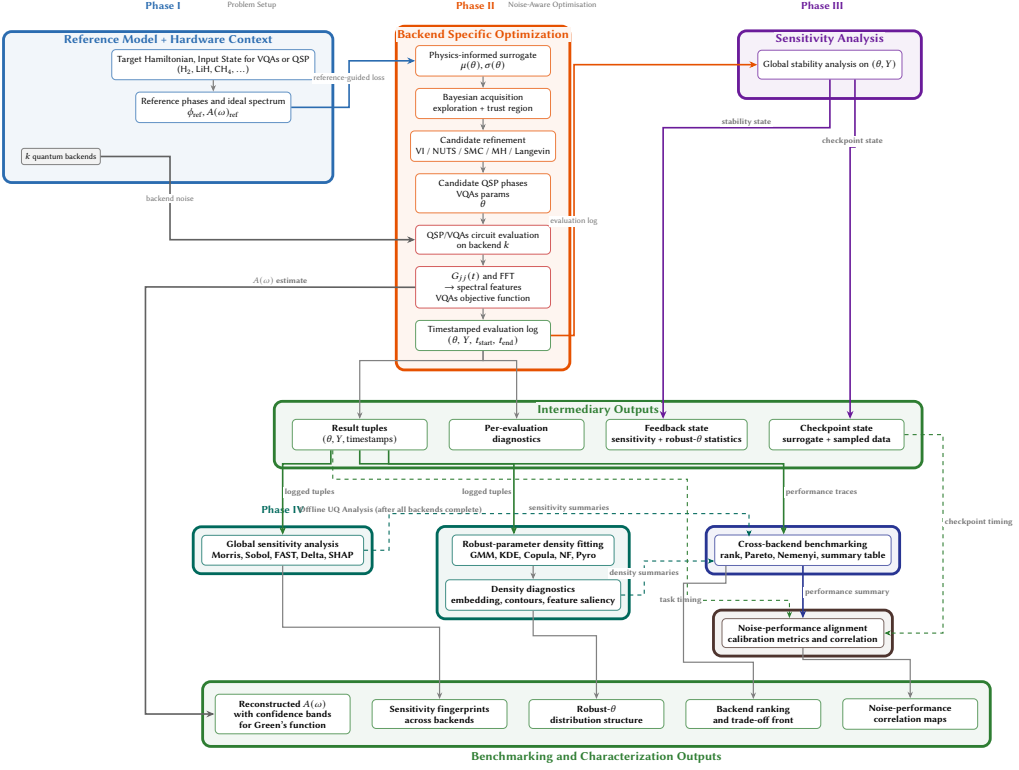}}
\caption{Shared UQ workflow used throughout this work. The same closed loop
generates posterior updates, sensitivity summaries, parameter densities,
backend rankings, and optional noise alignment products for both the
Green's/QSVT branch and the VQA branch.}
\label{fig:workflow_appendix}
\end{figure}

Figure~\ref{fig:workflow_appendix} gives the shared UQ workflow used
for both branches. The workflow has four phases: reference and backend setup,
backend specific closed loop optimization, offline sensitivity analysis, and
post processing for spectra, robust regions, rankings, resources, and noise
alignment.

\begin{algorithm}[t]
\caption{Unified UQ workflow for VQA and Green's/QSVT backend characterization}
\label{alg:unified_uq_workflow}
\begin{algorithmic}[1]
\Require Workload and reference data for the VQA and Green's/QSVT branches, backend set $\mathcal{B}$, backend wise evaluation budgets $\{N_b\}_{b\in\mathcal{B}}$, posterior refinement mode $m$
\Ensure Backend wise task summaries, robust parameter region statistics, benchmark rankings, and noise performance correlations
\State construct branch specific ideal references, target observables, and task score maps
\For{$b\in\mathcal{B}$}
  \State initialize design $\mathcal{D}_b^{(0)}$ from a space filling design or warm start records
  \State initialize GP surrogate, trust region, acquisition rule, and posterior refinement state
  \For{$n=0,\dots,N_b-1$}
    \State fit/update surrogate on $\mathcal{D}_b^{(n)}$
    \State choose candidate $\bm{\xi}_{n+1}$ by trust region Bayesian optimization
    \State refine $\bm{\xi}_{n+1}$ using mode $m\in\{\mathrm{VI},\mathrm{MH},\mathrm{Langevin},\mathrm{none}\}$
    \State bind the candidate parameters or phases into the cached workload circuit
    \State execute the noisy VQA or Green's/QSVT circuit on backend $b$
    \State set $\mathcal{D}_b^{(n+1)}\gets\mathcal{D}_b^{(n)}\cup\{(\bm{\xi}_{n+1},Y_{n+1})\}$
    \State save tuple, timestamp, checkpoint, and derived task features
  \EndFor
  \State save backend local task summaries, timing data, and metadata
\EndFor
\State aggregate backend records into a unified evaluation dataset
\State fit sensitivity and robust region density models
\State compute backend benchmark summaries and align records with calibration/noise metadata
\end{algorithmic}
\end{algorithm}

Phase~I fixes the workload instance, the ideal reference, and the backend
noise context. For the Green's/QSVT branch this means the H$_2$ Hamiltonian,
reference QSP phases, and ideal spectrum; for the VQA branch it means the
configured task and its classical, noiseless, or target value. Phase~II runs
one optimization loop per backend. Each evaluation updates a Gaussian process
surrogate, selects a candidate through Bayesian optimization, applies the
chosen posterior refinement mode, executes the circuit, and saves the tuple
$(\theta,Y,t_{\mathrm{start}},t_{\mathrm{end}})$. Phase~III computes backend
local sensitivity profiles from the saved records. Phase~IV replays the saved
records offline to produce spectra, sensitivity fingerprints, robust parameter
regions, backend rankings, resource summaries, and optional calibration/noise
association maps. This separation keeps quantum execution confined to Phase~II;
all later analysis can be reproduced from the saved evaluation records.
Figure~\ref{fig:workflow_appendix} and Algorithm~\ref{alg:unified_uq_workflow}
map one to one onto the main text: Phase~II produces the evaluation histories
scored in \S\ref{sec:results_restructured}.1 and
\S\ref{sec:results_restructured}.2, Phase~III feeds the sensitivity claims of
\S\ref{sec:results_restructured}.3, and Phase~IV produces the density,
ranking, resource, and noise products reported in
\S\ref{sec:results_restructured}.3-\S\ref{sec:results_restructured}.5.

For each backend, the loop fits a Gaussian process surrogate to the accumulated
objective evaluations, chooses the next candidate with a Bayesian optimization
acquisition rule, refines that candidate with VI, Metropolis Hastings, or a
Langevin variant, executes the backend circuit, and appends the resulting
objective value to the backend history. Standard references give the Gaussian
process, expected improvement, Bayesian optimization, and variational inference
formulations used by this loop
\cite{rasmussen2006gp,snoek2012practical,jones1998ego,shahriari2016bo,balandat2020botorch,blei2017variational}.
The candidate generator uses a local trust region in the spirit of
TuRBO~\cite{eriksson2019turbo}; the Langevin robustness branch follows the
stochastic gradient Langevin formulation of Welling and Teh~\cite{welling2011langevin}.
Algorithm~\ref{alg:unified_uq_workflow} gives the merged online and offline
workflow. It keeps the backend local guided loop explicit while avoiding a
separate wrapper algorithm for the downstream characterization stages.

The loop saves every evaluation with the control vector, objective value,
timestamps, and derived spectral or task features needed downstream. This fixed record
structure supports sensitivity analysis, density estimation, benchmarking, and
noise alignment without rerunning the quantum workload. Optional variants of
the same loop include sparse lengthscale priors, fully Bayesian surrogate
marginalization, periodic kernels for angular parameters, and cost weighted
fidelity search. These variants fit the same
Bayesian optimization framework
\cite{rasmussen2006gp,snoek2012practical,eriksson2019turbo,balandat2020botorch};
the Fourier structure of parameterized circuits motivates periodic angle aware
kernels~\cite{schuld2021fourier}. Backend hit rates could also be placed in a
small beta binomial ranking model, but the present single task reports
the empirical application level ordering instead.

\section{Statistical Analysis}
\label{sec:offline}
The offline stage converts the saved evaluation histories into the supporting
characterization products used in the main paper. The main text defines the
reported metrics. This appendix identifies the estimator families and the role
each family plays in the analysis.

For global sensitivity analysis, we compute Morris, Sobol, FAST, Delta, and SHAP style sensitivity summaries on
surrogate models fitted to the backend evaluation records. Morris gives the
main text representative because it remains stable for the available sample
counts and gives an interpretable phase importance fingerprint. Sobol and FAST
provide variance based checks, while Delta and SHAP style summaries provide
distributional and feature contribution checks. Standard references give the
estimator definitions and sampling rules
\cite{morris1991ee,saltelli2010sobol,lundberg2017unified}. The main paper reports
Morris directly and uses the remaining indices as supporting diagnostics.
Figure~\ref{fig:results_sensitivity} applies these estimators to the QSVT
phase space and backs the coordinate concentration claims of
\S\ref{sec:results_restructured}.3.

For density estimation of robust parameter regions, we treat the high performing parameter set as a distribution rather than as a
single optimum. The offline pipeline fits Gaussian mixture, KDE, copula, and
normalizing flow models to that set and then compares their concentration,
multimodality, and backend overlap. The main text reports the density based
robust region summary; standard density estimation references give the generic
KDE construction~\cite{silverman1986kde}.
Figure~\ref{fig:results_density} and Figure~\ref{fig:app_vqa_parameter_summary}
apply these density models to the QSVT and VQA branches and support the robust
region geometry discussion in \S\ref{sec:results_restructured}.3.

The backend benchmark combines objective quality, convergence behavior, and
robustness summaries. For the Green's/QSVT branch, the key summaries are hit
rate, time to first good reconstruction, best observed objective value, and the
sensitivity and density diagnostics. For the VQA branch, the same aggregation
uses workload quality, workload rank, and evaluation count to the target
quality level. These definitions appear in the main text, so we avoid repeating
the equations here. Figure~\ref{fig:results_benchmark},
Figure~\ref{fig:vqa_results_benchmark}, and the cross paradigm panel of
Figure~\ref{fig:cross_paradigm_score} report the resulting rankings.

With regards to noise performance alignment, the final offline stage aligns task performance summaries with backend noise
metadata, including relaxation times, readout error, and two qubit error
proxies. We use Spearman correlation, mutual information, and distance
correlation summaries as exploratory diagnostics. This stage does not replace
QCVV measurements; it asks which backend noise features best explain the
application level behavior observed in this study.
Figure~\ref{fig:app_vqa_benchmark_noise} in \appVQABenchmark{} reports the
resulting Spearman summary for the VQA branch.

\section{Resource Estimation}
\label{sec:resource_estimation}
Resource estimation adds the cost dimension to the UQ benchmark. A backend that
reconstructs the Green's function accurately but requires a much larger routed
circuit footprint is qualitatively different from a backend that reaches
similar accuracy at lower depth and lower two qubit burden. The resource study
therefore separates statistical reliability from compiled execution cost. The resource workflow studies two Green's/QSVT circuit families: the QSVT time
evolution circuit $E_{\mathrm{QSVT}}(t)$ and the full Green's function Hadamard
test circuit used to estimate $\alpha_j(t)$. For backend $b$, the reported
resource vector is
\begin{equation}
  \mathcal{R}_b(d,t)
  =
  \bigl(
    w_b(d,t),\,
    D_b(d,t),\,
    N^{(1)}_b(d,t),\,
    N^{(2)}_b(d,t),\,
    \{N_b^{(g)}(d,t)\}_{g\in\mathcal{G}}
  \bigr),
  \label{eq:resource_vector}
\end{equation}
where $w_b$ is width, $D_b$ is depth, $N^{(1)}_b$ and $N^{(2)}_b$ are one and
two qubit gate counts, and $\mathcal{G}$ is the backend basis gate set. We retain both pre transpiled and post transpiled counts so that we can
separate algorithmic cost from backend routing overhead.

\begin{figure}[tbp]
\centering
\begin{subfigure}[t]{0.48\textwidth}
\centering
\includegraphics[width=\linewidth]{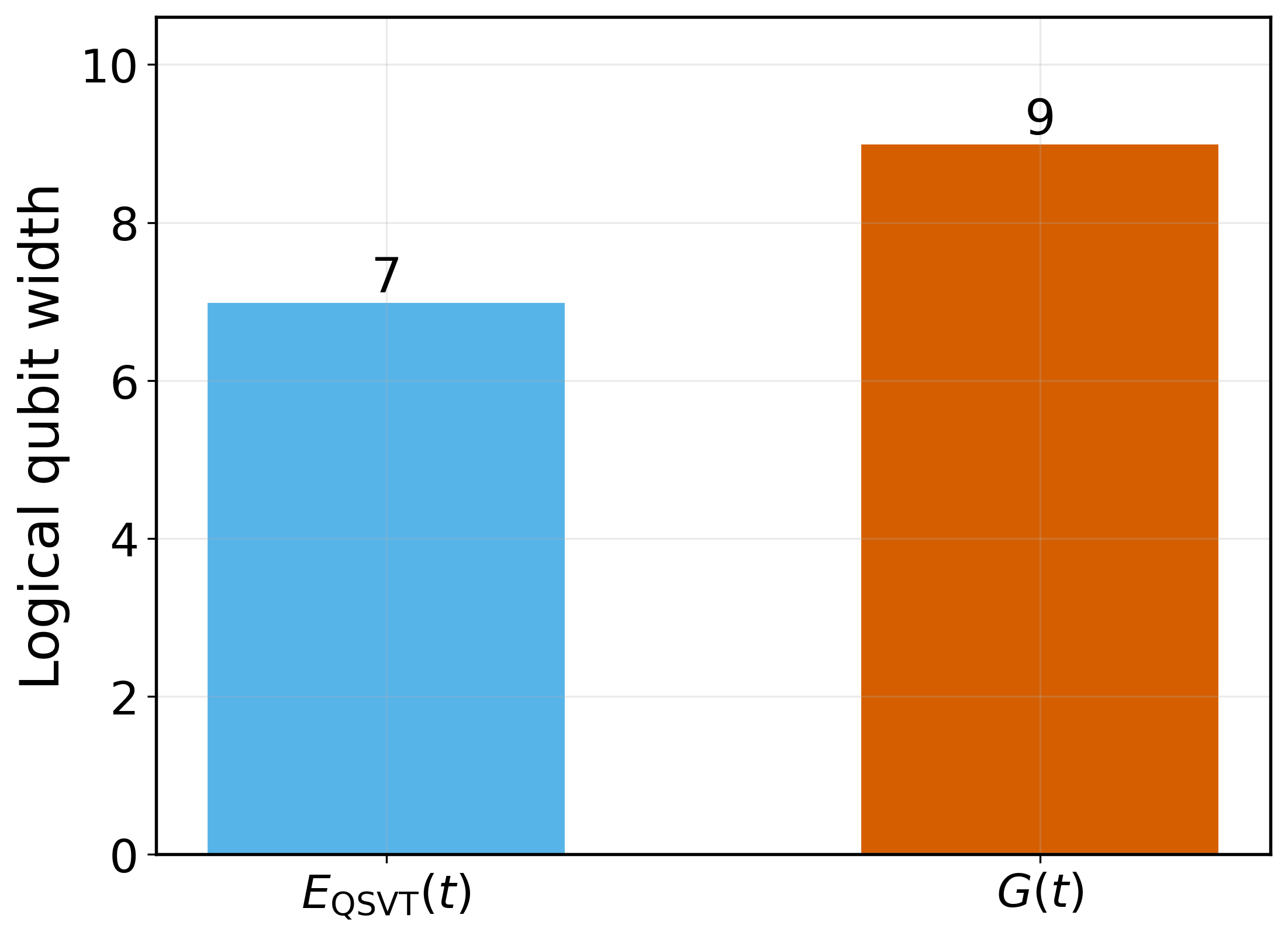}
\caption{Width comparison.}
\end{subfigure}
\hfill
\begin{subfigure}[t]{0.51\textwidth}
\centering
\includegraphics[width=\linewidth]{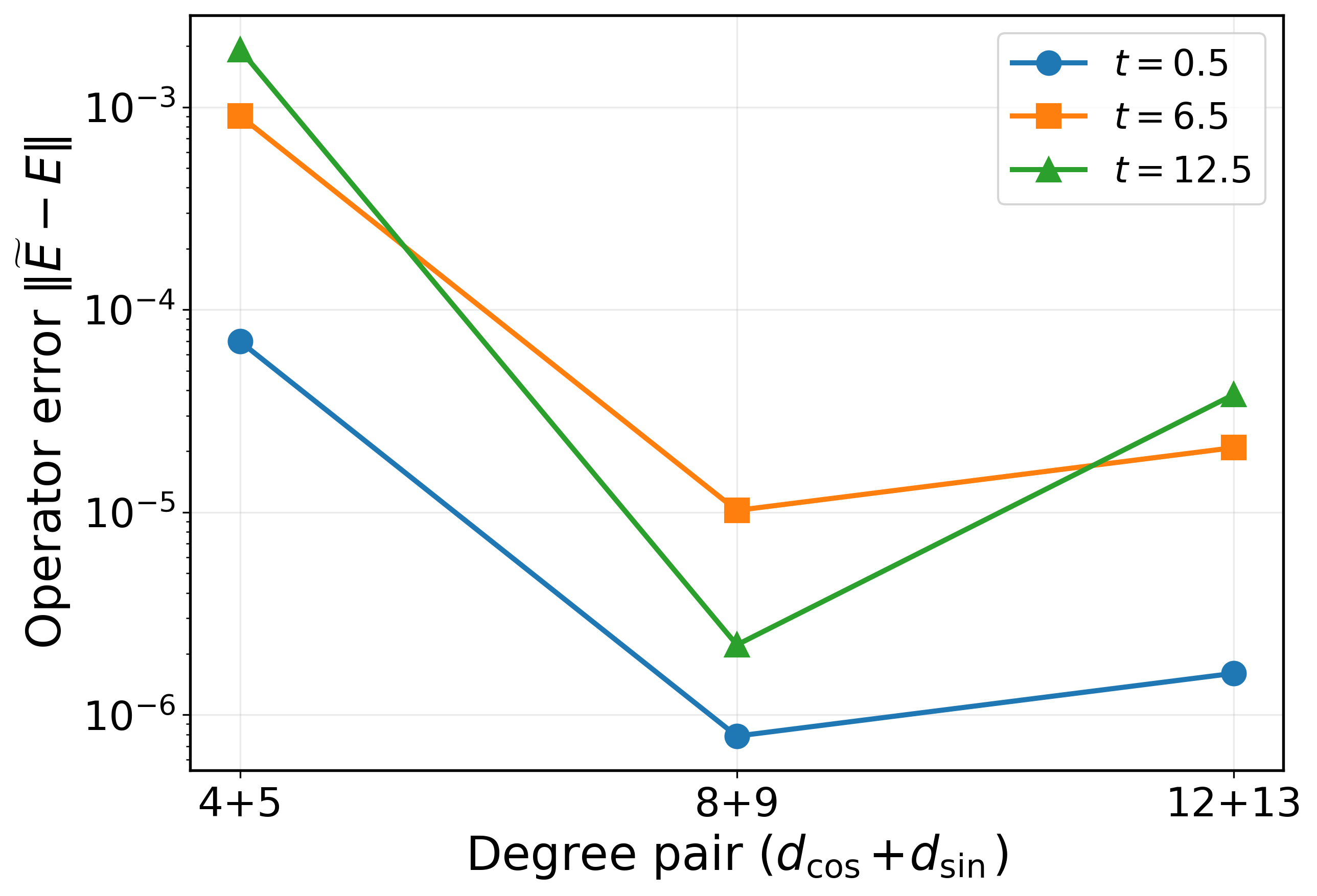}
\caption{QSVT operator error versus sampled degree pairs.}
\end{subfigure}

\vspace{0.6em}
\begin{subfigure}[t]{0.6\textwidth}
\centering
\includegraphics[width=\linewidth]{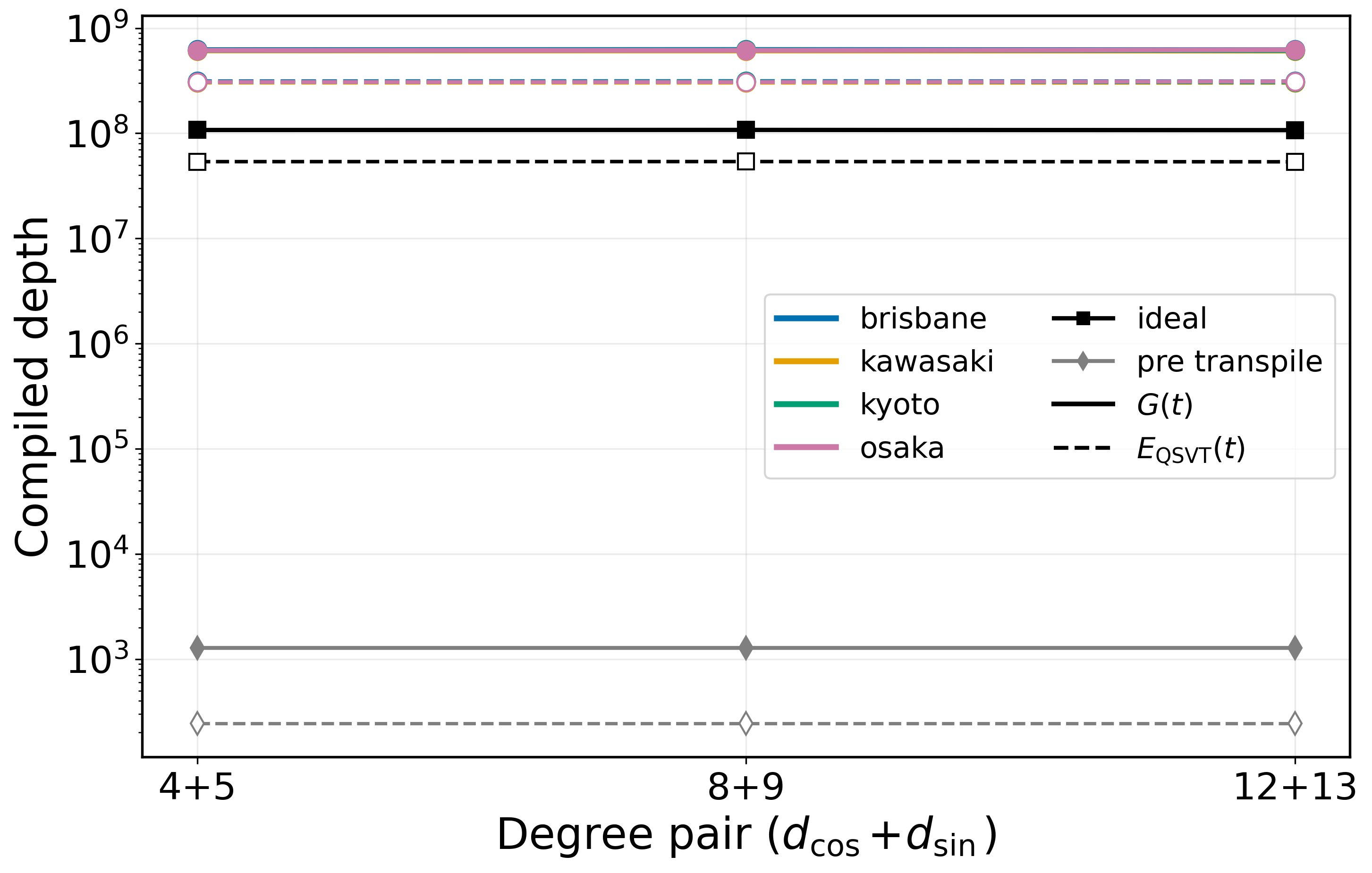}
\caption{Compiled depth versus degree at fixed $t=12.5$; one plot combines
$G(t)$ (solid, filled markers) and $E_{\mathrm{QSVT}}(t)$ (dashed, open
markers).}
\end{subfigure}
\caption{Degree sweep summary for the QSVT propagator and Green's function
circuits. Width is structural, while routed depth is dominated by backend
mapping overhead over the sampled degree window.}
\label{fig:resource_degree_summary}
\end{figure}

Figure~\ref{fig:resource_degree_summary} shows that the QSVT propagator block
uses $7$ qubits and the full Green's function Hadamard test circuit uses $9$
qubits for the H$_2$ instance. The dominant practical variable is therefore not
width but routed depth and gate volume. Over the sampled degree pairs, backend
to backend offsets dominate the within backend degree variation, while the
middle degree pair gives the smallest observed operator norm approximation
error. Panel (a) separates structural width from routed cost: the seven and
nine qubit counts follow Eq.~\eqref{eq:width_main_restructured} and do not
change with backend, so any backend difference in the other panels comes from
routing rather than register size. Panel (b) justifies the degree pair
$(12,13)$ used throughout \S\ref{sec:results_restructured}, because it sits
at the observed error minimum of the sampled window. Panel (c) combines both
circuit families in one plot: at fixed $t=12.5$ the routed depth curves of
the four backends stay nearly parallel across degrees, the Green's function
circuit sits about a factor of two above the propagator block, and both
families keep the same distance to the ideal and pre transpiled references,
which is the degree side counterpart of the overhead band reported in
Figure~\ref{fig:greens_resource_main_restructured}.

The Green's/QSVT uses a transpile once and bind phases policy. The cached
transpilation fixes the logical to physical layout, routing schedule, native
basis decomposition, and optimization pass choices for a fixed QSVT topology.
Each evaluation then binds a new QSP phase vector to the same parametrized
circuit and runs it with the Aer noise model constructed from the same fake
backend snapshot. This cache is noise faithful for the deterministic IBM fake
backend snapshots used here. Live QPUs would require a refresh policy, such as
periodic retranspilation or calibration triggered cache invalidation.

% \subsection{Degree and time sweeps}
At fixed physical time $t$, the degree sweep records
\begin{equation}
  \mathcal{R}_b(d_{\cos},d_{\sin};t),
  \label{eq:resource_degree_sweep}
\end{equation}
while at fixed degree the time sweep records
\begin{equation}
  \mathcal{R}_b(d;t).
  \label{eq:resource_time_sweep}
\end{equation}
Because the implementation uses segmented evolution, the segment count is
approximately
\begin{equation}
  R(t)
  =
  \left\lceil
    \frac{\lambda t}{\tau_{\max}}
  \right\rceil,
  \label{eq:segment_count}
\end{equation}
where $\tau_{\max}$ is the maximum allowed scaled time step.

\begin{figure}[tbp]
\centering
\begin{subfigure}[t]{0.48\textwidth}
\centering
\includegraphics[width=\linewidth]{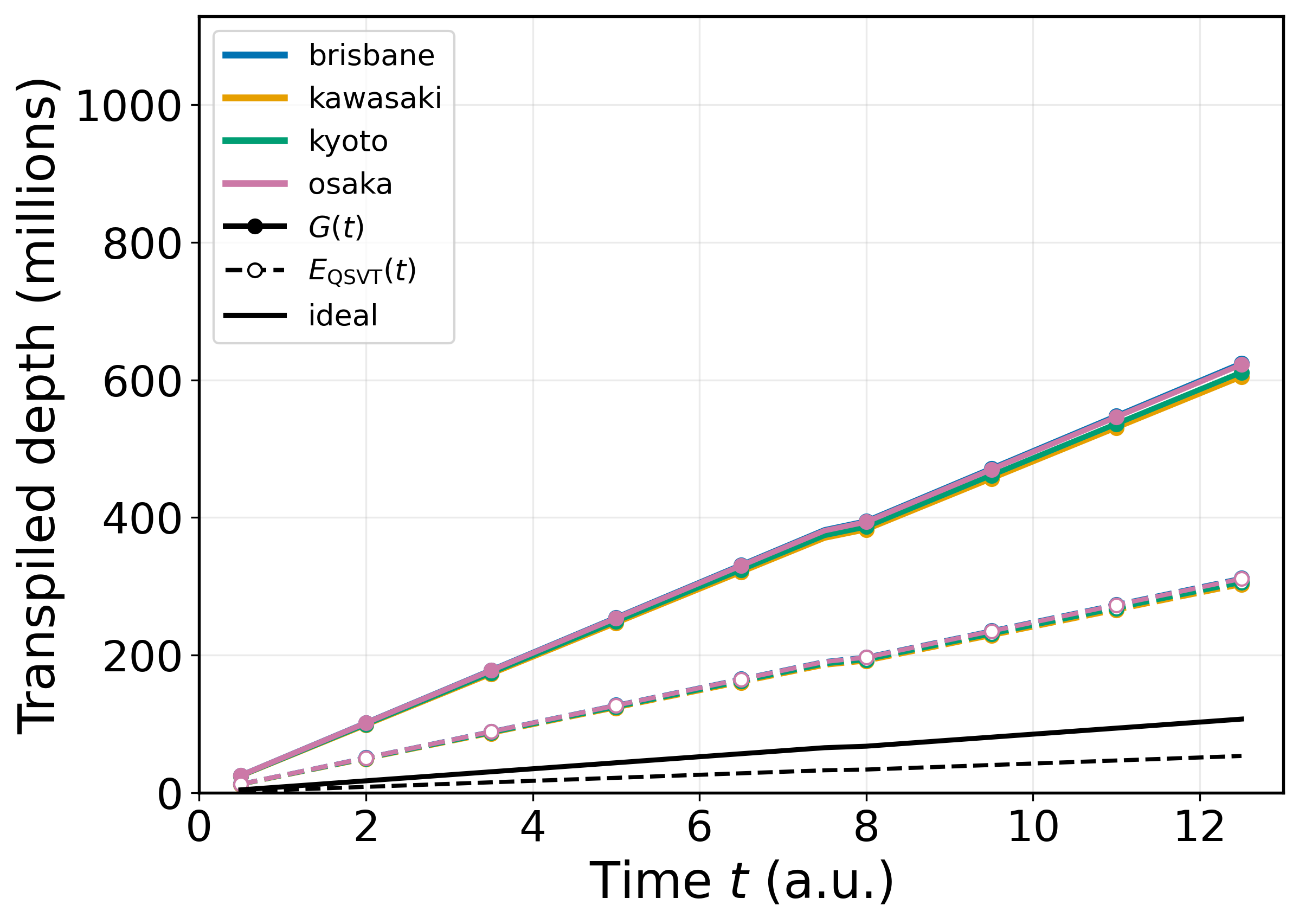}
\caption{Transpiled circuit depth versus time.}
\end{subfigure}
\hfill
\begin{subfigure}[t]{0.48\textwidth}
\centering
\includegraphics[width=\linewidth]{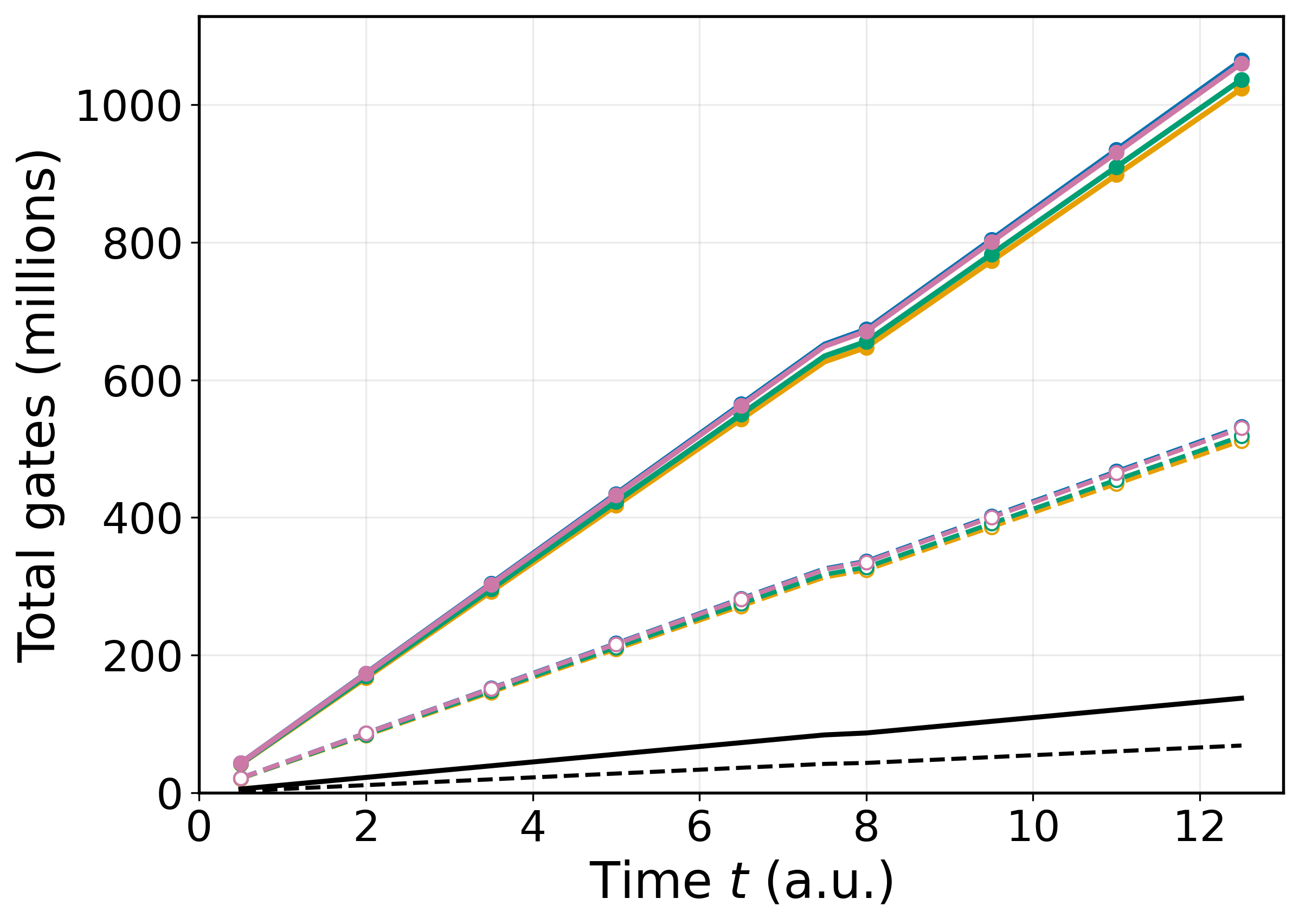}
\caption{Total gate count versus time.}
\end{subfigure}

\vspace{0.6em}
\begin{subfigure}[t]{0.48\textwidth}
\centering
\includegraphics[width=\linewidth]{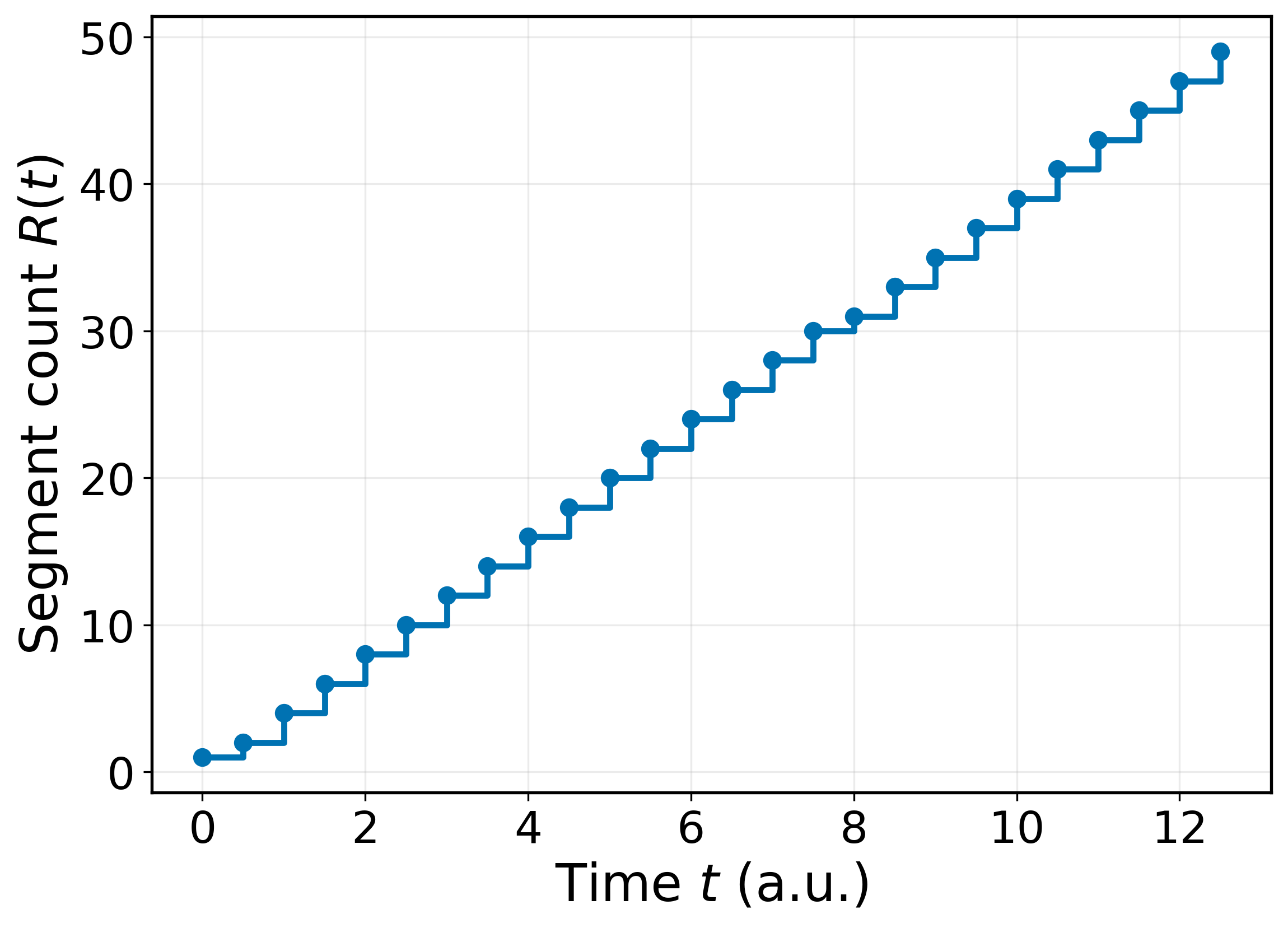}
\caption{Segment count growth versus time.}
\end{subfigure}
\caption{Time sweep summary at fixed degree $(12,13)$. Panels (a) and (b)
share identical linear axis ranges and report the metric values directly in
millions, with $G(t)$ solid, $E_{\mathrm{QSVT}}(t)$ dashed, black lines for
the ideal references, and the legend of panel (a) applying to both. Panel
(c) shows the segment count staircase of Eq.~\eqref{eq:segment_count}. The
fixed degree resource curves are close to linear in $t$ because each
additional segment repeats the same QSVT motif.}
\label{fig:resource_time_summary}
\end{figure}

Figure~\ref{fig:resource_time_summary} explains why the fixed degree curves in
the main paper are nearly linear in time: segmentation repeats a common circuit
motif, and backend routing contributes a roughly persistent per segment
penalty. The four backend curves nearly coincide at this scale, the Green's
function circuit tracks about a factor of two above the propagator block, and
the small residual ordering places Brisbane and Osaka at the upper edge with
Kawasaki the lightest of the four. Panel (c) makes the mechanism explicit: the segment count
of Eq.~\eqref{eq:segment_count} grows in steps with time, and each added
segment appends one more copy of the QSVT motif of
Figure~\ref{fig:alphaj_components}(d). This staircase is why the depth and
gate curves in panels (a) and (b) grow linearly and why the multimetric curves of
Figure~\ref{fig:greens_resource_multimetric_main} remain parallel: routing
multiplies a repeated motif instead of restructuring the circuit. The panel
therefore supports the main text claim that the routed cost penalty acts as a
persistent backend dependent multiplier.

\begin{figure}[tbp]
\centering
\begin{subfigure}[t]{0.52\textwidth}
\centering
\includegraphics[width=\linewidth]{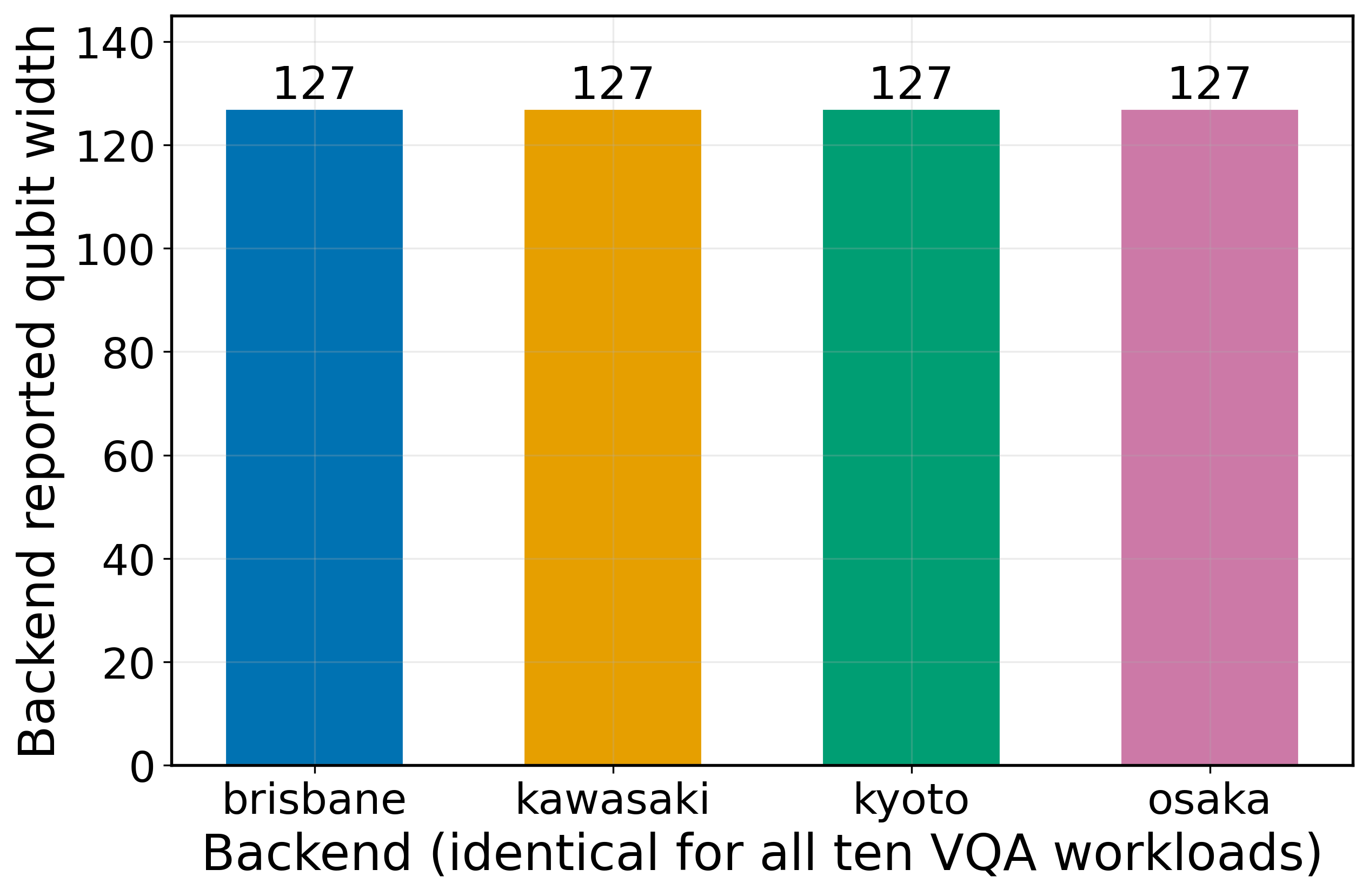}
\caption{Backend reported qubit width, identical for all ten VQA workloads.}
\end{subfigure}
\hfill
\begin{subfigure}[t]{0.47\textwidth}
\centering
\includegraphics[width=\linewidth]{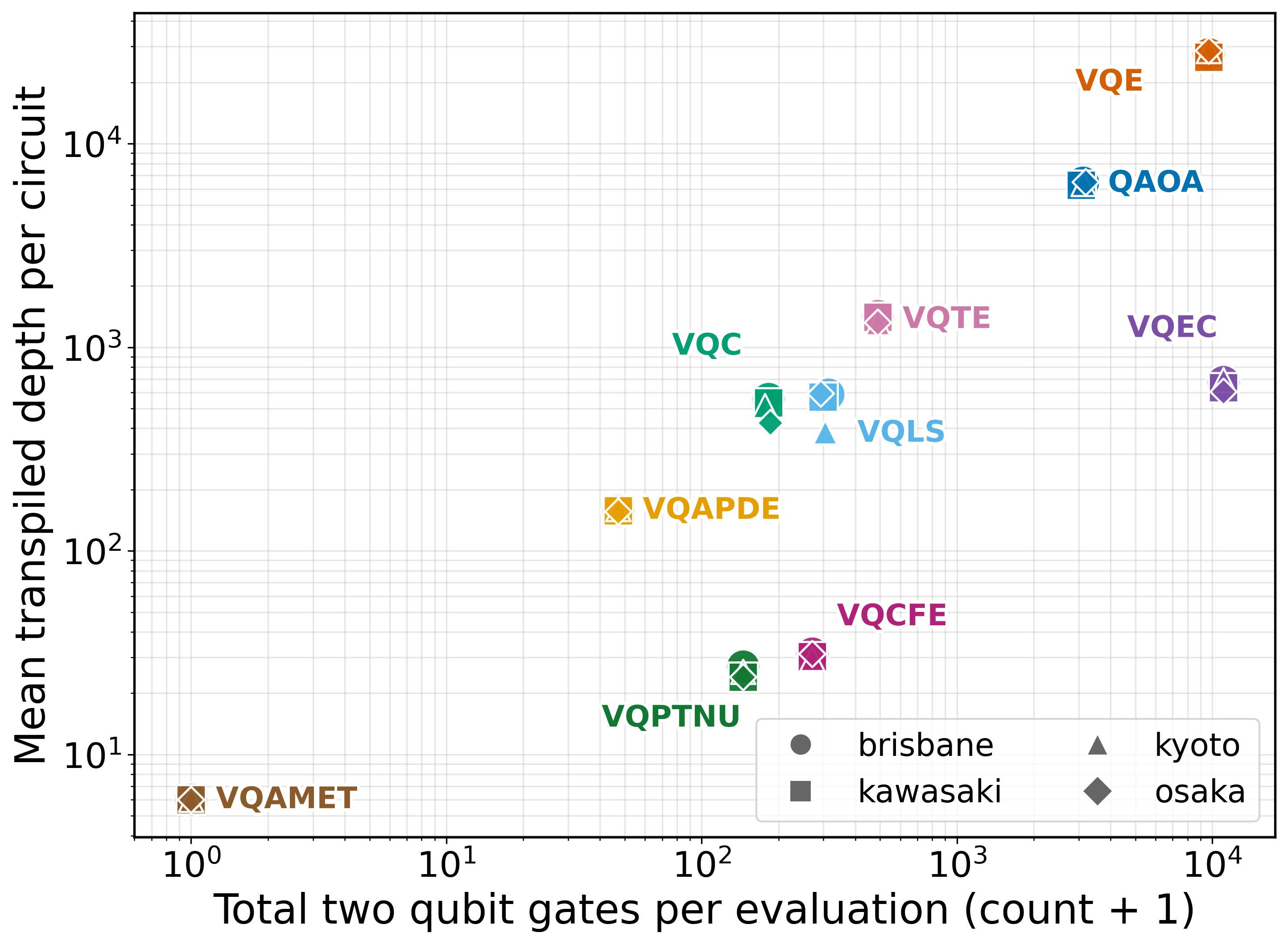}
\caption{Mean depth versus total two qubit burden per evaluation.}
\end{subfigure}
\caption{Supplemental VQA resource view. Every workload reports the same
width of $127$ on every backend because the transpiled circuit object
carries the full device register, so the top panel shows one bar per
backend instead of repeating the same value across the ten workloads; the
width should not be read as logical problem size. In the scatter, each
point is one workload and backend pair, colored by workload with the marker
shape giving the backend; nested marker sizes keep exactly overlapping
backend points visible. The scatter separates single circuit heavy
workloads from multi circuit characterization workloads.}
\label{fig:resource_appendix_vqa_extra}
\end{figure}

With regards to backend aware transpilation and VQA resources, for each circuit family and backend $b$, backend aware transpilation produces
\begin{equation}
  \mathcal{R}_b^{\mathrm{post}}(d,t)
  =
  T_b\!\left(\mathcal{C}(d,t)\right),
  \label{eq:post_resource}
\end{equation}
where $\mathcal{C}(d,t)$ is the abstract circuit family and $T_b$ denotes the
backend specific transpilation and routing map. The main paper reports the
compact multimetric QSVT resource panel and the main VQA resource heatmaps.
Figure~\ref{fig:resource_appendix_vqa_extra} provides one supplemental VQA view:
backend reported width and the depth versus two qubit burden scatter. We omit secondary degree count, native basis, and opaque versus decomposed
plots because they repeat the routed cost ordering already shown in the main
resource panels.

Figure~\ref{fig:resource_appendix_vqa_extra}(a) reports the backend visible
register width per workload; this width includes idle physical qubits
introduced by layout, so it measures placement footprint rather than logical
problem size. Panel (b) plots each workload in the plane of mean depth versus
total two qubit burden and separates the cost regimes of
\S\ref{sec:results_restructured}.4 at a glance: VQE and QAOA sit in the deep
single circuit corner, VQEC and the tomography style workloads sit in the
shallow but high total burden corner, and VQAMET stays near the origin. The
scatter is the visual evidence for the claim that hard and expensive are
different axes.

\section{Extended Results and Discussion on VQAs and Green's/QSVT}
\label{sec:results_discussion}
This section keeps only the secondary result views that add information beyond
the compact main text figures. The Green's/QSVT part uses the H$_2$ task,
27 QSVT phase parameters, four IBM fake backends, and 100 evaluations per
backend. The VQA branch uses the same four backends set across ten
workloads. The two branches stress different capabilities: the Green's/QSVT
branch probes matrix function execution and spectral reconstruction, while
the VQA branch probes optimizer interaction, ansatz robustness, and workload
specific objective stability. The shared UQ machinery lets both act as
application level backend probes without claiming that the two algorithmic
families are physically interchangeable.

\begin{figure}[tbp]
\centering
\begin{subfigure}[t]{0.48\textwidth}
\centering
\includegraphics[width=\linewidth]{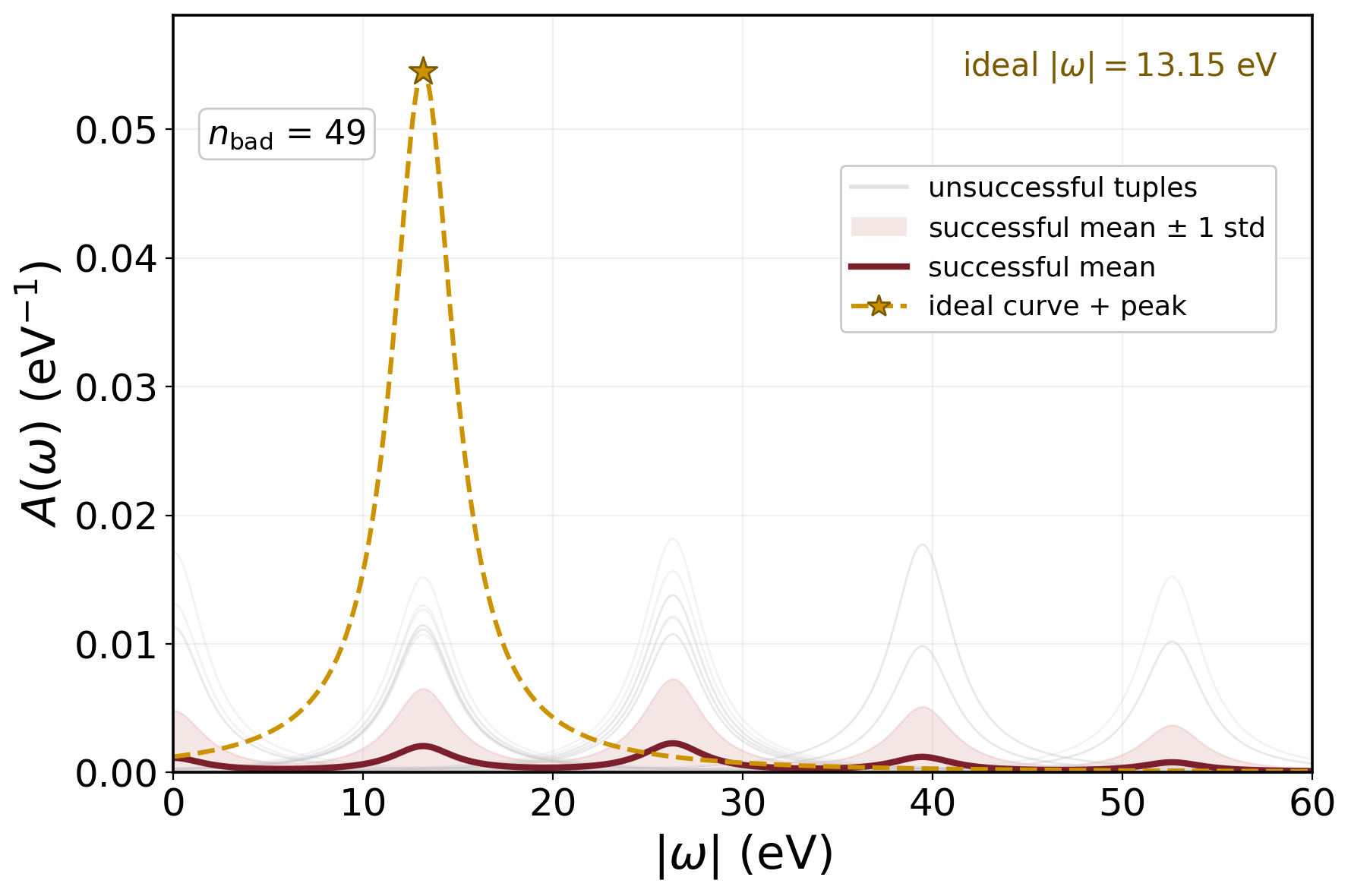}
\caption{\texttt{ibm\_brisbane}.}
\end{subfigure}
\hfill
\begin{subfigure}[t]{0.48\textwidth}
\centering
\includegraphics[width=\linewidth]{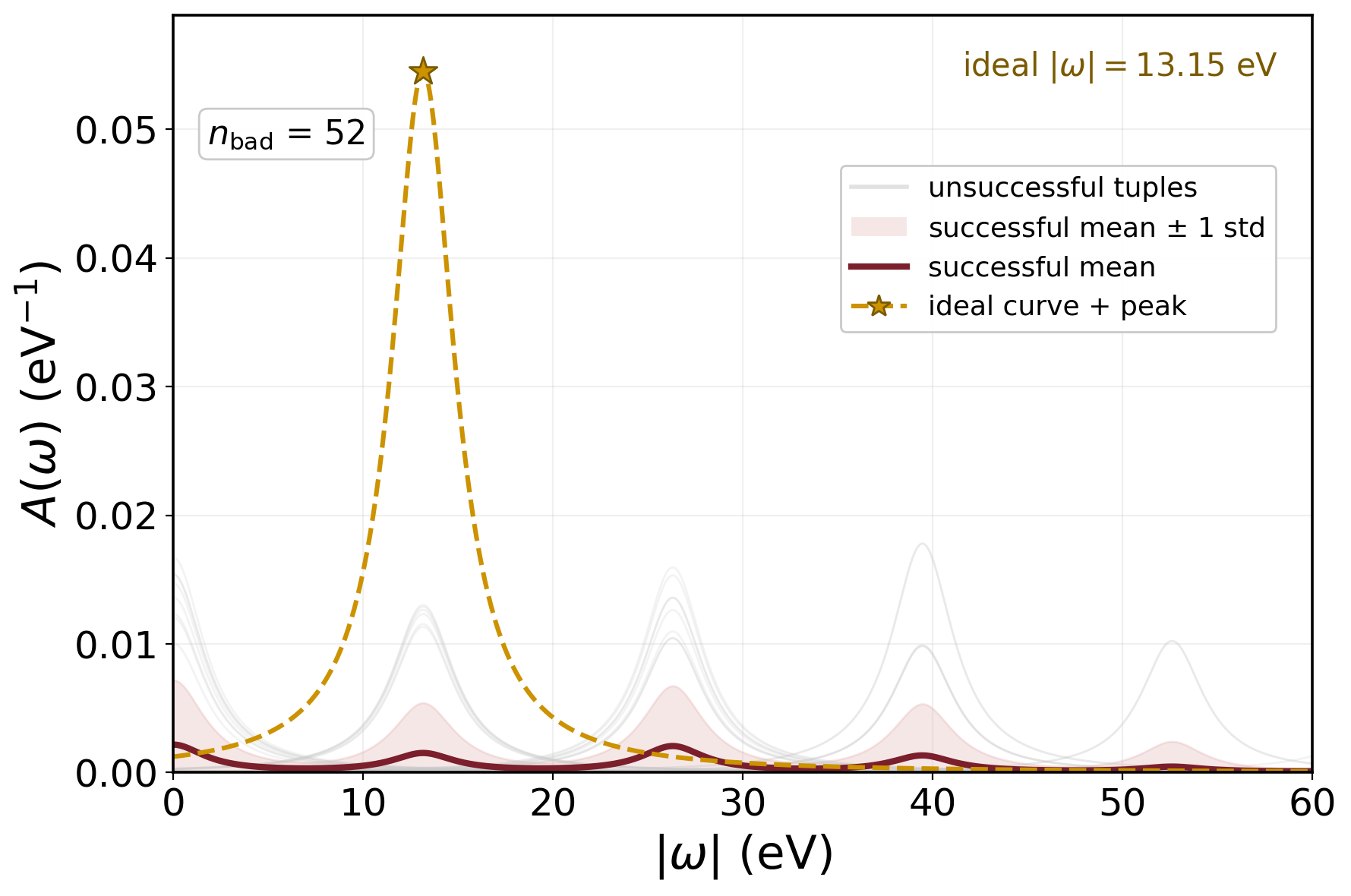}
\caption{\texttt{ibm\_osaka}.}
\end{subfigure}
\\[0.8em]
\begin{subfigure}[t]{0.48\textwidth}
\centering
\includegraphics[width=\linewidth]{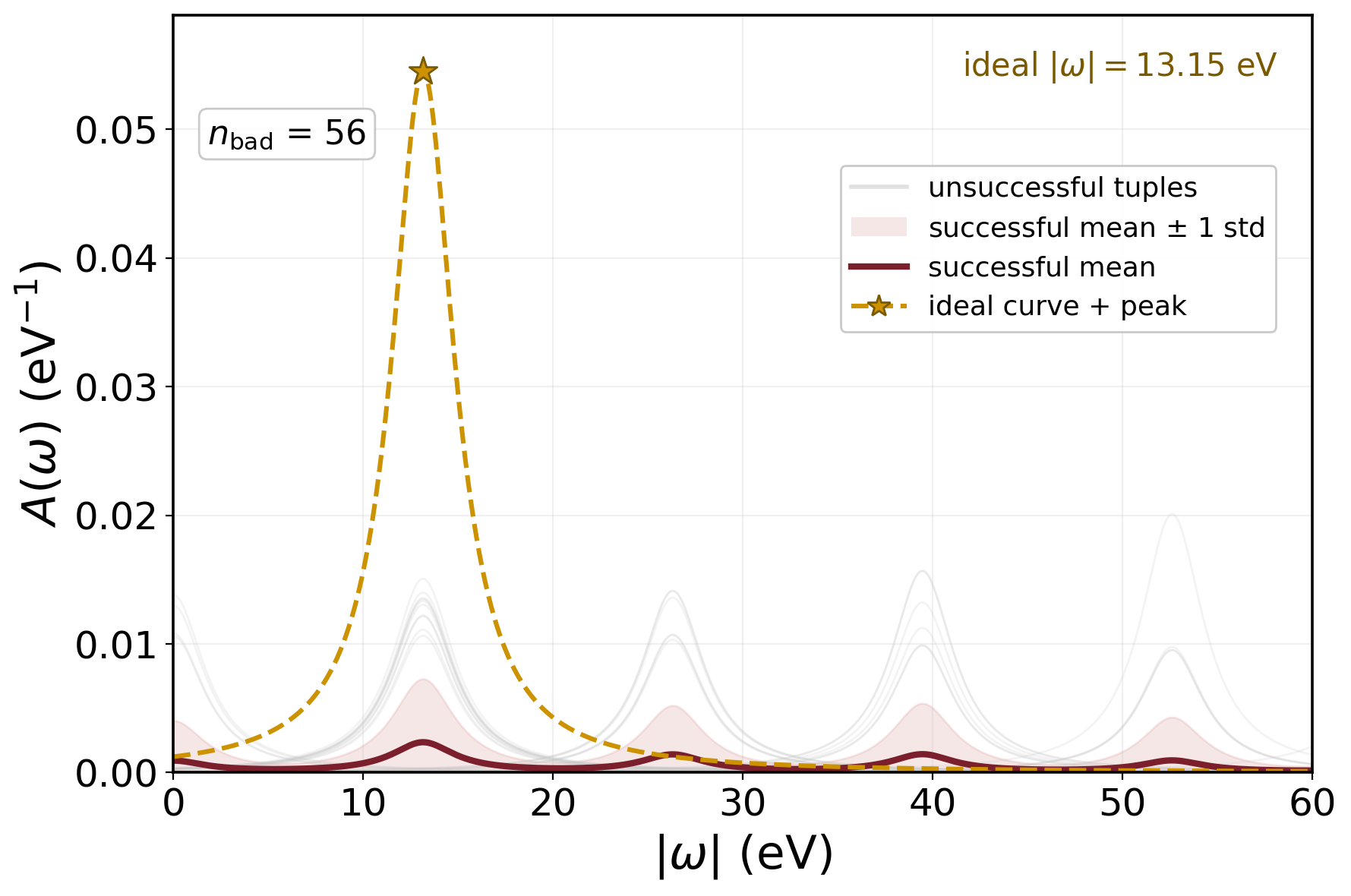}
\caption{\texttt{ibm\_kawasaki}.}
\end{subfigure}
\hfill
\begin{subfigure}[t]{0.48\textwidth}
\centering
\includegraphics[width=\linewidth]{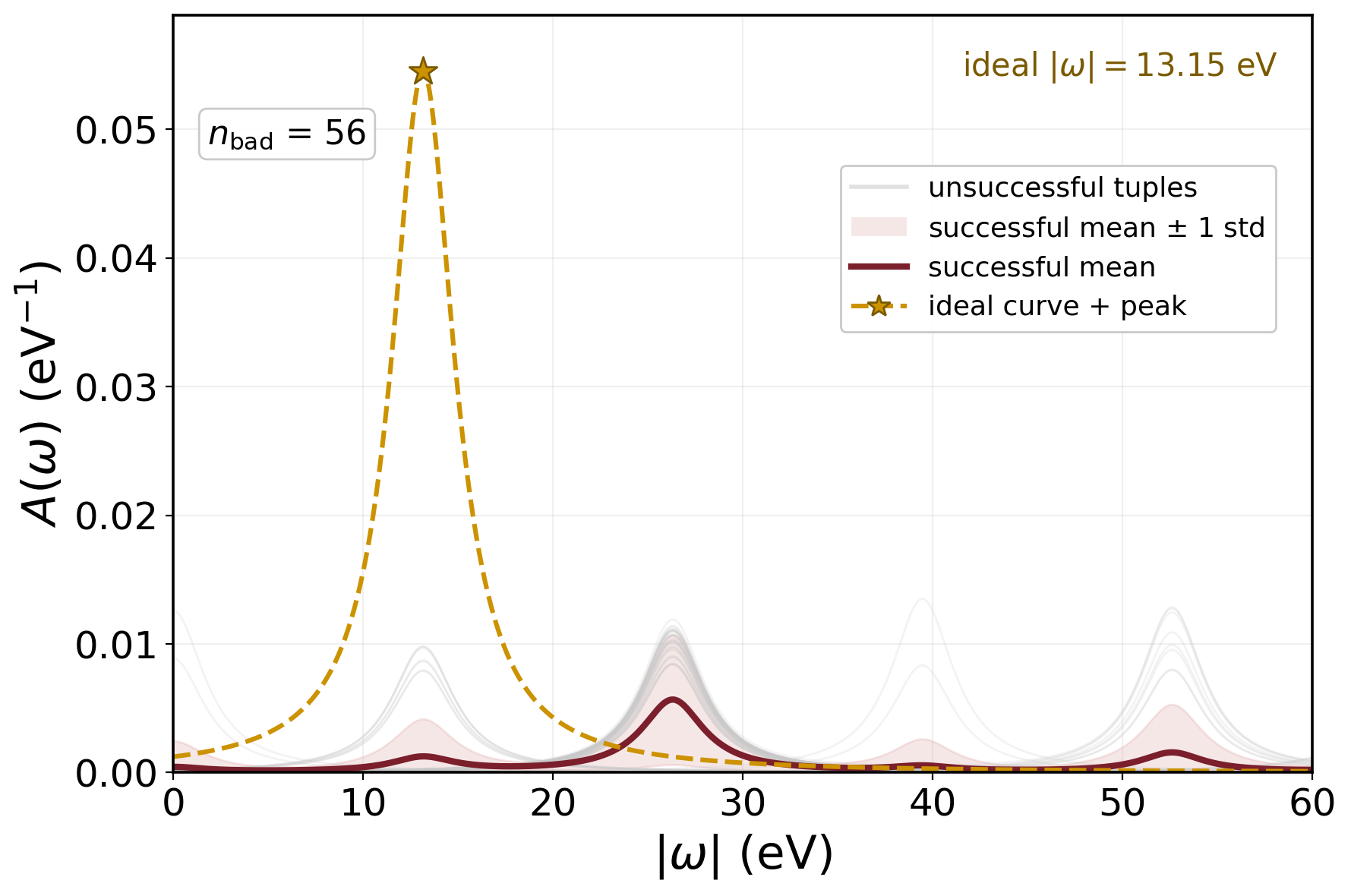}
\caption{\texttt{ibm\_kyoto}.}
\end{subfigure}
\caption{Unsuccessful tuple spectral diagnostics. Gray curves denote individual
unsuccessful reconstructions, the dark red curve and band show the mean and
one standard deviation envelope, and the gold reference marks the ideal peak at
$\abs{\omega}=13.15\,\mathrm{eV}$.}
\label{fig:results_spectral_bad_uq}
\end{figure}

% \subsection{Green's/QSVT Reliability and Spectral Diagnostics}
All four Green's/QSVT backends reach the target at least once, so the best
value alone would tie them. Hit rate and time to first good reconstruction
separate the backends: Brisbane has 51 hits, Osaka 48, Kawasaki 44, and Kyoto
44 out of 100 evaluations; Kyoto ranks below Kawasaki because its first good
reconstruction appears at evaluation 15 rather than evaluation 1. On successful
evaluations, all four backends recover the H$_2$ ionization peak near
$\abs{\omega}=13.15\,\mathrm{eV}$, but the peak weights remain below the ideal
orbital 2 height of about $0.0545~\mathrm{eV}^{-1}$. Figure~\ref{fig:results_spectral_bad_uq}
shows the complementary unsuccessful spectra omitted from the main text.
The gray traces show what failure looks like on each backend: unsuccessful
phase vectors displace the reconstructed peak away from
$\abs{\omega}=13.15\,\mathrm{eV}$, suppress its weight into the background,
or scatter spectral mass across spurious frequencies. The dark red mean and
band show that these failures are not mild perturbations of the good spectra
but a qualitatively different regime, which is why the good or bad
classification of \S\ref{sec:metrics_restructured} works as a meaningful
discriminator rather than as an arbitrary threshold. Comparing panels also
supports the tie break of \S\ref{sec:results_restructured}.2: Kawasaki and
Kyoto share the same hit rate and their failure clouds look similar, so the
discriminating information sits in when each backend first reaches the good
regime rather than in how its failures look.

\begin{figure}[tbp]
\centering
\begin{subfigure}[t]{0.44\textwidth}
\centering
\includegraphics[width=\linewidth]{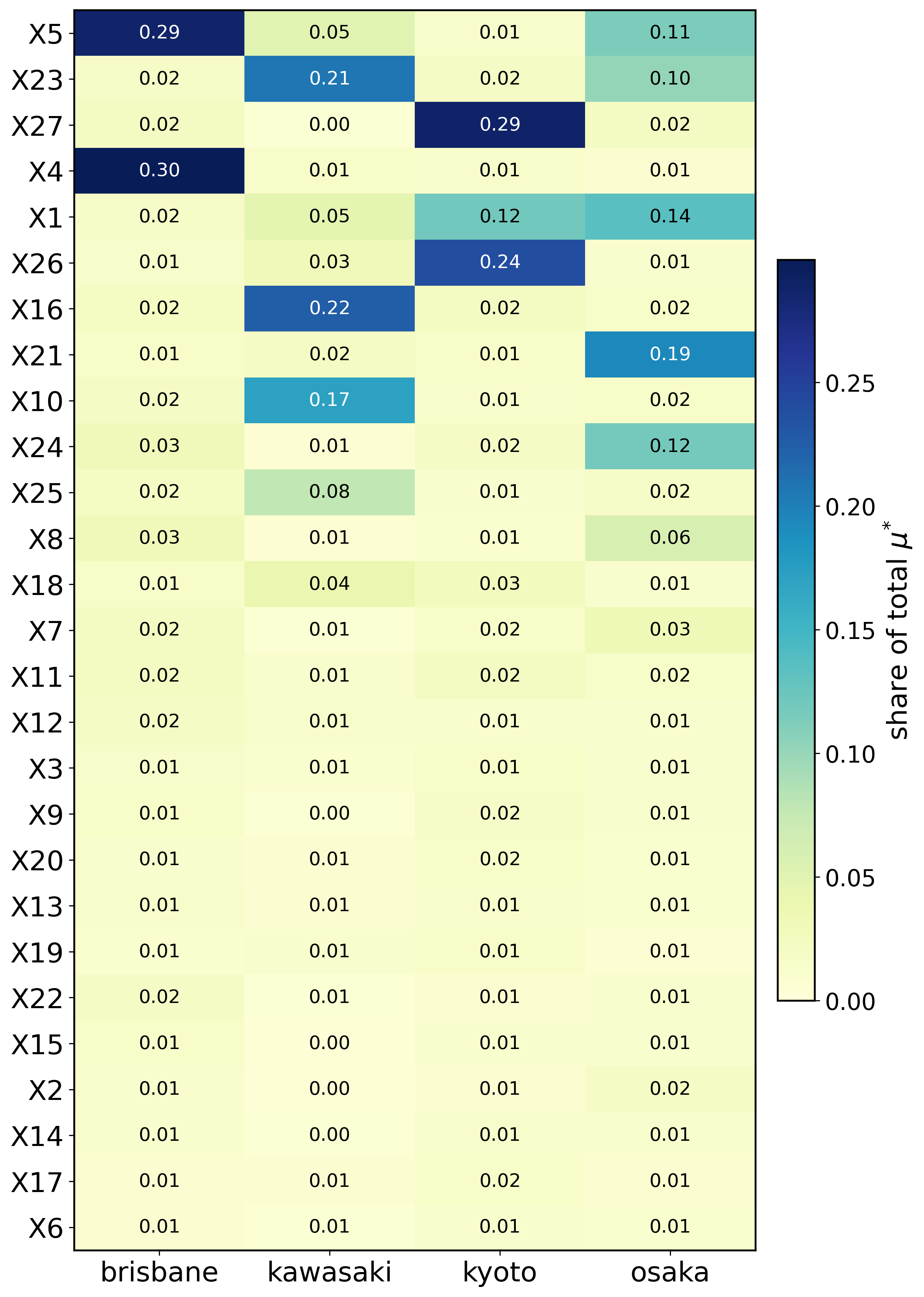}
\caption{Morris ($\mu^*$).}
\end{subfigure}
\hfill
\begin{subfigure}[t]{0.44\textwidth}
\centering
\includegraphics[width=\linewidth]{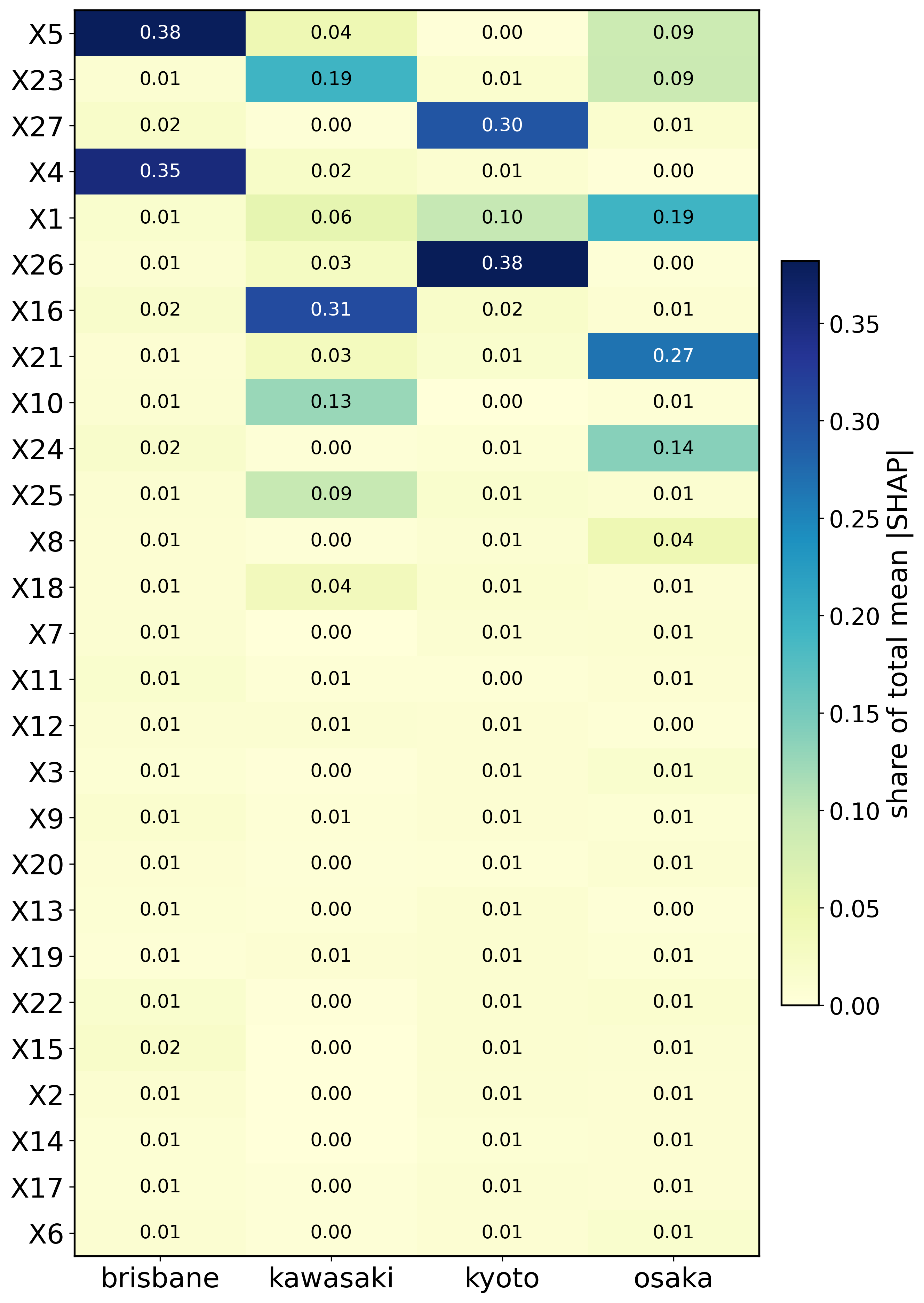}
\caption{SHAP importance.}
\end{subfigure}

\vspace{0.6em}
\begin{subfigure}[t]{0.44\textwidth}
\centering
\includegraphics[width=\linewidth]{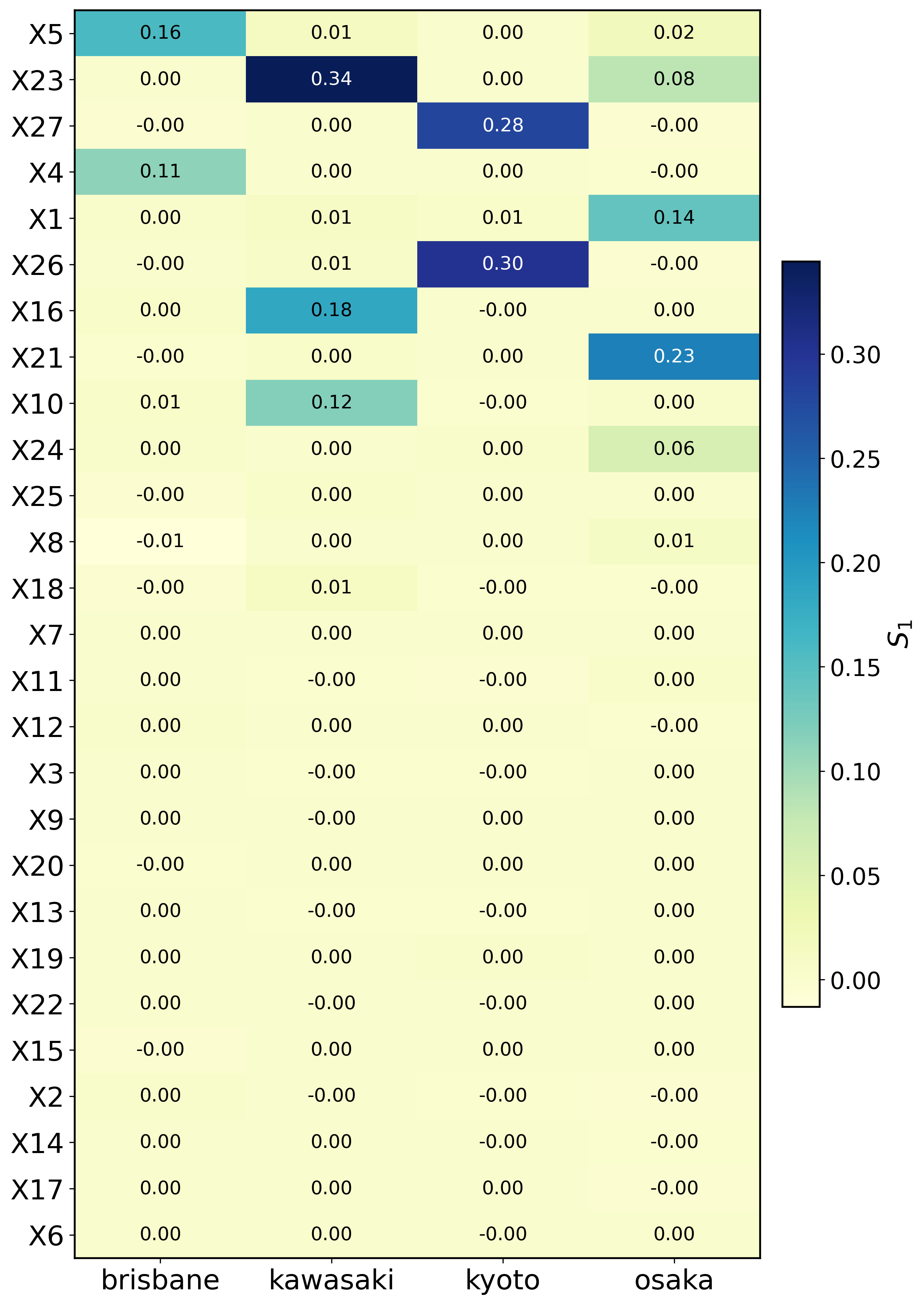}
\caption{Sobol first order indices ($S_1$).}
\end{subfigure}
\hfill
\begin{subfigure}[t]{0.44\textwidth}
\centering
\includegraphics[width=\linewidth]{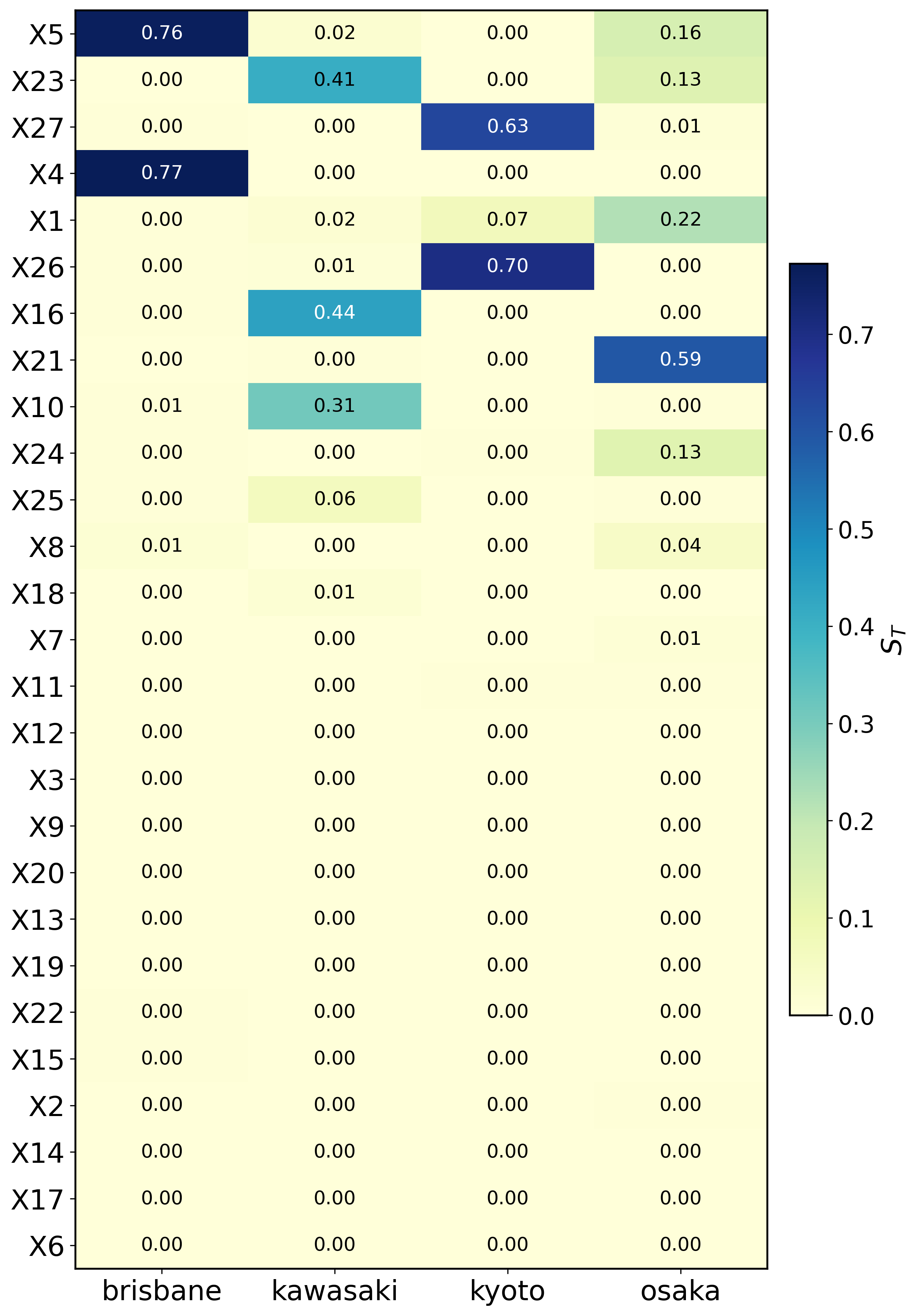}
\caption{Sobol total order indices ($S_T$).}
\end{subfigure}
\caption{Cross backend sensitivity summaries. The four heatmaps share one
row ordering, sorted by the mean Morris share across backends, so a
parameter occupies the same row in every panel. A recurring phase subset
appears across several backends, but the ordering and relative importance
remain backend specific.}
\label{fig:results_sensitivity}
\end{figure}

% \subsection{Sensitivity, Density, and Ranking Structure}
The dominant QSVT phase coordinates differ by backend. Brisbane emphasizes
$X4$, $X5$, $X8$, and $X24$; Kawasaki emphasizes $X16$, $X23$, $X10$, $X25$,
and $X5$; Kyoto emphasizes $X26$, $X27$, and $X1$; Osaka emphasizes $X21$,
$X1$, $X24$, $X23$, and $X5$. The top five parameters carry about $66\%$ to
$73\%$ of the Morris sensitivity mass, so the search behaves as effectively
low dimensional but not backend independent. Figure~\ref{fig:results_sensitivity}
compares this structure across Morris, SHAP, and Sobol summaries. The three
estimator families agree on the headline structure while disagreeing in
detail, which matches their construction in \appOffline{}: Morris and Sobol
emphasize variance driving coordinates, while SHAP redistributes importance
across correlated features. Coordinates that stay on top across all three
summaries for one backend, such as $X26$ and $X27$ on Kyoto, mark reliable
calibration targets for that backend, and coordinates that recur across
several backends, such as $X1$, $X5$, $X23$, and $X24$, mark the shared low
dimensional core claimed in \S\ref{sec:results_restructured}.3. The figure
therefore gives the direct evidence behind the statement that the search is
effectively low dimensional in a backend specific way.

\begin{figure}[tbp]
\centering
\begin{subfigure}[t]{0.36\textwidth}
\centering
\includegraphics[width=\linewidth]{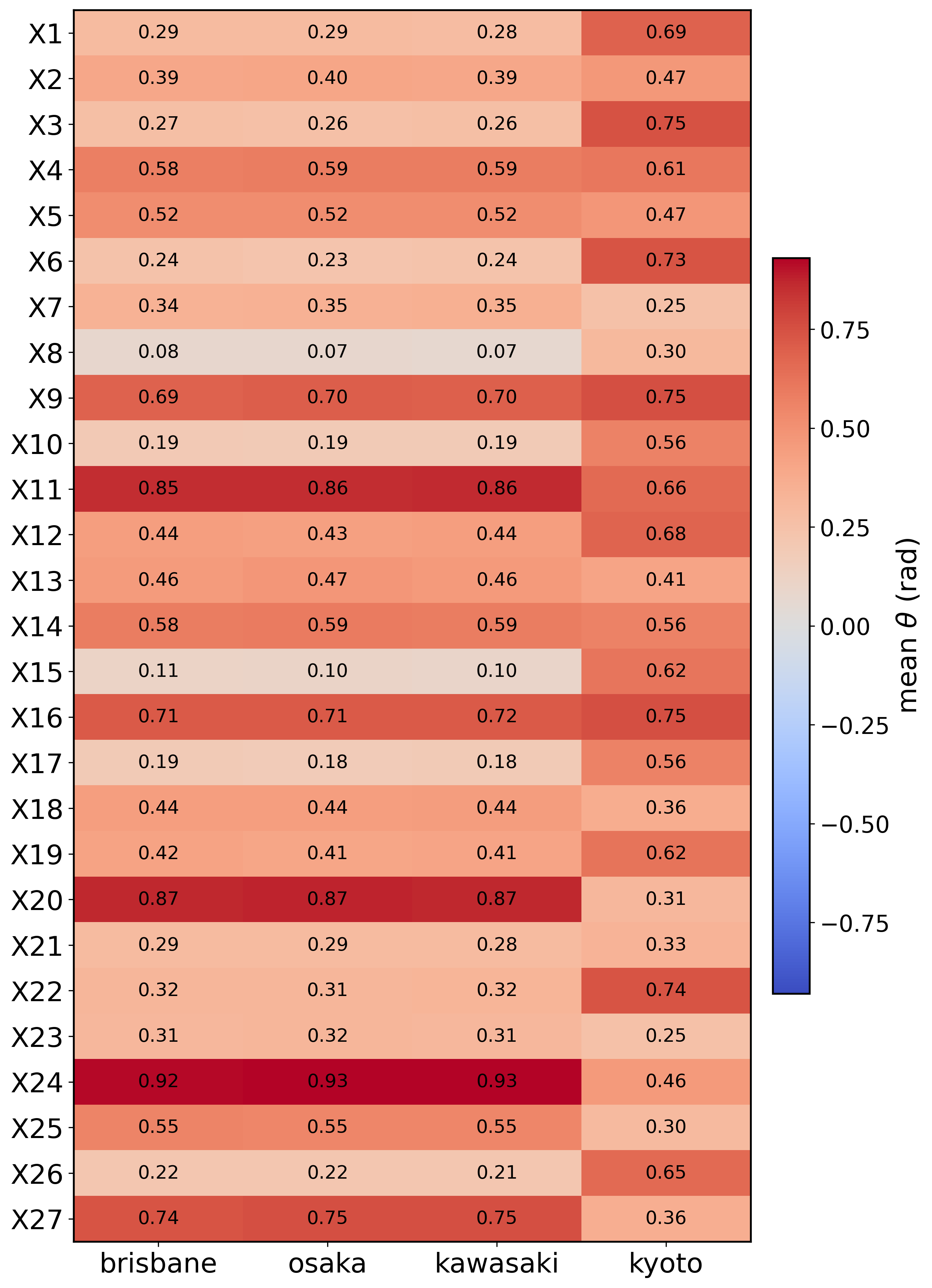}
\caption{Cross backend parameter means}
\end{subfigure}
\hfill
\begin{subfigure}[t]{0.63\textwidth}
\centering
\includegraphics[width=\linewidth]{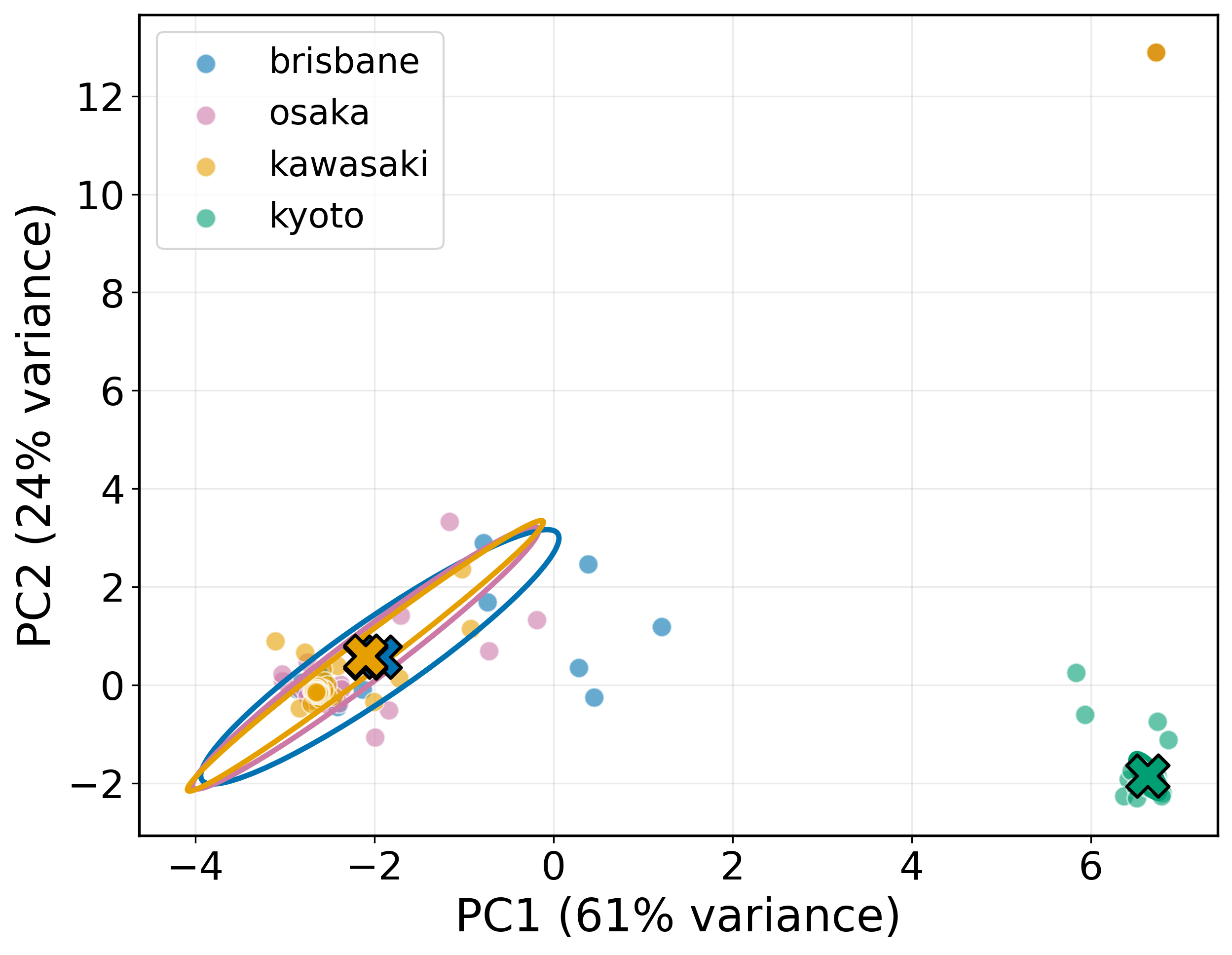}
\caption{PCA embedding of robust phase regions; crosses mark backend
centroids and ellipses the one standard deviation contour of each backend's
good tuple cloud}
\end{subfigure}
\caption{Density model summaries. Backend robust regions have different centers
and partial overlap rather than one universal robust basin.}
\label{fig:results_density}
\end{figure}

The density analysis shows that the robust region is not a single isolated
point. Brisbane, Kawasaki, and Osaka have similar concentration values
($0.762$ to $0.765$) and peak to mean ratios near $10$. Kyoto has lower
concentration ($0.740$) and a smaller peak to mean ratio ($7.49$), consistent
with its slower access to the good regime. Figure~\ref{fig:results_density}
shows the corresponding backend centers and robust region embedding. The
parameter means heatmap in panel (a) shows that the four backends settle on
visibly different phase centers, so noise compensation is not one universal
correction. The PCA embedding in panel (b) shows partially overlapping but
distinct robust regions: the centroid crosses and one standard deviation
ellipses make the cluster geometry explicit, with Kyoto's cloud sitting
apart from the Brisbane,
Kawasaki, and Osaka cluster; this separation matches Kyoto's lower
concentration ($0.740$ versus $0.762$ to $0.765$) and its smaller peak to
mean ratio ($7.49$). Together the two panels back the
\S\ref{sec:results_restructured}.3 claim that the robust region is backend
specific in both location and geometry: a phase setting calibrated on one
backend does not transfer unchanged to another, which is exactly the kind of
backend characterization signal this benchmark is designed to expose.

\begin{figure}[tbp]
\centering
\begin{subfigure}[t]{0.53\textwidth}
\centering
\includegraphics[width=\linewidth]{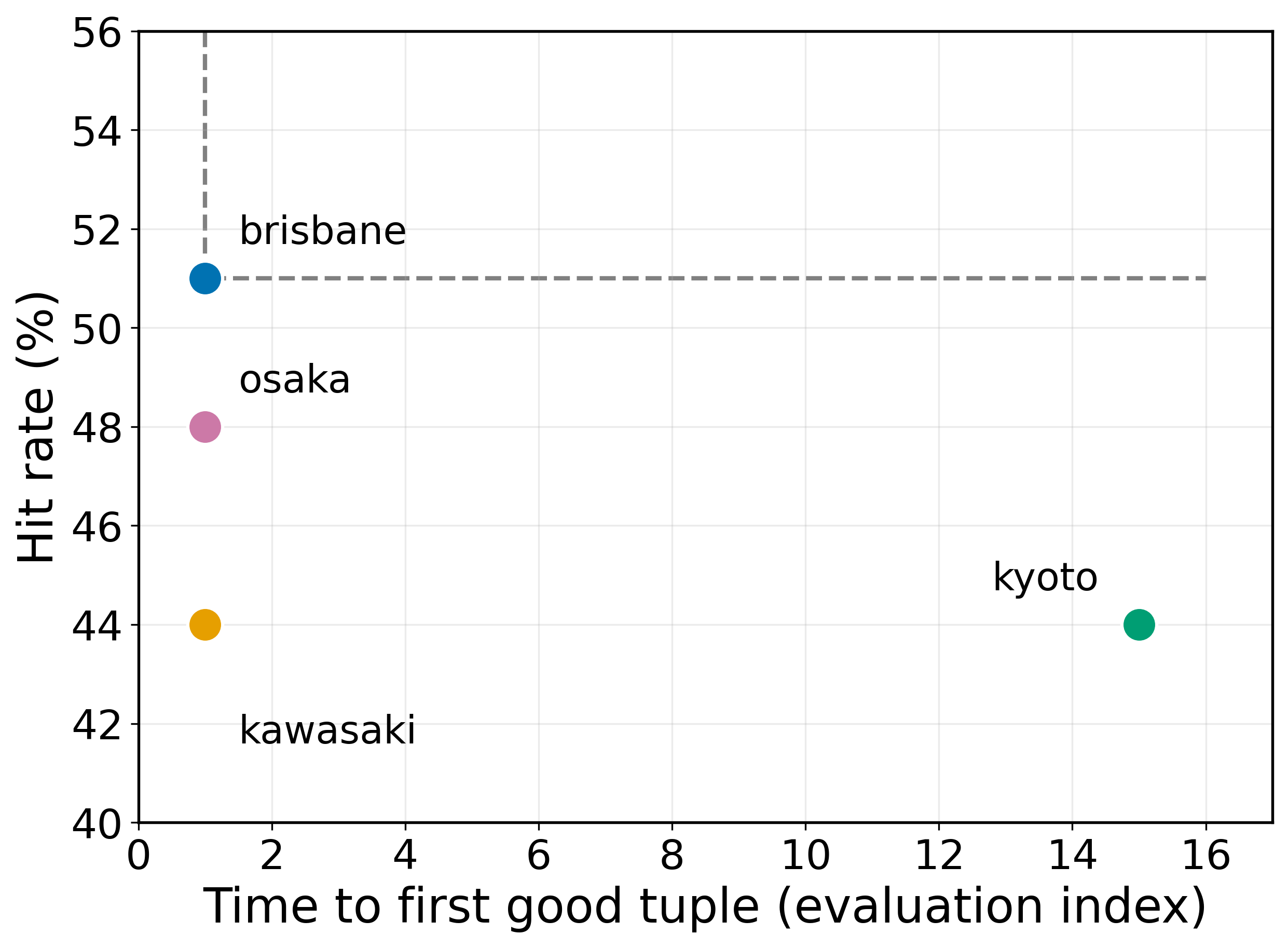}
\caption{Quality versus speed Pareto view.}
\end{subfigure}
\hfill
\begin{subfigure}[t]{0.46\textwidth}
\centering
\includegraphics[width=\linewidth]{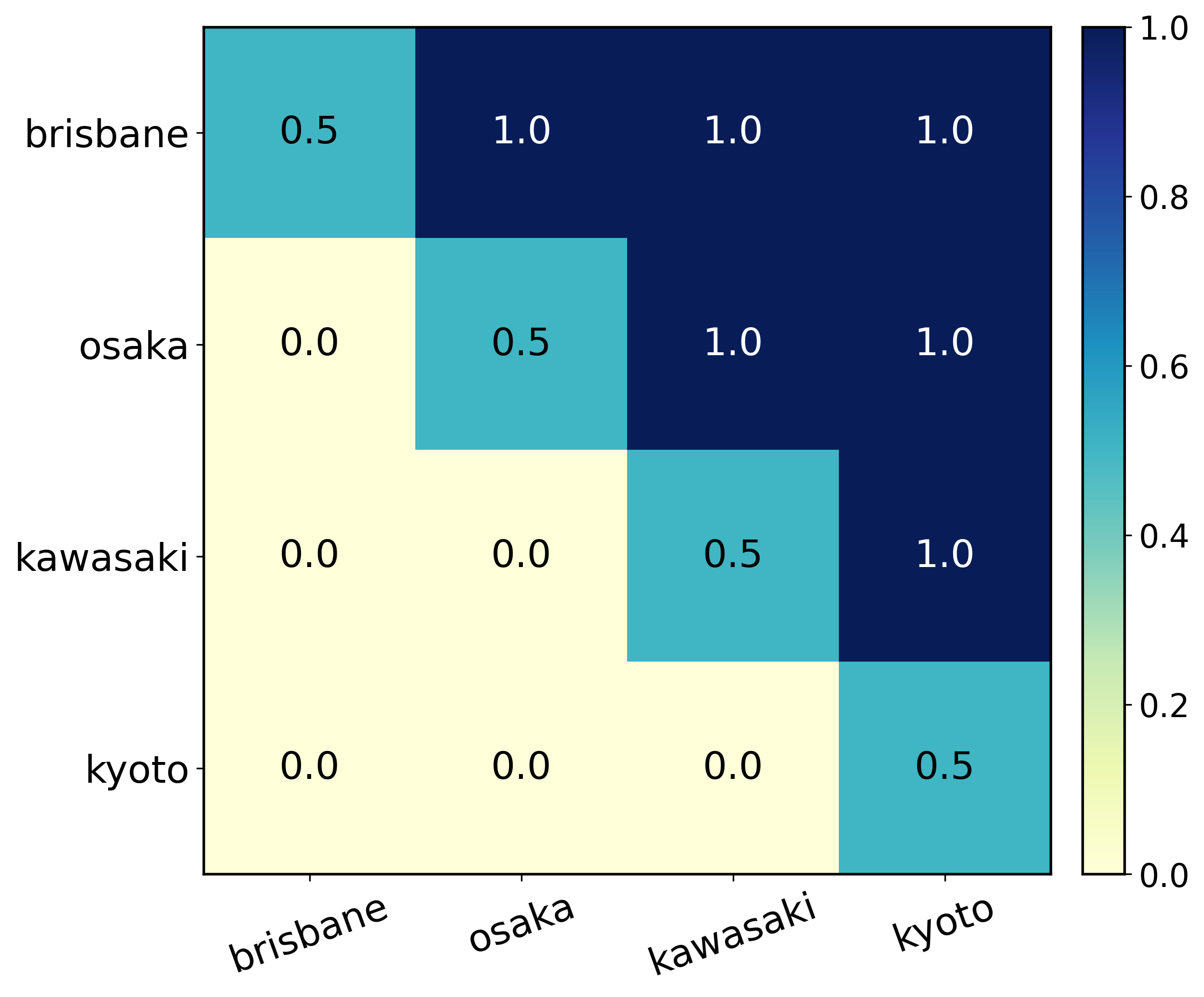}
\caption{Pairwise win probabilities.}
\end{subfigure}
\caption{Application level Green's/QSVT benchmark views supporting
Table~\ref{tab:greens_backend_rank}. Brisbane is strongest for this task,
Osaka is the closest competitor, Kawasaki is intermediate, and Kyoto is
limited by delayed access to the good regime.}
\label{fig:results_benchmark}
\end{figure}

The Green's/QSVT application ranking combines hit rate, time to first good
reconstruction, spectral recovery, sensitivity, and density.
Table~\ref{tab:greens_backend_rank} states the resulting order, Brisbane,
Osaka, Kawasaki, and Kyoto, together with the two reliability numbers that
produce it: Brisbane leads on hit rate, and Time-to-Good separates the equal
hit rates of Kawasaki and Kyoto. Figure~\ref{fig:results_benchmark} adds the
two graphical views behind the table. Panel (a) plots the quality versus
speed tradeoff and shows that Kyoto's delayed first success places it behind
Kawasaki even at equal hit rate. Panel (b) reports pairwise win
probabilities, which quantify how confidently one backend outperforms
another across the full evaluation histories rather than at a single summary
number. Together the table and the two panels form the evidence base for the
application level ordering and the tie break between Kawasaki and Kyoto
stated in \S\ref{sec:results_restructured}.2.

\begin{table}[tbp]
\centering
\caption{Application level Green's/QSVT backend ranking over the $100$
evaluations per backend. Hit Rate is the fraction of evaluations classified
as good tuples under Eq.~\eqref{eq:guided_main_restructured}, and
Time-to-Good is the evaluation index of the first good tuple; Time-to-Good
breaks the tie between the equal hit rates of Kawasaki and Kyoto.}
\label{tab:greens_backend_rank}
\begin{tabular}{lccc}
\toprule
\textbf{Backend} & \textbf{Hit Rate (\%)} & \textbf{Time-to-Good} & \textbf{Rank}\\
\midrule
Brisbane & 51 & 1 & 1\\
Osaka & 48 & 1 & 2\\
Kawasaki & 44 & 1 & 3\\
Kyoto & 44 & 15 & 4\\
\bottomrule
\end{tabular}
\end{table}

\section{VQA Parameter Density Summaries}
\label{app:vqa_parameter_summaries}
This section records the cross VQA summaries produced by the
theta distribution stage of the workflow when VI is used as the fixed
posterior refinement baseline. The purpose of this section is not to replace
the main text benchmark figures, but to show what the parameter distribution
analysis contributes beyond the aggregate quality plots. In particular, it
separates three questions that are easy to conflate:
\begin{enumerate}[leftmargin=5mm]
  \item how the ``good'' set is defined and how large it is,
  \item how predictive the retained theta samples are for task quality,
  \item what kinds of marginal parameter families best fit the robust region.
\end{enumerate}

\begin{figure}[tbp]
\centering
\begin{subfigure}[t]{0.45\textwidth}
\centering
\includegraphics[width=\linewidth]{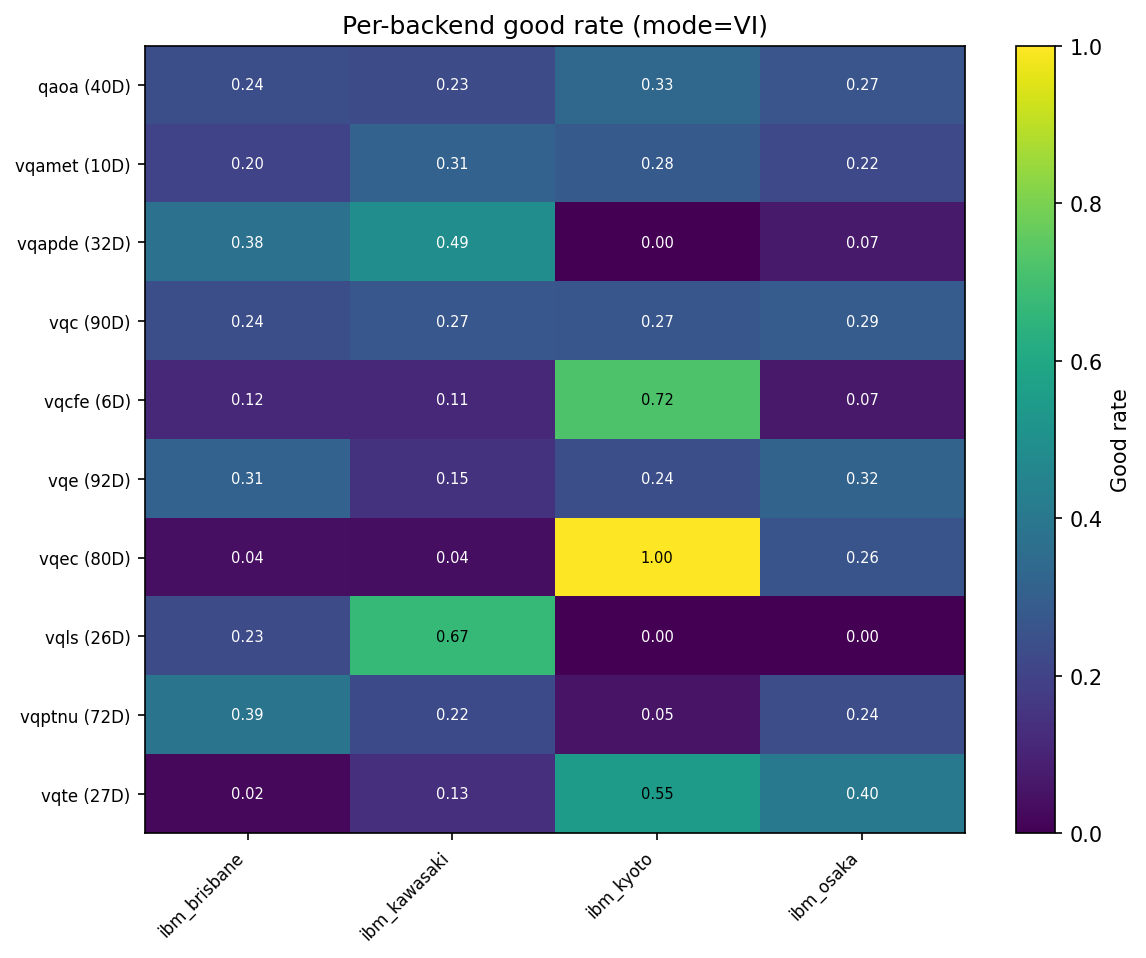}
\caption{Per backend good rate heatmap under VI.}
\end{subfigure}
\hfill
\begin{subfigure}[t]{0.54\textwidth}
\centering
\includegraphics[width=\linewidth]{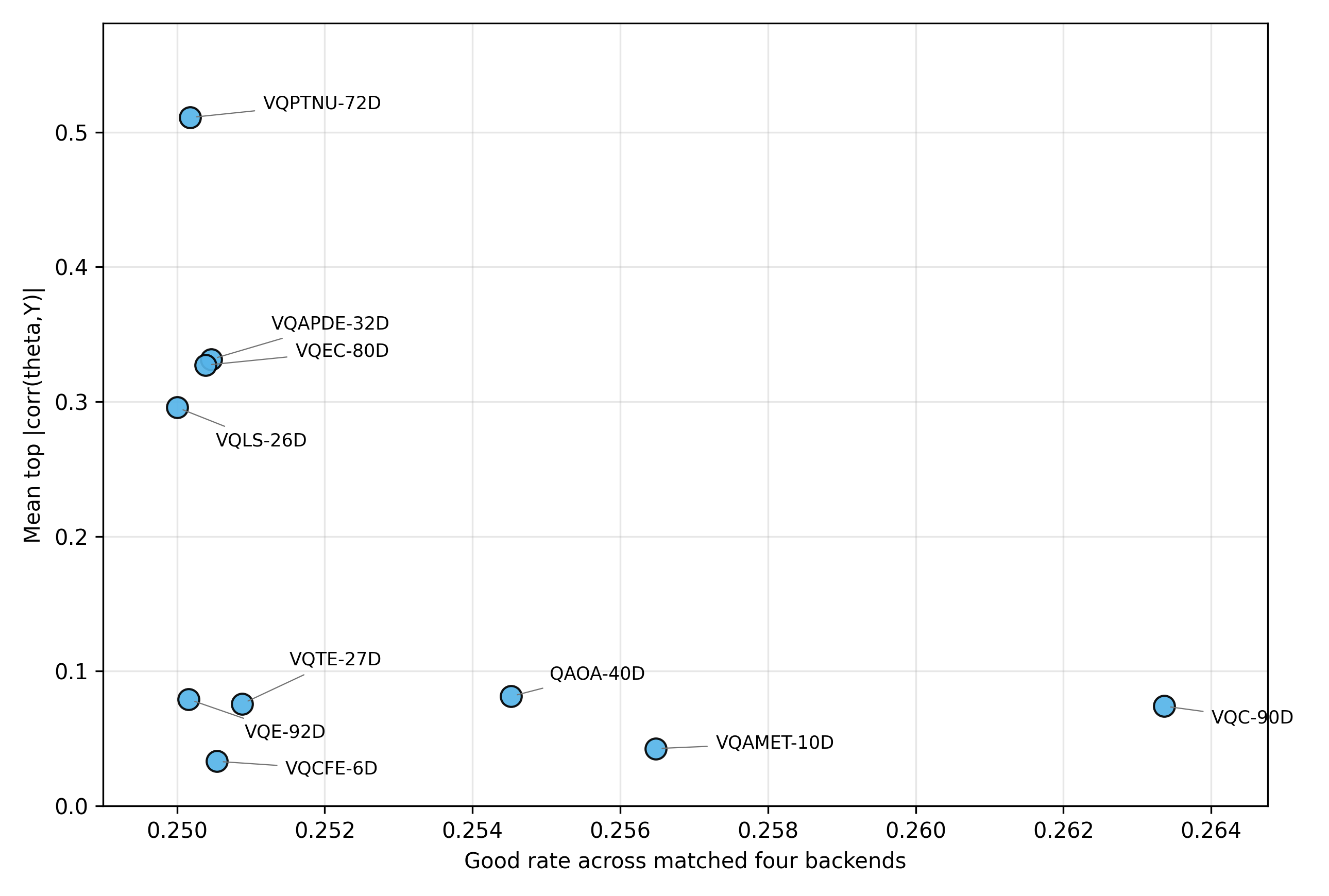}
\caption{Predictive quality / correlation characterization map.}
\end{subfigure}

\vspace{0.6em}
\begin{subfigure}[t]{0.70\textwidth}
\centering
\includegraphics[width=\linewidth]{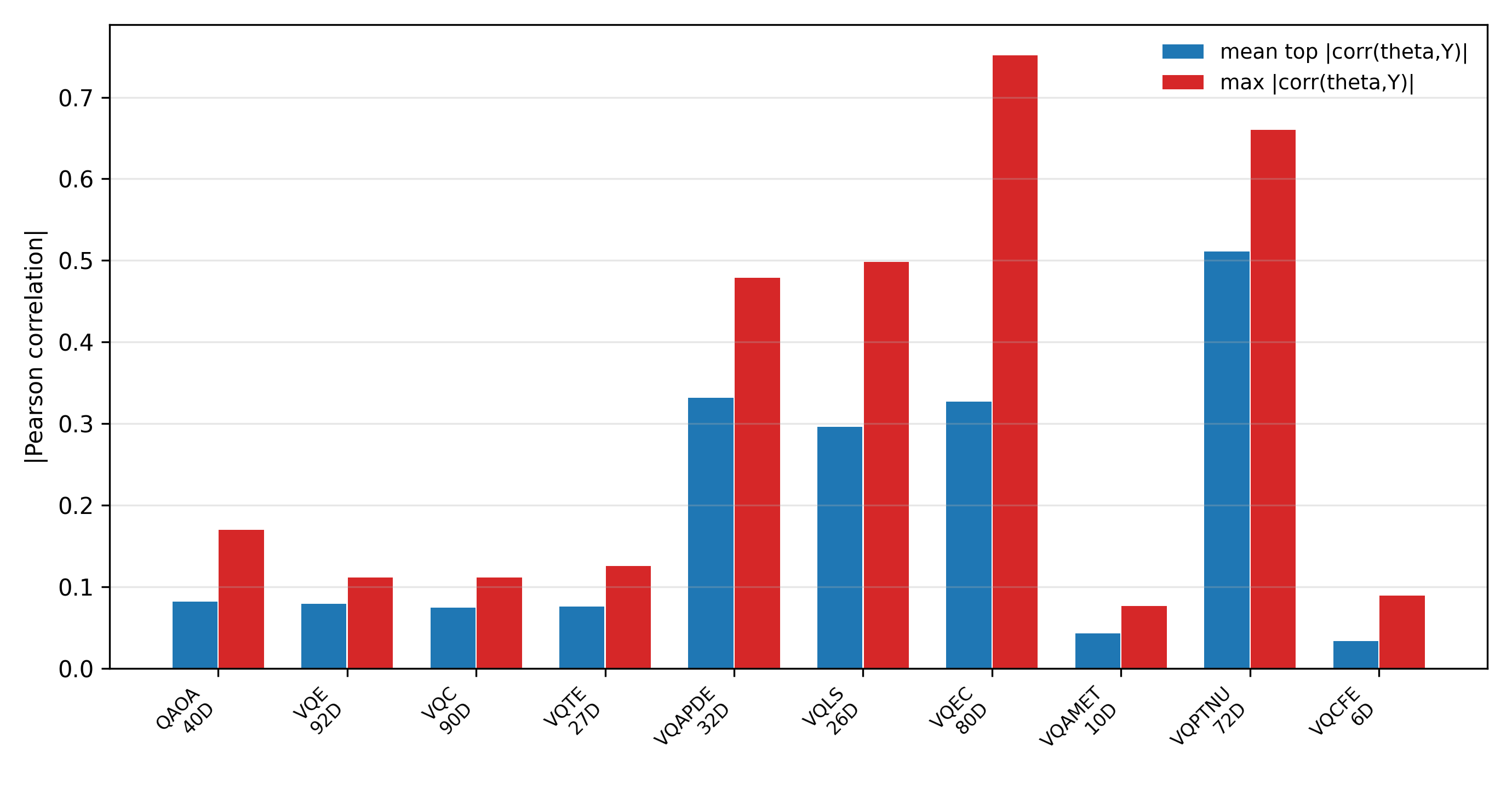}
\caption{Mean and maximum parameter output correlation strength.}
\end{subfigure}
\hfill
\begin{subfigure}[t]{0.70\textwidth}
\centering
\includegraphics[width=\linewidth]{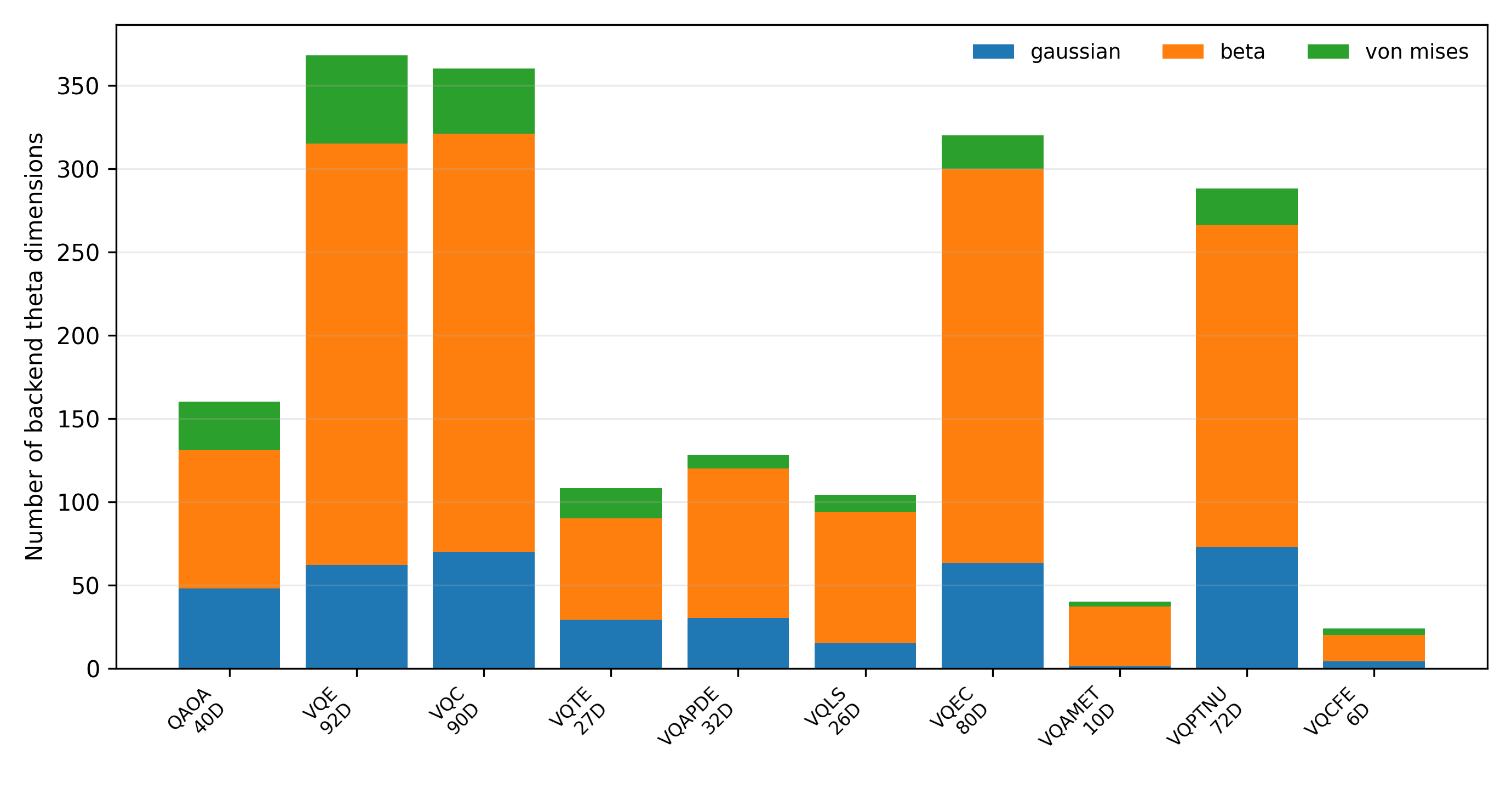}
\caption{Best marginal family counts within the robust region.}
\end{subfigure}
\caption{VI based VQA parameter distribution summaries. Panel (a) reports the
backend local good rate under the common robust region rule. Panel (b) shows
predictive quality and correlation diagnostics. Panel (c) reports mean and
maximum parameter output correlation strengths. Panel (d) counts the marginal
families selected for robust region fits.}
\label{fig:app_vqa_parameter_summary}
\end{figure}

Figure~\ref{fig:app_vqa_parameter_summary} should be read as the VQA analogue
of the robust phase density discussion in the Green's/QSVT branch. Because the
good set is defined by a common quantile rule, each workload keeps an
overall good rate close to $0.25$. The backend local heatmap then shows where
that retained mass actually sits: Kyoto dominates VQCFE, VQEC, and VQTE,
Kawasaki dominates VQLS, and the easier workloads remain more diffuse across
the four backends. Once that normalization is fixed, the predictive and
correlation plots become informative. VQPTNU sits at the highest mean
$|\mathrm{corr}(\theta,Y)|$ value ($0.190$), followed by VQEC, VQAPDE, and
VQTE, whereas VQE and VQC remain much weaker. The marginal family counts tell a
complementary story: beta marginals dominate many workloads, but the periodic family remains prominent for VQC and VQTE, which is consistent with
their more structured angular parameterizations. These panels supply the
evidence for two main text statements. The correlation strengths in panel (c)
back the \S\ref{sec:results_restructured}.3 claim that VQPTNU and VQEC show
the strongest parameter to output coupling, and the good rate heatmap in
panel (a) is the density level counterpart of the workload winners of
\S\ref{sec:results_restructured}.1, because the backends that lead a
workload also hold most of its retained robust mass.

\section{VQA Benchmark and Noise Views}
\label{app:vqa_benchmark_supplement}
\begin{figure}[tbp]
\centering
\begin{subfigure}[t]{0.43\textwidth}
\centering
\includegraphics[width=\linewidth]{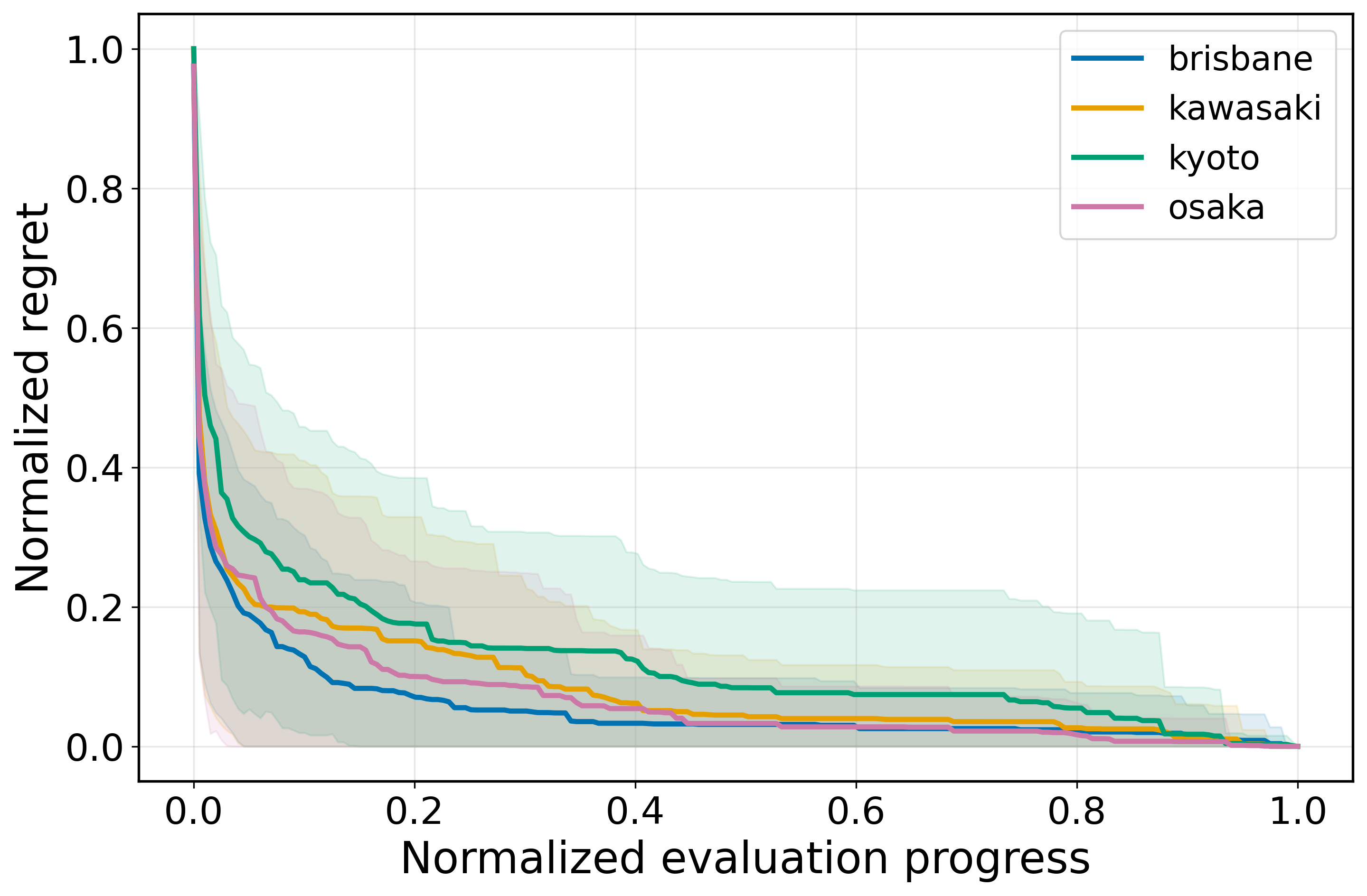}
\caption{Aggregated backend regret trajectories.}
\end{subfigure}
\hfill
\begin{subfigure}[t]{0.56\textwidth}
\centering
\includegraphics[width=\linewidth]{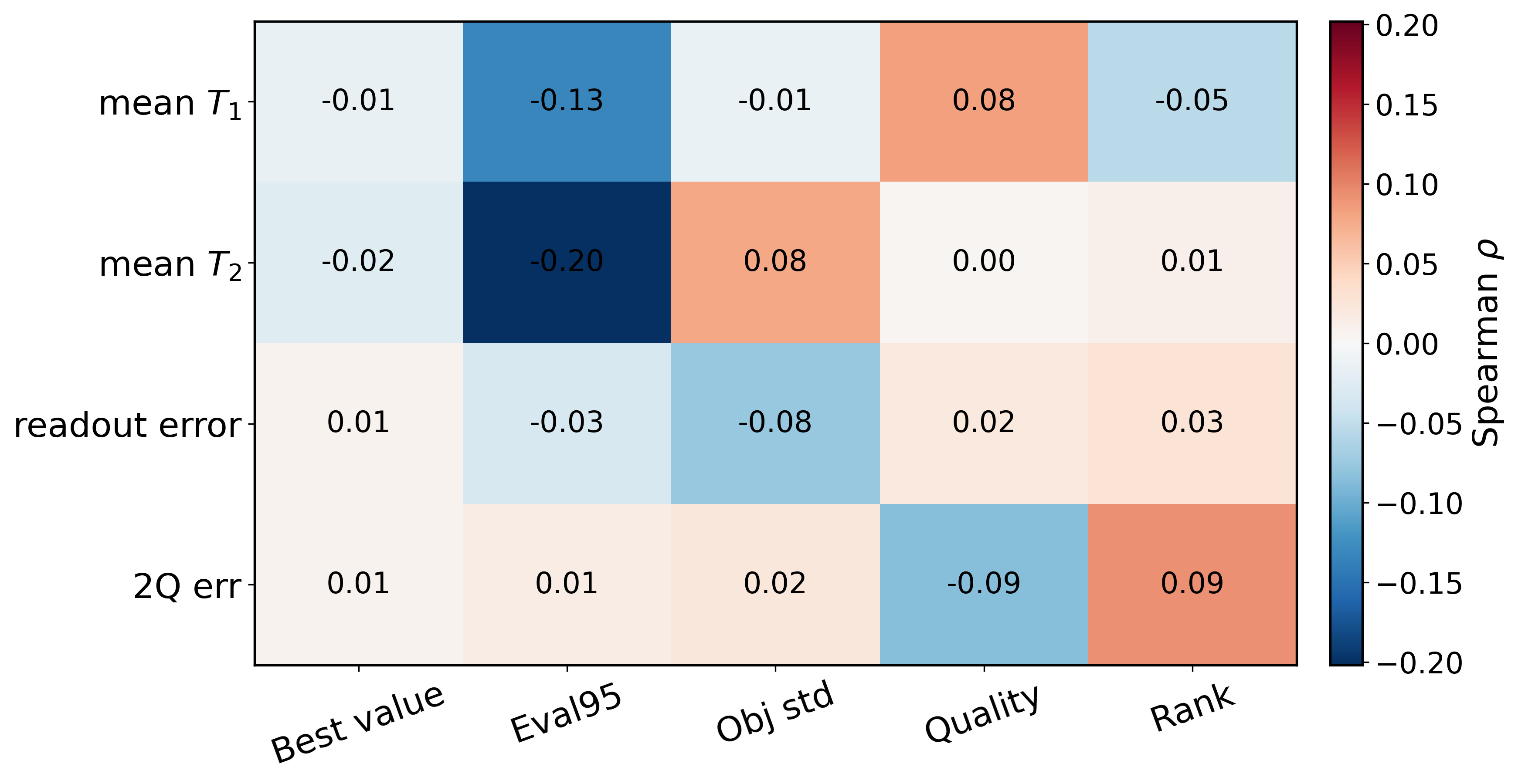}
\caption{Spearman summary from static fake backend property proxies.}
\end{subfigure}
\caption{Backend level summaries for the VQA branch. The regret
panel provides the aggregate optimization view behind the quality speed
benchmarking discussion, showing that the mean regret drops rapidly for all
backends before separating in the long tail. The noise panel summarizes
Spearman correlations between static fake backend property proxies and
task level performance metrics over the four backends. The correlations
are weak overall, with the clearest visible trend being a modest negative
association between mean $T_2$ and time to target; the panel is therefore
best read as qualitative context rather than as a stand alone ranking
explanation. In the heatmap, `Best value' is the best observed objective,
`Eval95' is \texttt{eval\_to\_target\_95}, `Obj std' is the objective
standard deviation, `Quality' is normalized quality, `Rank' is the within task
backend rank, and `2Q err' denotes the available two qubit gate error proxy
from the fake backend metadata.}
\label{fig:app_vqa_benchmark_noise}
\end{figure}

This appendix gathers two auxiliary outputs of the VQA benchmark stage that are
useful for interpretation but too detailed for the main narrative. The first is
the aggregated regret view across backends. The second is the
noise performance alignment heatmap produced from the four fake backends
property summaries and the corresponding performance table. Figure~\ref{fig:app_vqa_benchmark_noise}
collects both views. The regret trajectories in panel (a) give the convergence
evidence behind the \S\ref{sec:results_restructured}.1 statement that VQE,
QAOA, and VQC carry the longest optimization tails: mean regret falls quickly
for every backend and then separates in the tail, where the slow workloads
keep the curves from flattening. The Spearman panel in (b) shows that no
single static noise proxy explains the application level rankings; the
strongest visible association is a modest negative link between mean $T_2$
and time to target, consistent with the qualitative reading stated in
\S\ref{sec:limitations_restructured}. The weakness of these correlations is
itself a result, because it supports the paper's position that application
level benchmarking adds information that device level summaries do not
carry.

\section{VQA Posterior Engine Sensitivity and Benchmark Views}
\label{app:vqa_posterior_engine_sensitivity}
\begin{figure}[tbp]
\centering
\includegraphics[width=1.0\textwidth]{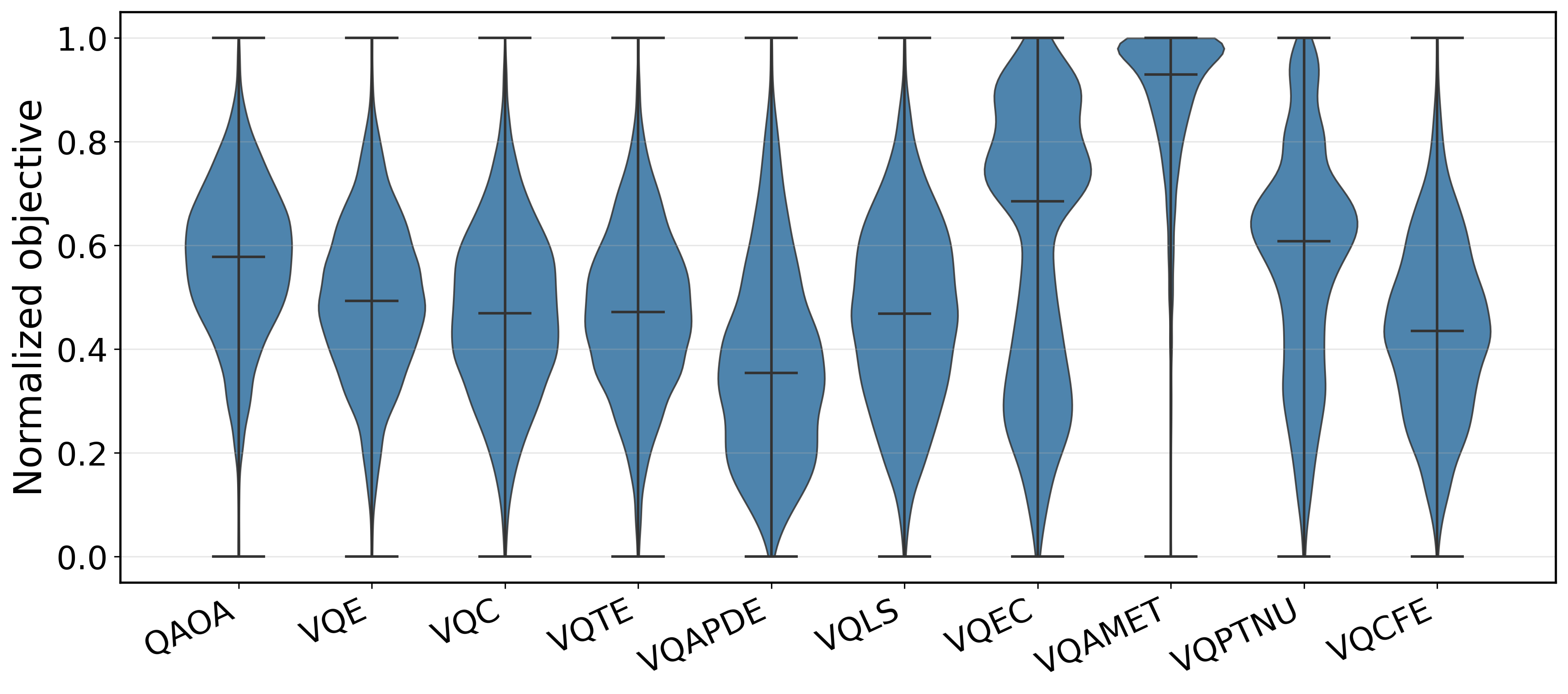}
\caption{Objective distributions for the VQA workloads under the VI
posterior refinement baseline. Each violin summarizes normalized objective
values over the four backends; larger values indicate better task
performance. The other three refinement modes (none, Langevin, and
Metropolis-Hastings) produce closely similar distributions for every
workload, so we show one representative panel and report the small engine
differences in Table~\ref{tab:vqa_posterior_engine_summary}.}
\label{fig:vqa_objective_distributions}
\end{figure}

\begin{figure}[tbp]
\centering
\begin{subfigure}[t]{0.49\textwidth}
\centering
\includegraphics[width=\linewidth]{figures/vqa_results_common4/fig5_backend_ranking.png}
\caption{Mean rank by workload.}
\end{subfigure}
\hfill
\begin{subfigure}[t]{0.5\textwidth}
\centering
\includegraphics[width=\linewidth]{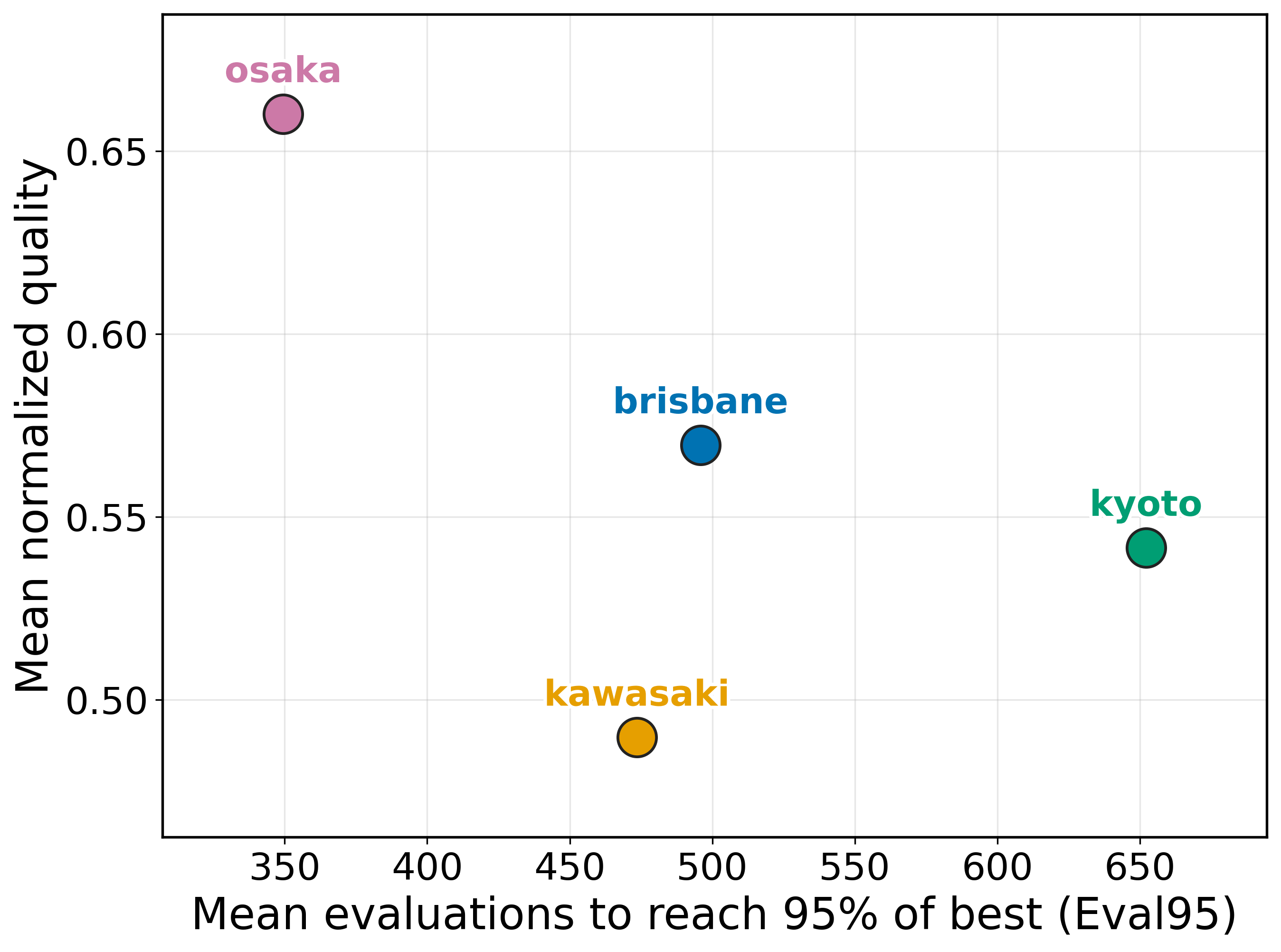}
\caption{Quality speed Pareto view.}
\end{subfigure}
\caption{Application level VQA backend benchmarks. Brisbane has the best mean
rank, while Osaka gives the strongest quality speed tradeoff on the four backends set.}
\label{fig:vqa_results_benchmark}
\end{figure}

This section records the workload by workload posterior engine comparison
that complements the VI based main text narrative. Its purpose is not to
replace the fixed VI baseline used in the main text, but to document how the
preferred posterior refinement engine changes across workloads when VI,
Metropolis-Hastings (MH), and Langevin variants are compared directly on
paired checkpoint sets. The comparison uses the best observed tuple value $y$
per common run, with larger $y$ interpreted as better because the BOpt loop is
written in maximization form. The main conclusion is simple: posterior engine
preference is workload dependent. Here, a "common run" means a workload/backend/run combination for which all compared posterior engines are
available, and the reported mean best $y$ is the arithmetic average across
those paired units. Figure~\ref{fig:vqa_objective_distributions} motivates
the comparison at the distribution level before
Table~\ref{tab:vqa_posterior_engine_summary} reports it numerically, and
Figure~\ref{fig:vqa_results_benchmark} closes the section with the benchmark
views that aggregate across engines.

Figure~\ref{fig:vqa_objective_distributions} shows the full outcome spread
per workload rather than only the mean quality. VQAMET stays tightly
concentrated near the best normalized value, while VQCFE retains broad low
quality support, so refinement choice cannot rescue an intrinsically
selective workload. The remaining modes reproduce these shapes closely;
the engine dependence that does exist is small, concentrates in QAOA, VQE,
VQC, and VQTE, and appears numerically in
Table~\ref{tab:vqa_posterior_engine_summary}. The distributions therefore
justify reading the table row by row instead of declaring one engine
globally best.

\begin{table}[tbp]
\centering
\scriptsize
\setlength{\tabcolsep}{2pt}
\caption{Appendix level posterior engine sensitivity summary for the VQA
branch. The reported statistic is the mean best observed tuple value $y$ on
paired common runs across VI, MH, and Langevin checkpoints. We include this
table as a robustness comparison only; the main text continues to use VI as
the common baseline for cross paradigm consistency with the Green's QSP/QSVT
branch.}
\label{tab:vqa_posterior_engine_summary}
\begin{tabular}{>{\raggedright\arraybackslash}p{0.17\linewidth} >{\centering\arraybackslash}p{0.16\linewidth} >{\centering\arraybackslash}p{0.15\linewidth} >{\centering\arraybackslash}p{0.18\linewidth} >{\raggedright\arraybackslash}p{0.22\linewidth}}
\toprule
Workload & Common runs & Best engine & Mean best $y$ & Interpretation \\
\midrule
QAOA & 24 & MH & 1.1471 & Essentially tied with VI; no strong engine separation. \\
VQAMET & 24 & VI & 0.000867 & VI gives the strongest paired run score. \\
VQAPDE & - & - & - & No paired VI/MH/Langevin checkpoint set across common runs. \\
VQC & 22 & Langevin & 0.28684 & Langevin is slightly stronger than MH and VI. \\
VQCFE & 25 & Langevin & 0.13430 & Langevin gives the highest fidelity style score. \\
VQE & 20 & Langevin & 1.37799 & Langevin and MH are close; VI is slightly lower. \\
VQEC & 24 & VI & 1.86123 & VI is the clearest winner in this workload. \\
VQLS & - & VI only & - & Only the VI branch exists; no three way comparison is possible. \\
VQPTNU & 17 & MH & 9.50590 & MH gives the strongest tomography mismatch score. \\
VQTE & 23 & VI & 0.11444 & VI leads over Langevin and MH. \\
\bottomrule
\end{tabular}
\end{table}

Table~\ref{tab:vqa_posterior_engine_summary} lists all ten workloads. VQLS
and VQAPDE appear with dashes in place of a comparison result because their
checkpoints do not support the same paired three way comparison used for the
other eight workloads: VQLS currently has only the VI branch, and VQAPDE
lacks a clean paired VI/MH/Langevin checkpoint set across common runs.
Keeping both rows in the table makes the coverage of the comparison explicit
while restricting engine conclusions to the workloads where the evidence is
balanced across engines. Across the eight comparable rows, VI leads three
workloads, Langevin leads three, and MH leads two, which confirms that
posterior engine preference is workload dependent and supports the fixed VI
baseline of the main text as a consistency choice rather than as a
performance claim.

Figure~\ref{fig:vqa_results_benchmark}(a) repeats the mean rank view that
also appears in Figure~\ref{fig:vqa_main_restructured} so that this section
stays readable on its own: Brisbane holds the best average rank, with Osaka,
Kyoto, and Kawasaki close behind. Panel (b) adds the quality versus speed
Pareto view, in which Osaka offers the strongest tradeoff between mean
normalized quality and evaluations to target. Read together with the violins
and the engine table, the two panels connect the posterior engine analysis
back to the aggregate conclusions of \S\ref{sec:results_restructured}.1:
engine choice moves individual workload distributions, but the backend level
ordering stays stable under the aggregation.

\section{Extended Scope, Limitations, and Threats to Validity}
\label{app:extended_scope}
The main body states the two boundaries that scope our claims, namely the use
of IBM fake backend noise models in place of live QPUs and the qualitative,
monotonic reading of the noise to performance correlations
in the main text. We add two threats to validity that the
appendix material raises directly.

First, the Green's function and VQA branches are complementary rather than
identical. We do not claim that a non-variational matrix function workload and a
family of variational workloads should produce numerically identical backend
rankings; their value lies in exposing both backend behavior and workload
selectivity, so a divergence between the two rankings reflects workload
specificity rather than an inconsistency in the method.

Second, the VQA branch is broad by design, and that breadth carries a tradeoff.
Spanning chemistry, optimization, simulation, PDEs, metrology, tomography, and
error correction style tasks gives the benchmark its generality, but full
algorithmic derivations for every workload would obscure the benchmarking
narrative. We therefore give a compact objective map, a shared circuit template, and
canonical citations for the algorithm specific derivations.

These limitations do not weaken the central result. They sharpen the
contribution: a shared uncertainty quantification methodology for application
level backend benchmarking across both variational and non-variational quantum
workloads.

% \end{appendix}

\end{document}